\documentclass[12pt]{article}
\usepackage{epsfig,amssymb} 
\usepackage{amsmath}

\usepackage{amscd,color}
\usepackage{amsmath,graphicx}
\usepackage{graphicx}
\usepackage{verbatim}
\usepackage{latexsym}
\usepackage{amsmath,amsfonts,amssymb,amsthm}
\usepackage{amsmath,amsthm}

\newcommand{\be}{\begin{equation}}
\newcommand{\bea}{\begin{eqnarray}}
\newcommand{\eea}{\end{eqnarray}}
\newcommand{\ee}{\end{equation}}

\begin{document}
\setcounter{page}{0}
\begin{titlepage}
\titlepage
\begin{center}
\LARGE{\Huge Scattering Amplitudes and Toric Geometry}
\end{center}
\vskip 1.5cm \centerline{
{\bf Antonio Amariti\footnote{\tt amariti@lpt.ens.fr}  and
Davide Forcella\footnote{\tt davide.forcella@ulb.ac.be}
}
}
\vskip 1cm
\footnotesize{
\begin{center}
$^1$ Laboratoire de Physique Th\'eorique de l'\'Ecole Normale Sup\'erieure and\\
 Istitute de Physique Th\'eorique Philippe Meyer, 24 Rue Lhomond, Paris 75005, France\\
\medskip
$^2$
Physique Th\'eorique et Math\'ematique and International Solvay Institutes\\
Universit\'e Libre de Bruxelles, C.P. 231, 1050 Bruxelles, Belgium\\
\end{center}}
\bigskip

\begin{abstract}
  In this paper we provide a first attempt towards a toric geometric
  interpretation of scattering amplitudes.  In recent investigations
  it has indeed been proposed that the all-loop integrand of planar
  $\mathcal{N}=4$ SYM can be represented in terms of well defined
  finite objects called on-shell diagrams drawn on disks. Furthermore
  it has been shown that the physical information of on-shell diagrams
  is encoded in the geometry of auxiliary algebraic varieties called
  the totally non negative Grassmannians. In this new formulation the
  infinite dimensional symmetry of the theory is manifest and many
  results, that are quite tricky to obtain in terms of the standard
  Lagrangian formulation of the theory, are instead manifest.  In this
  paper, elaborating on previous results, we provide another picture
  of the scattering amplitudes in terms of toric geometry. In
  particular we describe in detail the toric varieties associated to
  an on-shell diagram, how the singularities of the amplitudes are
  encoded in some subspaces of the toric variety, and how this picture
  maps onto the Grassmannian description.  Eventually we discuss the
  action of cluster transformations on the toric varieties.  The hope
  is to provide an alternative description of the scattering
  amplitudes that could contribute in the developing of this
  fascinating field of research.
\end{abstract}

\vfill
\begin{flushleft}
{\today}\\
\end{flushleft}
\end{titlepage}

\newpage

\tableofcontents

\section{Introduction}

In the last years a deeper understanding of scattering amplitudes in
QFTs with a high degree of symmetry was pursued.  The most studied and
understood example is the planar limit of $\mathcal{N}=4$ SYM in four
dimensions, where, thanks to the large symmetry group, many efficient
methods have been developed
\cite{Witten:2003nn,Drummond:2006rz,Alday:2007hr,ArkaniHamed:2009dn,ArkaniHamed:2010kv}.
In particular an hidden symmetry of this theory has been recently
discovered: the dual superconformal invariance
\cite{Drummond:2006rz,Alday:2007hr,Drummond:2010uq,Drummond:2010qh,Beisert:2010gn},
that together with the usual superconformal symmetry, gives origin to
the infinite dimensional Yangian symmetry of the theory.

The presence of such infinite dimensional symmetry hidden in the
Lagrangian formulation of the theory, and the unexpected simplicity of
some 
scattering amplitudes, foster the research
for a "dual" formulation for its $S$-matrix, where the full
symmetry group is manifest and the scattering processes and their
singularities have a simple form.  In
\cite{ArkaniHamed:2009dn, ArkaniHamed:2010kv, ArkaniHamed:2012nw} a
dual formulation was proposed in terms of the geometry of an auxiliary
algebraic geometric variety called the totally non negative
Grassmannian.  In this formulation the information needed to
reconstruct the scattering amplitudes are encoded in the combinations
of residues of contour integrals of some well defined invariant form
on the Grassmannian
\cite{ArkaniHamed:2009dn,ArkaniHamed:2010kv,ArkaniHamed:2012nw}.  The
space-time locality and unitarity of the theory are emergent, non
explicitly imposed, properties, while the full symmetry of the theory
is manifest and encoded in the algebraic structure of the Grassmannian
and its decomposition in subspaces.

Quite interestingly the totally non negative Grassmannians are
intimately related to a set of bipartite graphs (collections of edges
connecting black and white nodes) drawn on disks
\cite{PostnikovLungo}.  These graphs are the so called on-shell
diagrams describing the scattering process: all the particles in the
diagrams are on-shell.  Very similar bipartite graphs, but living on
tori instead of disks, have played a central role in the understanding
of the AdS/CFT correspondence \cite{Hanany:2005ve}.  They describe the
field theory living on a stack of D3 brane probing some Calabi-Yau
singularity with at least $U(1)^3$ symmetries: the so called toric
Calabi-Yau singularities.  Recently there was some interest in the
study of the bipartite graphs on Riemann surfaces other that the torus
and the field theory associated to them
\cite{Franco:2012mm,Xie:2012mr,Heckman:2012jh,Franco:2012wv,Franco:2013pg,Cremonesi:2013aba}.

The field theories defined by bipartite graph on disks, relevant for
scattering processes, are special examples of this new class of
theories. In particular it was shown that, as in the case of the torus,
it is always possible to associate a toric variety to these field
theories \cite{Franco:2012mm}.  It is natural to wonder if these two
parallel lines of research based on bipartite graphs on disks could
produce interesting contributions one to the other.

Indeed for example in \cite{Nimatalk} it was observed that the
mathematical structure of the two lines of research is exactly the
same, supporting the idea that it is worthwhile to investigate this
relationship in more detail.  An important step was done in
\cite{PostnikovCorto} where it was shown how to associate a
toric variety to a given bipartite graph on the disk, that can be  
smoothly mapped to a specific subspace of 
the non negative Grassmannian, relevant for the scattering
amplitudes.  This important result opens the possibility of an explicit
reformulation of the scattering process in terms of toric varieties.
This is the line of thought that we will pursue in this paper.  Indeed,
motivated by this connection, in this paper we study the
relation among bipartite graphs, the Grassmannian and its subspaces, 
and the toric geometry.
Taking advantage of some well developed techniques available
for the case of the bipartite diagrams on the tori we find a very simple
global description of the toric geometry associated to the bipartite
graphs on the disk and its map to the non negative Grassmannian.  We
study the local and global coordinates and the
associated map between toric and Grassmannian geometry, the
decomposition of the Grassmannian in subspaces and the associated
decomposition of toric varieties in their toric subspaces, and the
cluster transformations of the coordinates among different patches.

The rest of the paper is organized as follows: in section
\ref{Sec:ReviewNima} we give a general overview of the connection
between the scattering amplitudes and the bipartite graphs on the disk,
Grassmannian geometry and toric geometry.  To clarify some aspects of
these connections we study two simple examples.  In section
\ref{Bipsec} we start a deeper analysis of the bipartite diagrams on
the disks. We define the main objects necessary to our investigations,
the perfect matchings and the perfect orientations and we 
discuss the two polytopes, the matching and the matroid polytope,
that play a key role in the relation with the amplitudes.  In section
\ref{mainres} we derive one of the main results of our paper. We show
that there is a simple system of coordinates describing the points in
the Grassmannian obtained from the bipartite diagrams.  In section
\ref{toriccell} we review the relation between the bipartite graphs
and toric geometry and explain the difference between the toric
varieties associated to the scattering amplitudes and the ones studied
in the AdS/CFT literature.  In section \ref{ESE} we revisit the
examples under this new light, and study the amplitudes in terms of
their relation with the toric varieties.  In section \ref{CLUSTER} we
analyze the transformation of our coordinates under a set of movement
on the diagrams, usually referred as square moves (Seiberg duality in
the physics literature), and we describe in this picture the
associated cluster transformations.  In section \ref{CONC} we conclude
and discuss some further directions of research in this field.  In
appendix \ref{APPE} and \ref{COPLANAR} we review some mathematical
material related to the Grassmannian and some further property of
the toric geometric varieties associated to the on-shell diagrams.

\section{An overview: amplitudes and toric geometry}
\label{Sec:ReviewNima}

The standard formulation of quantum field theory is based on the
concepts of manifest locality and unitarity.  However this description
forces to introduce redundancies associated to the gauge symmetries and it
hides some of the important structures of the field theory
itself. This observation is supported by the simple final results of
various complicated scattering amplitudes processes, the hidden dual
conformal symmetry and the infinite dimensional Yangian symmetry of planar
$\mathcal{N}=4$ SYM, and, more generally, by the existence of various
gauge/string dualities and gauge/gauge dualities hidden in the
standard formulation of quantum field theory
\cite{Drummond:2006rz,Alday:2007hr,Drummond:2010uq,Drummond:2010qh,Beisert:2010gn, Seiberg:1994rs,Seiberg:1994pq,Maldacena:1997re}.
Inspired by these motivations there was a recent effort in trying to
reformulate the description of some of the physical processes of a
quantum field theory, namely scattering amplitudes, in a setup where
the concepts of locality, unitarity, Lagrangian, etc $\dots$, do not
play any fundamental role, but they are instead emergent structures of
a description where all the symmetries of the theory are manifest and
where all the quantities that are computed are on-shell and well defined.  

The idea to reconstruct the scattering amplitudes of a field theory
starting from an on-shell formulation where the symmetries are 
manifest and the basic objects are well defined and free of divergences is somehow a
natural paradigm for a theoretical physicist, but that has been a bit
obscured by all the well know difficulties of standard quantum field
theory.

Along this general line of reasoning, a systematic way to reconstruct
the all loops scattering amplitudes of planar $\mathcal{N}=4$ SYM was
proposed in \cite{ArkaniHamed:2012nw}.  The main insight is to focus
on a certain class of objects: the so called on-shell diagrams. These
diagrams are obtained by gluing together elementary three particle
processes, and are well defined functions of the external momenta of a
scattering process. Pictorially they are very similar to the usual
Feynman diagrams, but physically they are very different: all the
momenta are on-shell and complex in general.  The underlying intuitive
understanding of this reformulation is a different way to think about
permutations.  Indeed it was observed that the four dimensional
scattering amplitudes behave somehow similarly to the $1+1$ integrable
two dimensional systems, where the scattering matrix can be
reformulated in terms of permutations only.  In the four dimensional
case the relation between scattering amplitudes and permutations is
slightly different from the case of two dimensional systems, but the
bottom line is that a permutation, or a set of permutations, can be
associated to a scattering process.  The integrand of the scattering
amplitudes can be computed from the permutations and hence the full
amplitude can be reconstructed.

The first step is to observe that also in four dimensions permutations
can be represented graphically by classes of on-shell two dimensional
diagrams.  These on-shell diagrams are unoriented bicolored graphs
drawn on a disk with black and white vertices connected by edges (see
for example (\ref{beeq})), and they are supposed to contain the
physical information of the scattering of $n$ massless particles: $k$
with 
negative helicity and $n-k$ with 
positive one\footnote{What one is actually computing in the case of $\mathcal{N}=4$ SYM is a super-amplitude, where the 
highest
order component is the amplitude for $k$ gluons 
with 
negative helicity and $n-k$ gluons with 
positve helicity. In all the paper we will keep on focusing on the
highest
component of the super-amplitudes, with the hope that the
reader will not be confused by this abuse of notation.}.  The diagram
has indeed $n$ external edges that end on the boundary and intuitively
represent the $n$ massless particles in the scattering process.  The
edges represent the on-shell particles involved in the process and the
coloring of the vertices are related to the helicities.  At
computational level one should integrate over all the internal edges
(complex internal momenta) and impose momentum conservation at every
vertex.

In \cite{ArkaniHamed:2012nw} a systematic way to deal with these
on-shell diagrams was explained. It unifies the description of any
scattering processes and it is manifestly invariant under the full
symmetry of the theory.  This strategy reformulates the scattering
process in terms of permutations and auxiliary algebraic geometric
varieties, and provides a nice new point of view to look at the
quantum field theory dynamics.

The on-shell diagrams are associated to the permutations as follows.
Let us first introduce a notion of path in the diagram: a path is a
collection of consecutive edges in the graph; if a path turns
maximally right at the black nodes and maximally left at the white
ones it is called \emph{left-right} path \footnote{If the diagram is
  bipartite, i.e. every edge connects a black and a white node, the
  left-right paths are called zig-zag paths.}. Given an on-shell
graph, the permutation is obtained by connecting the external vertices
of the diagram with the left-right paths. These paths uniquely provide
a map among the $n$ external particles.  This map is a permutation of
$n$ objects. The permutation provides also the number $k$ of external
particles with negative helicities: this is the number of particles
that are sent "beyond" $n$ by the permutation.  See for example
(\ref{beeq})
\begin{center}
\begin{tabular}{cc}
\begin{tabular}{c}
\includegraphics[width=3cm]{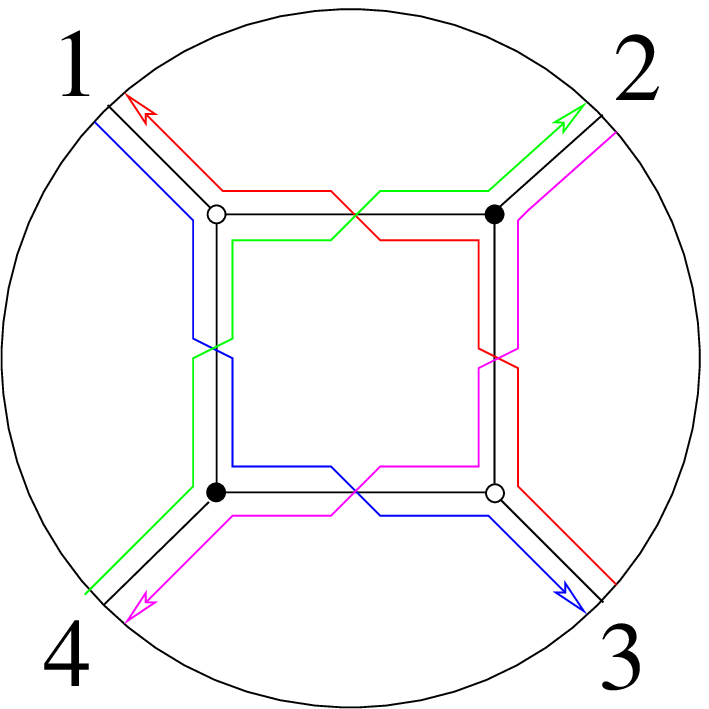}
\end{tabular}
&
\begin{tabular}{c}
$
\left(
\begin{array}{cccc}
1&2&3&4\\
\downarrow&\downarrow&\downarrow&\downarrow\\
3&4&1&2
\end{array}
\right)
$
\end{tabular}
\end{tabular}
\end{center}
\begin{equation}
\label{beeq}
\end{equation}
More formally the integer $k$ is defined as follows.  Define a
permutation of $n$ objects $a_i$ as a map
${\Sigma}:\{1,\dots,n\}\rightarrow \{1,\dots,2n\}$ that permutes $a_i$
with $a_j$: $\Sigma(a_i)=a_j$ if $i<j$ or $\Sigma(a_i)=a_{j+n}$ if
$i>j$.  The integer $k$ counts the number of elements that are sent
\emph{beyond} $n$ by the permutation. These kind of permutations are
called \emph{decorated permutation} of type $(k,n)$.

At this level we have just qualitatively discussed the relation
between the permutations and the on-shell diagrams. At the same time
we know that the on-shell diagrams should also represent scattering
processes. For this reason the permutations should also encode the
quantitative information of this scattering.  The second step of this
reformulation of scattering amplitudes is a mathematical theorem that
states that a given $(k,n)$ permutation is in one to one
correspondence with a specific subspace (called a "cell") in the space
of $k$-dimensional planes in $\mathbb{R}^n$: the so called
Grassmannian $Gr(k,n)$ \cite{PostnikovLungo}.

The appearance of the Grassmannian $Gr(k,n)$ in the description of a
scattering amplitude process should not be a surprise. Indeed for
example the space of two planes in $n$ dimensions, $Gr(2,n)$, is the
natural Lorentz invariant way to describe the scattering process of
$n$ massless particles \cite{ArkaniHamed:2009dn}.

We will review most of the terminology about the Grassmannian in the
next sections, but here we need just a few definitions. A $k$-plane in
$\mathbb{R}^n$ is specified by $k$ vectors of $n$ components (a $n
\times k$ matrix) modulo the action of the $GL(k)$ group, that moves
the vectors, but it leaves the $k$-plane invariant.  The $k$-minors of
the $n \times k$ matrix are invariant under the $SL(k)$ group and they
are simply multiplied by a common factor by the action of $GL(k)$.
They provide indeed a good set of projective coordinates to describe
$Gr(k,n)$: the so called Pl\"ucker coordinates.  The totally non
negative Grassmannian $Gr(k,n)^{tnn}$ (the main actor in this
reformulation of the scattering amplitudes) is the subspace of
$Gr(k,n)$ such that all the $k$-determinants are non negative and the
$GL(k)$ group is restricted to the positive numbers.  The
$Gr(k,n)^{tnn}$ admits a simple decomposition in open subspaces,
called cells, that are obtained by sending (in a precise way to be
explained in the following sections) some of the Pl\"ucker coordinates
to zero and forcing the others $k$-determinants to be positive.  This
process is called the cell decomposition of the $Gr(k,n)^{tnn}$ and it
reconstructs $Gr(k,n)^{tnn}$ as the disjoint union of its cells, in a
similar way as, the probably more familiar, decomposition of the
projective space in the union of some of its subspaces \footnote{ It
  is important to observe from the very beginning that the cell
  decomposition we will speak about in this paper is not the usual
  Schubert cells decomposition but an alternative, much simpler one,
  called positroid decomposition
  \cite{PostnikovLungo,PostnikovLecture} relevant for the scattering
  amplitudes \cite{ArkaniHamed:2012nw}}.

This decomposition of $Gr(k,n)^{tnn}$ is somehow the algebraic
geometric counterpart of the reconstruction of the scattering
amplitude of $n-k$ massless particles with positive  helicity and $k$
massless particles with negative helicity, from its singularities (the
BCFW procedure is a famous example).

The main claim of this reformulation of scattering amplitudes
is that the scattering amplitude process of $n$ particles, with $k$ of
negative helicities, is somehow encoded in the decorated permutations
of type $(k,n)$ that are themselves in $1-1$ correspondence with the
cells in the totally non negative Grassmannian $Gr(k,n)^{tnn}$.
Namely to every permutation ${\Sigma}$, relevant to the scattering
process, can be indeed associated a contour integral of the top
invariant form of $Gr(k,n)^{tnn}$ along some appropriate contour \cite{ArkaniHamed:2012nw}:
\begin{equation}
\label{dafare}
\mathcal{F}_{\Sigma}=\int
\frac{1}{\text{Vol}(GL(k))}
\frac{d^{k \times n}A_{ij}}{(12...k)...(n,1,...,k-1)}
\prod_{i=1}^{k} \delta(A_{ij} Z_j)
\end{equation}
where the integral should be interpreted as a contour integral and one
should compute its residues.  The $n \times k$ matrix $A$
parameterizes the cells in $Gr(k,n)^{tnn}$, in the denominator there
are the $n$ consecutive $k$-minors of the $A$ matrix, while $Z_j$ are
the twistor superspace variables containing the momenta of the
external particles, and the $\delta$-functions impose the momentum
conservation of the scattering process.  The specific parametrization
of $A$ that allows to associate a cell to the permutation is obtained
(as explained in the following sections) by using the on-shell diagram
representation of the permutation.

The full set of singularities of the top form in (\ref{dafare}) have in
general a quite complex structure and they are encoded in the cell
decomposition of $Gr(k,n)^{tnn}$, as shown in
\cite{ArkaniHamed:2012nw}.
A careful parametrization of $Gr(k,n)^{tnn}$ in terms of $n(n-k)$
local coordinates transforms the integrand form in (\ref{dafare}) in a
product of $n(n-k)$ $d$-logarithms: $d f_i/f_i$, explicitly showing that the only
singularities are of the logarithmic type and that they occur at
the boundary of the positive part of $Gr(k,n)^{tnn}$ .
However this is just a local description and it can be shown that it
is not possible to represent all the singularities in a single patch
\cite{ArkaniHamed:2012nw}. For this reason the full geometry of
$Gr(k,n)^{tnn}$, the gluing procedure of the various patches or a
global description in terms of embedding coordinates, is required to
obtain the scattering amplitudes and is the full variety that encodes
the complete set of physical information.

In this paper we want to provide a complementary description of the
scattering amplitudes in terms of toric geometry (a branch of
algebraic geometry where the varieties can be simply described in
terms of polytopes).  We will elaborate on previous results about the
existence of a map between Grassmannians and certain toric varieties
in the mathematical literature \cite{PostnikovCorto}, and the known
relation between bipartite graphs and toric geometry in the physical
literature \cite{Hanany:2005ve}.  Our aim is to provide an alternative
description of the scattering process that could hopefully help in the
progress of this fascinating field of research.  We will show how to
associate a toric variety to the scattering process and how the global
description of this geometry encodes the necessary informations to
reconstruct the amplitude.

Toric varieties will be introduced in the following sections.  For the
moment it should be enough to state that the main actor in this
alternative description of the scattering amplitudes are the non
negative part of real toric varieties.  Roughly speaking a complex
toric variety of dimension $d$ is an algebraic geometric variety that
has at least a $U(1)^d$ isometry group, its algebraic equations are
of the kind monomial = monomial and it can be represented as a $d$
dimensional polytope.  The real non negative part of these varieties
is just the restriction of the complex coordinates to their non
negative real values.  Toric geometry is usually a very interesting
tool that provides a simple description of quite complex systems. This
simple description revealed to be very useful for example in the domain of the
gauge/gravity correspondence and in  algebraic geometric in
general. We hope that it could be useful also in the domain of the 
scattering amplitudes.

Toric geometry can be introduced thanks to the relation between
decorated permutations and bicolored graphs.  We previously stated
that a permutation can be represented in terms of on-shell diagrams
thanks to the left-right paths.  However this representation is not
unique and different on-shell diagrams are in general associated to
the same permutation.  Namely one
class of on-shell diagrams is associated
to one permutation.  At a first look an infinite number of on-shell graphs
could be associated to every permutation, obtained
for example just by keeping on adding internal faces.  However we
should distinguish between "reduced" and "non-reduced" graphs. The
reduced graphs are the graphs that have somehow the minimum amount of
faces for a given permutation. The "non-reduced" graphs could then be
obtained for example by simply  adding faces to the reduced ones. The
class of reduced diagrams will then contain only a finite set of
graphs with the same permutation.

More formally reduced diagrams are defined as follows.  First of all
we should define a set of moves on the diagram.  The moves are
\begin{eqnarray}
\label{mosse}
\begin{array}{lc}
\begin{array}{c}
\text{\bf Square move}
\end{array}
&
\begin{array}{c}
\includegraphics[width=5cm]{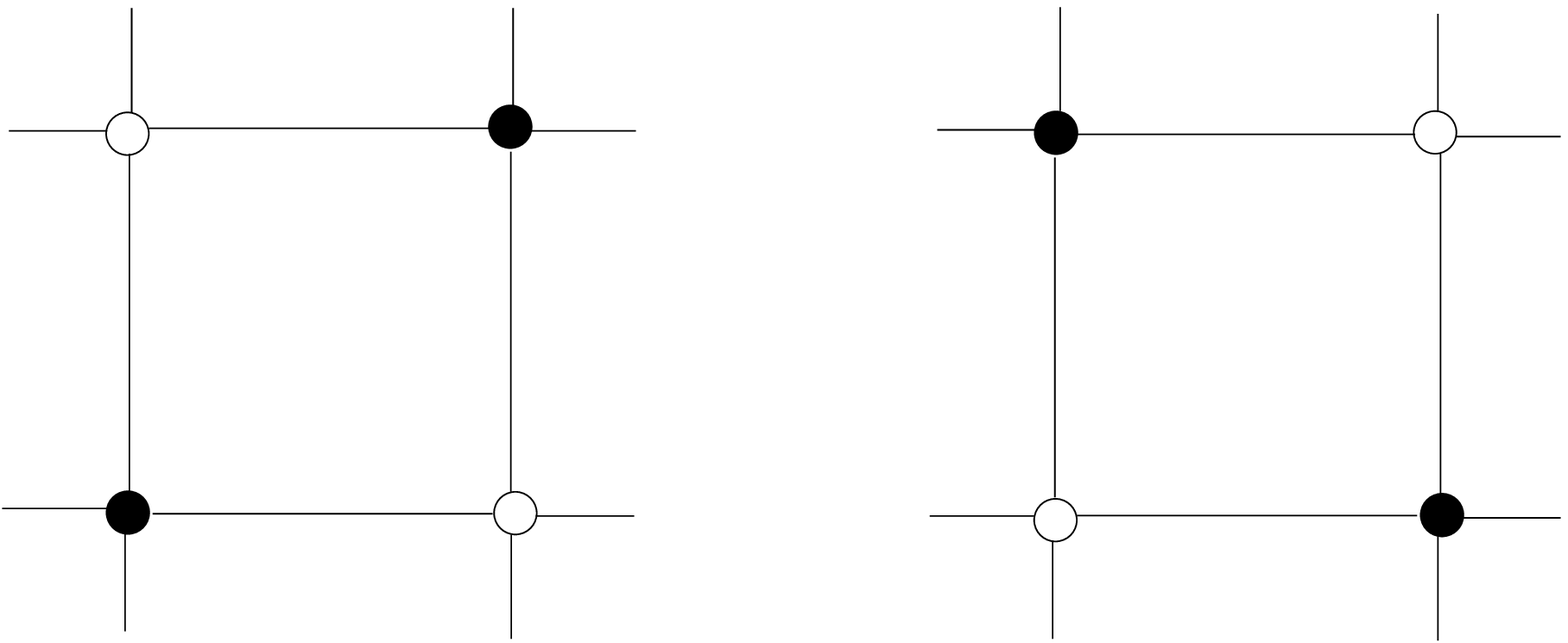}
\end{array}
\\
\begin{array}{c}
\text{\bf Bubble reduction}
\end{array}
&
\begin{array}{c}
\includegraphics[width=5cm]{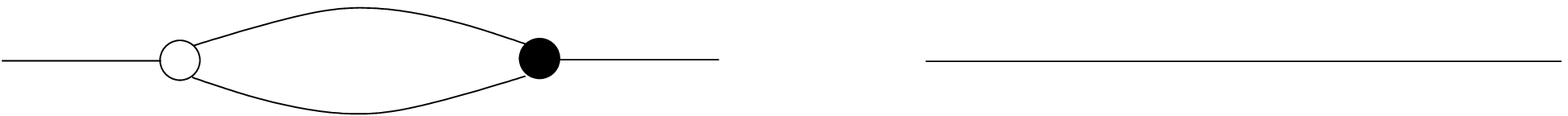}
\end{array}
\\
\begin{array}{c}
\text{\bf Mergers}
\end{array}
&
\begin{array}{c}
\includegraphics[width=5cm]{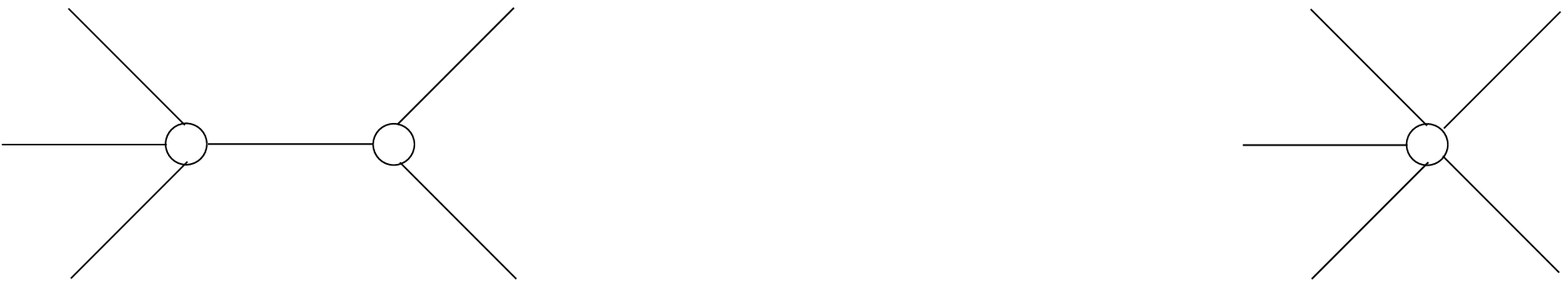}
\end{array}
\end{array}
\end{eqnarray}
If we apply mergers and square moves to an on-shell diagram one can in
principle generate a bubble on some internal line of the diagram.
This bubble can be eliminated thanks to the bubble reduction.  This
process can be iterated until no bubble can be created anymore.  The
diagram obtained by this procedure is called reduced.

Among these reduced diagrams there exists a finite set associated to
the same permutation and hence the same physical process.  The class
of reduced diagrams related to the same permutation and hence to the
same cell in the Grassmannian is obtained from a reduced diagram by
applying mergers and square-moves.

The basic elements associated to a scattering amplitude are the
reduced diagrams modulo the action of square move and mergers\footnote{It is however important to state that the non-reduced graphs play 
an important role too in particular in relation to loop amplitudes, but that they can however be reconstructed from 
reduced on-shell graphs.}.
By applying those transformations each bicolored graph can be put in a
bipartite form, i.e. every edge connects a white vertex with a black
one. The toric varieties are constructed from these bipartite graphs.

The correspondence between bipartite graphs and toric geometry may
look not surprising, especially due to some important developments in
the AdS/CFT correspondence.  Indeed in this area there exists a class
of superconformal field theories (SCFTs) that can be represented and
engineered in terms of branes as bipartite diagrams drawn on a torus
$\mathbb{T}^2$.  A toric variety can be associated to each bipartite
graph, modulo the action of movements
similar to the square move and the mergers \cite{PostnikovCorto}.
  
In the present discussion the bipartite diagrams are drawn on a disk.
However recently these diagrams have been studied by mathematicians
\cite{PostnikovLungo} and their connection with toric geometry was
firstly investigated in \cite{PostnikovCorto}.

In \cite{PostnikovCorto} it was shown that bipartite diagrams on the
disk can be associated to projective toric varieties. First one
defines the perfect matchings, PMs (from now on $\sigma_i$, with
$i=1,...,c$): collections of edges minimally covering all the vertices
in the diagram (for every vertex there is only one edge ending on
it). The PMs are not generically independent, but they satisfy a set
of $s$ linear relations among themselves.  To every PM one associates
a projective coordinate (cfrom now on $\pi_i$, with $i=1,...,c$)
in $\mathbb{RP}_{> 0}^{c-1}$ where $c$ is the total number of PMs.
The linear relations among the PMs become product relations among the
associated projective coordinates (the exponential of the PM linear
relations).

The closure in $\mathbb{RP}^{c}_{\geq0}$ of these product relations,
namely the addition of the limit points such that the coordinates can go to
zero, defines the projective toric variety.  To every PM one can also
associate a vector in $\mathbb{R}^{c-s}$ ($v_{\sigma_i}$ with
$i=1,...,c$). 
This map is obtained by solving  the set of $s$ linear
relations in terms of $c-s$ basis PMs.  The convex hull formed by
these $c$ $s$-vectors defining the projective toric variety is a
polytope, called the matching polytope.  The singularities of the form
in the integrand in (\ref{dafare}), described as cell decomposition in
the $Gr(k,n)^{tnn}$ picture, are instead described as a "partial
facets decomposition" of the toric matching polytope. Namely to every
singularity in the form in (\ref{dafare}) is associated at least a
facet in the polytope describing the toric projective variety.

More concretely let us introduce the notation $p_I$ for the previously
introduced Pl\"ucker coordinates of $Gr(k,n)^{tnn}$. $I$ is the set of
all possible $k$ edges among the possible $n$ external edges of the 
on-shell diagram, namely the set of all the possible $k$-minors of the $n
\times k$ matrix defining a point in the Grassmannian: there are $n
\choose k$ of them.  These coordinates are not free, but satisfy a set
of algebraic equations that define the Pl\"ucker embedding of
$Gr(k,n)^{tnn}$ into $\mathbb{RP}^{{n \choose k }-1}_{\geq0}$.

Elaborating on some previous results \cite{Talaska} in this paper we
will show that the map from the projective toric variety to the
totally non negative Grassmannian, and hence to the integrand in
(\ref{dafare}) and its singularities, is provided by the following
linear map between the embedding coordinates of the two varieties:
\begin{equation}
\label{Grpi}
p_I = \sum_{j \in \text{internal loop}} \pi_I^{(j)}
\end{equation}
where $\pi_I^{(j)}$ is the coordinate associated to the $j$-th PM and
we are summing over the PMs that only differ by internal lines. This
map stands at this point at the intuitive level, and it will be
formalized in detail in the following sections. As we will show later
indeed PMs differing by internal lines are associated to the same set
of $k$ elements that is parameterized by $I$.  It is important to
underline from the very beginning that the map from the non negative
part of the real projective toric variety to $Gr(k,n)^{tnn}$ or a
closure of one of its cell, is an homeomorphism between the positive
part of both varieties, but it is in general not an isomorphism on the
boundaries, namely for the zero values of some of the coordinates in
the toric varieties and in the Grassmannian.

This map is the global version, in a sense that will be explained later
in the paper, of the one in \cite{Talaska}.

The singularities of the on-shell diagram are obtained by the
knowledge of the polynomial relations among the $\pi$ coordinates of the
toric variety and the map from the $\pi$ to the coordinates $p$ of
the Grassmannian.  One starts from the polytope identified by the PMs
and projects on the facets, by sending some edges of the on-shell
diagram to zero.  The edges that can be sent to zero are called
$\emph{removable}$. An edge is removable if the two left-right paths
that intersect on it do not intersect on other edges in the diagram.
By removing an edge one removes also some PMs, the ones that contain
the edge, and reduces on some face of the polytope. Some $p$
coordinate vanishes as well.  The remaining non vanishing coordinates
on each facet represent the singularity.  This procedure can be
redundant: by sending to zero different sets of PMs we may describe
the same singularity.  This happens if the PMs that we send to zero
differ only by internal faces.  In the rest of this section we study
in detail two simple examples to elucidate the procedure that we just
reviewed.

\subsection{A simple example: three particles amplitude}
\label{3Gr}

In this section we use the toric approach to study the on-shell
diagram describing the tree level scattering of three particles, two
with positive and one with negative helicity: namely the maximal
helicity violating (MHV) tree level three particles amplitude.
The on-shell diagram for this amplitude is represented by one of the
three graphs in (\ref{threepm}), where we also explicitly show the
three PMs.
\begin{equation}
  \label{threepm}
\includegraphics[width=10cm]{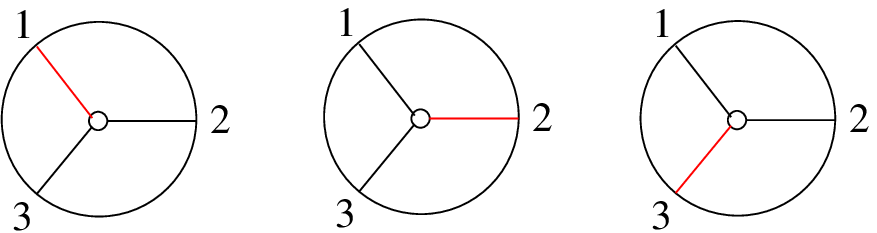}
\end{equation}
The corresponding anti MHV amplitude ($\overline{MHV}$) is obtained by
inverting the color of the vertex.  In both cases it should be
easy to see that there are three PMs and no relations among them
(every edge is a PM).

In this very simple case, we don't have internal faces, and we do not
have relations, and hence the map between the totally non negative
toric variety and the totally non negative Grassmannian (\ref{Grpi})
should be an isomorphism and the two descriptions are totally
equivalent.

Let us see explicitly how it works.  The diagram in (\ref{threepm}) is
associated, according to the map that we have previously explained, to
the totally non negative Grassmannian with $k=1$ and $n=3$,
$Gr(1,3)^{tnn}$.  This is the space of lines in $\mathbb{R}^3$ and it
is exactly the totally non negative part of the two dimensional real
projective space: $\mathbb{RP}^2_{\geq0}$.

This space is decomposed in cells as follows.  The top dimensional
subcell in the decomposition is $\mathbb{RP}_{>0}^2$, where all the
coordinates are non vanishing. The lower dimensional subspaces are
three copies of $\mathbb{RP}_{>0}^1$, where one of the original three
coordinates is vanishing, while the other two are strictly
positive. The lowest dimensional cells are three points in the
projective space: each one associated to a single non vanishing
coordinate.  Hence the full $Gr(1,3)^{tnn}$, namely
$\mathbb{RP}^2_{\geq0}$, can be reconstructed as the union of its
seven cells: $\mathbb{RP}_{\geq0}^2=\mathbb{RP}_{>0}^2 \cup 3$ $
\mathbb{RP}_{>0}^1 \cup 3$ points.

The same description can be obtained in terms of toric geometry.  One
starts from the three PMs: $\sigma_i$ for $i=1,2,3$, in
(\ref{threepm}).  There are not relations among them and the toric
diagram (that will be often called \emph{matching polytope} in the rest of the paper) 
is described by the three vectors
$v_{\sigma_i}$:
\begin{equation} \label{vp}
v_{\sigma_1} = (1,0,0) \quad,\quad
v_{\sigma_2} = (0,1,0) \quad,\quad
v_{\sigma_3} = (0,0,1) \quad,\quad
\end{equation}
We can associate a totally non negative toric
variety to this toric diagram in the following way.
To each PM $\sigma_i$ we associate a coordinate $\pi_i$ in
$\mathbb{RP}_{>0}^2$. The absence of relations among the $\sigma_i$
implies the absence of relations among the $\pi_i$, and we can
safely take the closure of the space. We obtain in this way that the
toric variety associated to the tree level three points MHV amplitude
is $\mathbb{RP}_{\geq0}^2$ that is exactly equal to $Gr(1,3)^{tnn}$.

The $\pi$ coordinates are in 1-1 correspondence with the $p$
coordinates of the Grassmannian.  From the toric diagram (\ref{tutto})
we can recover the cell decomposition of $Gr(1,3)^{tnn}$ in terms of
the decomposition into subpolytopes of the matching polytope.  The
various subspaces are obtained by sending to zero the various $\pi$
coordinates.

One can give a pictorial representation of this decomposition.  First
assign a vector on a three dimensional lattice to every PM as in
formula (\ref{vp}). This identifies a multidimensional polytope. These
vectors lie on a plane and we can restrict to the basis of the
polytope.  By eliminating a PM at each step we re-obtain the
decomposition of $Gr(1,3)^{tnn}$ in positive subspaces.  For example
in this case the positive part of the projective toric variety, the
interior part of the central diagram in (\ref{tutto}),
describes $\mathbb{ R P}_{>0}^2$. The boundaries are added by
eliminating some of the corners of the diagram and this corresponds to
a reduction on the (positive) facets.  The first elimination indeed
gives three lines that correspond to the three $\mathbb{ R P}_{>0}^1$.
The last step identifies the three single coordinates.
\begin{equation} \label{tutto}
\includegraphics[width=10cm]{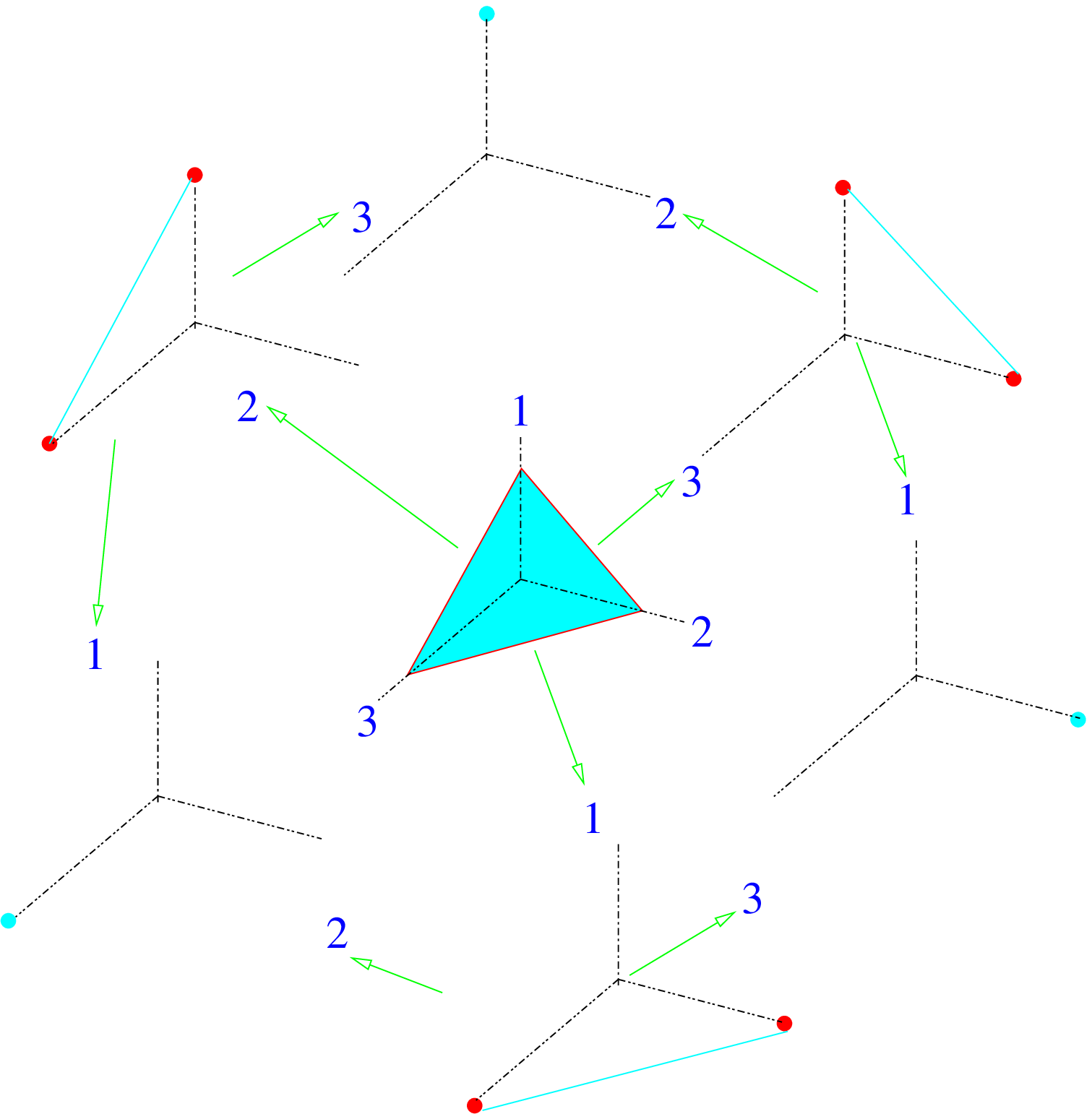}
\end{equation}
Observe that every reduction corresponds to the elimination 
of an edge in the diagram, and if
more than one PM must be eliminated at the same time one has to erase
the edges common to these PM. In this case every edge is associated to 
a single PM and the situation is simpler.

At every step the elimination  of a PM sets to zero the corresponding
coordinate. In  (\ref{tutto}) we marked in cyan the 
interior (positive) part of each toric diagram. 
The final step of this construction corresponds to the 
single points that identify each PM.

Summarizing, in the toric picture we started with $\mathbb{RP}_{\geq
  0}^2$ and we obtain its cell decomposition in positive subspaces by
projecting on the facets of the diagram, reproducing exactly the cell
decomposition of $Gr(1,3)^{tnn}$.

\subsection{The box diagram}
\label{4Gr}
In the previous example the on-shell diagram had no internal faces.
In this subsection we want to study a slightly more complicate case,
with one internal face.  The diagram is shown in (\ref{24top})
\begin{equation}
\label{24top}
\includegraphics[width=3cm]{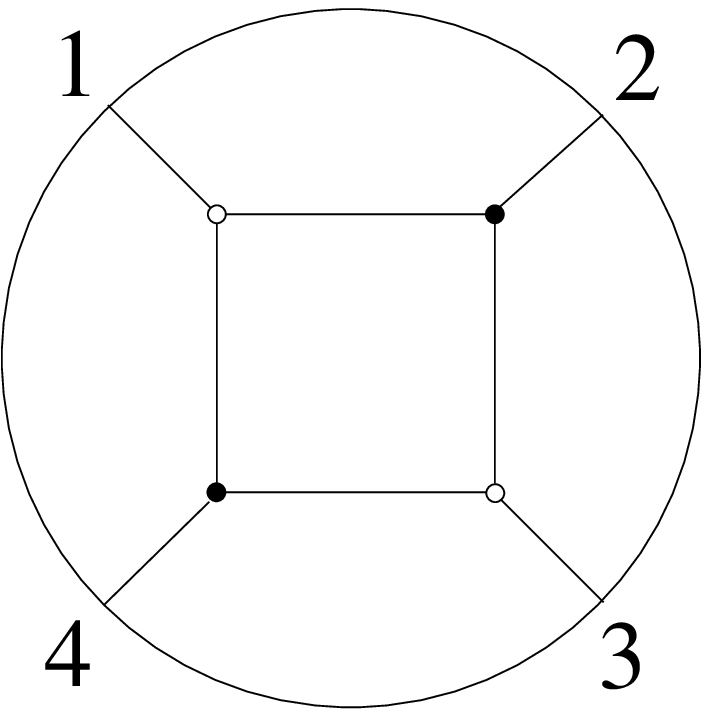}
\end{equation}
There are two novelties compared to the previous example: first the
PMs are not associated to single edges but to collections of edges;
second some PMs differ by internal faces and they are associated to
the same projective coordinate $p_I$.

The on-shell diagram in (\ref{24top}) is the box diagram, representing
the tree level scattering of four particles, two with positive and two
with negative helicity in the planar limit, another example of a tree
level MHV amplitude. In the formulation of \cite{ArkaniHamed:2012nw}
this on-shell diagram is associated to $Gr(2,4)^{tnn}$, the non
negative part of the space on two-planes in $\mathbb{R}^4$.  The
standard algebraic geometric description of this space is provided, as
previously explained, by the Pl\"ucker embedding, that in this case is
simply a quadratic equation among the $p_I$ \footnote{In the following
  $I$ will be associated to a specific subset of the $n$ external
  edges of the on-shell diagram. In this section for simplicity, with
  a clear abuse of notation, we will use $I$ as the number classifying
  the set of external edges $I$.  We hope that the reader will not be
  confused} coordinates of $\mathbb{RP}_{\geq 0}^5$:
\begin{equation}\label{Pl24}
p_1\, p_6 = p_2\, p_3 + p_4\, p_5
\end{equation}
Let us now try to show how the underlining toric geometry associated
to the box diagram in (\ref{24top}) maps to $Gr(2,4)^{tnn}$ and hence
it relates to the renewed understanding of scattering amplitudes.

The box diagram has seven PMs, $\sigma_i$, as shown in (\ref{tutt1}).
\begin{equation} \label{tutt1}
\includegraphics[width=10cm]{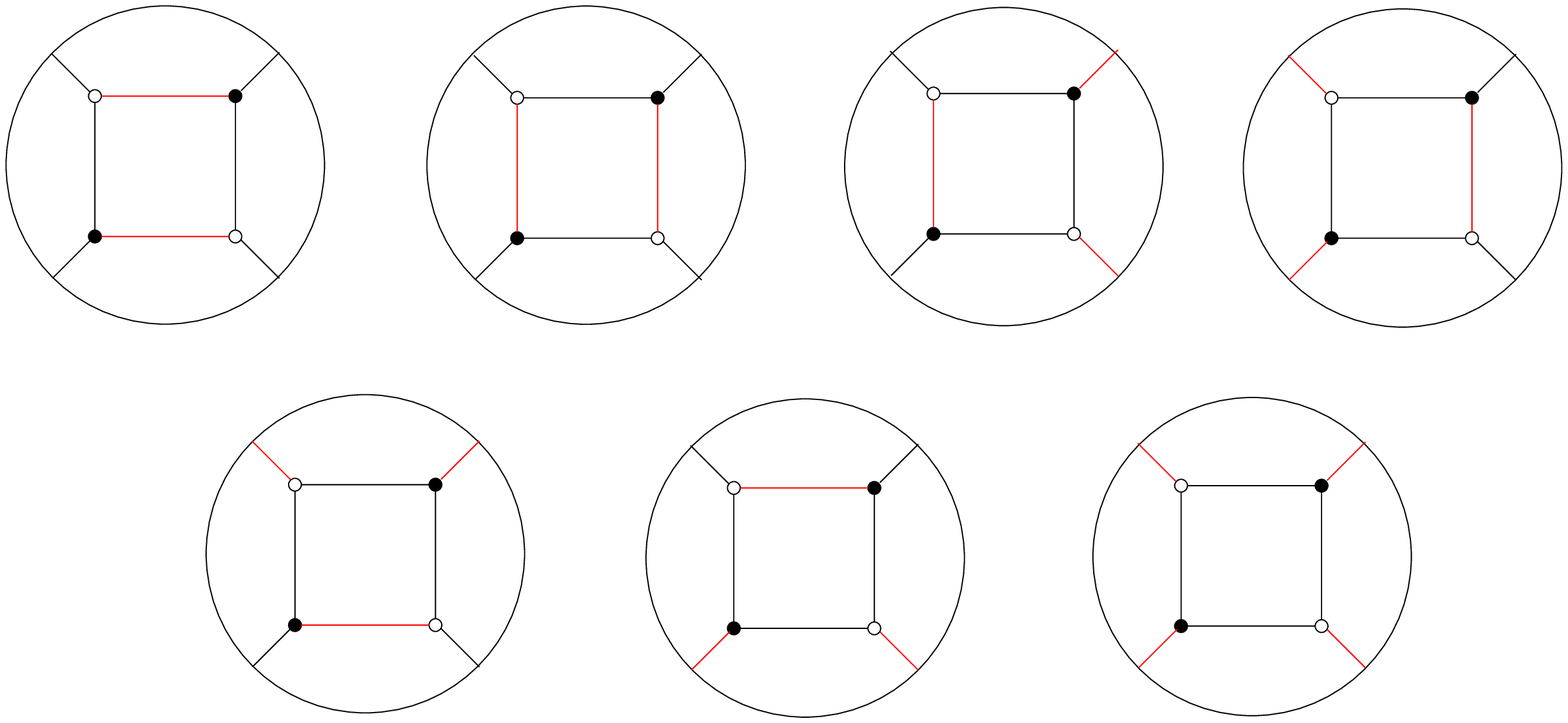}
\end{equation}
In this case the $\sigma_i$ are not free, but they satisfy two relations:
\begin{eqnarray}\label{PMs24}
\sigma_1 + \sigma_7 = \sigma_5 + \sigma_6 \quad \quad 
\sigma_2 + \sigma_7 = \sigma_3 + \sigma_4 
\end{eqnarray}
As previously explained we can associate a vector $v_{\sigma_i}$ to
every PM and obtain the polytope in $\mathbb{R}^5$ described by these
vectors by solving the two equations (\ref{PMs24}) in terms of five
vectors, for example $v_{\sigma_1}, v_{\sigma_2}, v_{\sigma_3},
v_{\sigma_4}, v_{\sigma_5}$. The convex hull described by these seven
vectors is the matching polytope of the box diagram and it can be
parameterized as follow:
\begin{eqnarray}\label{Mtch24}
&&
v_{\sigma_1}=(1,0,0,0,0),\quad
v_{\sigma_2}=(0,1,0,0,0),\quad
v_{\sigma_3}=(0,0,1,0,0),\quad
v_{\sigma_4}=(0,0,0,1,0) \nonumber \\
&&
v_{\sigma_5}=(0,0,0,0,1),\quad
v_{\sigma_6}=(1,-1,1,1,-1),\quad
v_{\sigma_7}=(0,-1,1,1,0)
\end{eqnarray}
The toric variety described by this polytope is  obtained
by associating to every vector $v_{\sigma_i}$ a coordinate $\pi_i$ in
$\mathbb{RP}_{> 0}^6$, by taking the exponential of the relations in
(\ref{PMs24}) and by interpreting them as multiplicative relations among the
$\pi_i$, and then by taking the closure of this map. We obtain in this
way the algebraic equations for the projective toric variety
associated to the box diagram.  The final result is a toric variety
described by two quadrics in $\mathbb{RP}_{\geq 0}^6$:
\begin{equation} \label{dualrel}
\pi_1 \pi_7 = \pi_5 \pi_6 \quad \quad 
\pi_2 \pi_7 = \pi_3 \pi_4
\end{equation}
Maybe it worthwhile to underline, as we will explain more in details
in the following sections, that the toric variety described by the
relations (\ref{dualrel}) is not the Master Space introduced in
\cite{Forcella:2008bb}, and further discussed in the case of bipartite
graphs on disks in \cite{Franco:2012mm}, even if the matching polytope
in (\ref{Mtch24}) is the same. The procedure to associate the
coordinates and relations to the polytope used here is indeed
different from the one used in the case of the Master Space.

We would like now to illustrate how the two four dimensional varieties
in (\ref{Pl24}) and (\ref{dualrel}) are related and hence how the
toric variety in (\ref{dualrel}) encodes the information of the four
particles scattering process.  The map from the toric variety to the
Grassmannian, and hence to the scattering amplitude, is given by the
equation (\ref{Grpi}) that we introduced in the previous section. This
equation essentially boils down to a linear map that relates some
combinations of the coordinates of the embedding space
$\mathbb{RP}_{\geq 0}^6$ of the toric variety, to the coordinates of
$\mathbb{RP}_{\geq 0}^5$: the standard Pl\"ucker embedding space for
$Gr(2,4)^{tnn}$.  This map will become much more clear when we will
explain it in detail in the following sections. For the time being
the important observation is that
the first two PMs: $\sigma_1$ and $\sigma_2$, in  (\ref{tutt1}),
differ by an internal face and hence, according to (\ref{Grpi}),
their sum should be associated to a single $p_I$.
The final map is:
\begin{equation}
p_1=\pi_1+\pi_2,\quad
p_2=\pi_3,\quad
p_3=\pi_4,\quad
p_4=\pi_5,\quad
p_5=\pi_6,\quad
p_6=\pi_7
\end{equation}
These $p_I$ coordinates defined in this way satisfy then the Pl\"ucker
embedding equation (\ref{Pl24}) for $Gr(2,4)^{tnn}$, once the toric
$\pi_i$ coordinates are restricted on the toric variety
(\ref{dualrel}).

In this case the toric diagram lives on a five dimensional lattice and
we cannot give a pictorial description of it.  However, as before, we
can decompose this totally non negative space in totally positive
subspaces associated to (sub)cells in $Gr(2,4)^{tnn}$ and hence to the
singularities required to reconstruct the scattering amplitude.  The
codimension one cells in $Gr(2,4)^{tnn}$ are associated to some facets
of dimension 4 in the matching polytope in (\ref{Mtch24}).  One can
end up on those facets by erasing some of the PMs. We will
carefully redo all this procedure in the next sections, once all
the required technology will be introduced.  The first step of this
reduction is
\begin{equation}
\label{first}
\begin{array}{c|c||c|c}
\text{Vanishing ~} \pi_I &\text{Surviving relation}&
\text{Vanishing ~}p_I& \text{Surviving relation}\\
\hline
\pi_1= \pi_5 = 0&\pi_2 \pi_7 = \pi_3 \pi_4 &p_4=0&p_1 p_6 = p_2 p_3 \\
\pi_1= \pi_6 = 0&\pi_2 \pi_7 = \pi_3 \pi_4 &p_5=0&p_1 p_6 = p_2 p_3 \\
\pi_2= \pi_3 = 0&\pi_1 \pi_7 = \pi_5 \pi_6 &p_2=0&p_1 p_6 = p_4 p_5 \\
\pi_2= \pi_4 = 0&\pi_1 \pi_7 = \pi_5 \pi_6 &p_3=0&p_1 p_6 = p_4 p_5
\end{array}
\end{equation}
where on the right we present the four codimension one cells of
$Gr(2,4)^{tnn}$, and on the left the corresponding toric codimension
one subspaces of the associated toric variety.  Every choice
corresponds to a different singularity in the amplitude.  In 
(\ref{first}) we identify the vanishing PMs, the vanishing coordinates
$p$ and the relations among the other surviving coordinates. It is
important to remember that the surviving $p$ and $\pi$ coordinates
are constrained to have only positive and non zero values. The toric
sub dimensional varieties and the cells are isomorphic.  Further
reductions to higher codimensional cells and facets are possible.  For
example from the three dimensional cell and facet in in the first line
of (\ref{first}) one can reduce to the sub-facets by imposing either
$\pi_2= \pi_3=0$, or $\pi_2=\pi_4=0$, or $\pi_6=0$.  This corresponds
to set to zero other $p$ coordinates and hence go to subcells in the
cell decomposition.

By looking at all the possibilities we can summarize the first two
steps in the cells/facets decomposition in the following table
\begin{center}
\begin{tabular}{c|c}
First Step & Second Step \\
\hline
\begin{tabular}{cc}
$\pi_1 = \pi_5 = 0$&$p_4=0$ 
\end{tabular}
&
\begin{tabular}{cc}
$\pi_2 = \pi_3 = 0$ & $p_1=p_2=0$  \\
$\pi_2 = \pi_4 = 0$ & $ p_1=p_3=0$ \\
$\pi_7 = \pi_3 = 0$ & $ p_6=p_2=0$ \\
$\pi_7 = \pi_4 = 0$ & $p_6=p_3=0$  \\
$\pi_6=0$&$p_5=0$
\end{tabular}
\\\hline
\begin{tabular}{cc}
$\pi_1 = \pi_6 = 0$&$p_5=0$ 
\end{tabular}
&
\begin{tabular}{cc}
$\pi_2 = \pi_3 = 0$ & $p_1=p_2=0$  \\
$\pi_2 = \pi_4 = 0$ & $ p_1=p_3=0$ \\
$\pi_7 = \pi_3 = 0$ & $ p_6=p_2=0$ \\
$\pi_7 = \pi_4 = 0$ & $p_6=p_3=0$  \\
$\pi_5=0$&$p_4=0$
\end{tabular}
\\\hline
\begin{tabular}{cc}
$\pi_2 = \pi_3 = 0$&$p_2=0$ 
\end{tabular}
&
\begin{tabular}{cc}
$\pi_1 = \pi_5 = 0$ & $p_1=p_4=0$  \\
$\pi_1 = \pi_6 = 0$ & $ p_1=p_5=0$ \\
$\pi_7 = \pi_5 = 0$ & $ p_6=p_4=0$ \\
$\pi_7 = \pi_6 = 0$ & $p_6=p_5=0$  \\
$\pi_4=0$&$p_3=0$
\end{tabular}
\\\hline
\begin{tabular}{cc}
$\pi_2 = \pi_4 = 0$&$p_3=0$ 
\end{tabular}
&
\begin{tabular}{cc}
$\pi_1 = \pi_5 = 0$ & $p_1=p_4=0$  \\
$\pi_1 = \pi_6 = 0$ & $ p_1=p_5=0$ \\
$\pi_7 = \pi_5 = 0$ & $ p_6=p_4=0$ \\
$\pi_7 = \pi_6 = 0$ & $p_6=p_5=0$  \\
$\pi_3=0$&$p_2=0$
\end{tabular}
\end{tabular}
\end{center}
At the second step we can identify twenty possibilities but just ten
of them are different (in the other cases we are setting to zero the
same set of $\pi$ coordinates, but in a different order).  We can
then resume the two dimensional cells of $Gr(2,4)^{tnn}$ and the
associated two dimensional toric varieties as in the table
\begin{center}
\begin{tabular}{c|c|c}
$p_I=0$&$\pi_I=0$&relations\\
\hline
$(1,2,4)$&$(1,2,3,5)$&\\
$(1,3,4)$&$(1,2,4,5)$&\\
$(1,2,5)$&$(1,2,3,6)$&\\
$(1,3,5)$&$(1,2,4,6)$&\\
$(2,4,6)$&$(1,3,5,7)$ or $(2,3,5,7)$&\\
$(3,4,6)$&$(1,4,5,7)$ or $(2,4,5,7)$&\\
$(2,5,6)$&$(1,3,6,7)$ or $(2,3,6,7)$&\\
$(3,5,6)$&$(1,4,6,7)$ or $(2,4,6,7)$&\\
$(2,3)$&$(2,3,4)$&$p_1 p_6 = p_4 p_5$\\
$(4,5)$&$(1,5,6)$&$p_1 p_6 = p_2 p_3$
\end{tabular}
\end{center}
Let us observe that in the first four lines the projection of the
diagram identifies a single subspace.  In the next four cases there
are two possible reductions of the toric diagram that identify the
same subspace.  This is a common feature: by projecting on the faces
of the diagram we end up on a sub cell of the Grassmannian.  But
different projections can identify the same cell, if they are obtained
sending to zero $\pi$ associated to PMs differing by internal
faces. This is related to the fact that the map between the toric
variety and the Grassmannian is not in general an isomorphism.  By
iterating this procedure one obtains the other lower dimensional
cells.  The process finishes when one has only one non vanishing
$p$.

\section{Bipartite diagrams, polytopes and toric geometry}
\label{Bipsec}

Until this point we based our discussion on an intuitive
understanding. In the rest of the paper we will go through in
detail in the relation between toric geometry and the top form in
(\ref{dafare}) associated to the scattering amplitude.  In particular
we will deeply investigate the connection between the bipartite graphs
on the disk and the decomposition of the totally non negative
Grassmannian in positive subspaces, or cells. In addition we will analyze
the detailed relation between bipartite graphs on a disk and toric
geometry. We will then reformulate a relation between the toric and
the Grassmannian picture and, elaborating on previous results, we will
provide a toric geometrical understanding for the top form in
(\ref{dafare}) associated to scattering amplitude.
 
We start this analysis by reviewing some basic definitions about the
bipartite graphs on the disk.

A graph is a set of vertices connected by edges. If every edge
connects a black and a white vertex the graph is called bipartite. A
graph on the disk has two types of edges, internal and external edges.
The internal edges connect two vertices, while the external edges
connect a node to the boundary of the disk.  
The external edges do not intersect at the boundary.

Bipartite graphs have been extensively studied in the analysis of D$3$
branes probing toric CY$_3$ singularities in the AdS$_5$/CFT$_4$
correspondence \cite{Hanany:2005ve}. The difference with the case that
we discuss here is that the graphs do not live on a torus but on a
disk, where there are also external edges.

There is a useful quantity in the analysis of bipartite diagrams both
on the disk and on the torus, that we will often use in this
paper. This is the notion of perfect matchings (PMs).  A PM on a
bipartite diagram on a torus is a set of edges without common vertices
that covers all the vertices of the graphs.  Every edge in a PM
connects a white to a black node.  In the case of the disk there is a
similar definition of PMs. The difference is that the external edges
cover a single vertex.  One can see an example of PMs on the disk in
(\ref{tutt1}).

Closely related to the notion of PMs one can define the
perfect orientations, POs. This definition does not have any particular
role in the case of the torus but it becomes important in this
setup.  A  PO is defined by
assigning an arrow to every edge in the graph such that
\begin{itemize}
\item Every black vertex is incident to just one edge directed away
  from the vertex.
\item Every white vertex is incident to just one edge directed toward
  the vertex.
\end{itemize} As shown in
\cite{PostnikovLungo,PostnikovCorto,Franco:2012mm} there is a 1-1
correspondence between PM and PO. Given a PO the associated PM is
composed by the set of edges directed from the black to the white
nodes in the PO.  On the other hand, given a PM, one can construct a PO
by assigning arrows from black to white nodes to the internal edges in
the PM and arrows from white to black nodes to the internal edges not
in the PM.  For the external edges one assigns an ingoing (outgoing)
arrow to the edges in the PM ending on a white (black) node and an
ingoing (outgoing) arrow to the edges not in the PM ending on a black
(white) node.  One can observe an example of the correspondence
between the PMs and the POs in (\ref{24POPM}).

\subsection{Polytopes}
\label{poly}
Similarly to the case of the torus the PMs can be used to define a set of
polytopes associated to a bipartite diagram.  These polytopes
are used to associate toric varieties to these diagrams.
There are two polytopes naturally associated to a bipartite graph,
the matching and the matroid polytope.

\subsubsection*{The matching polytope}

The first polytope that is usually associated to a bipartite diagram
is the so called matching polytope.  The matching polytope can be
defined by using the $s$ linear relations among the $c$ perfect
matchings.  It is indeed the convex hull described by the $c$ vectors
of length $c-s$ that satisfy the $s$ linear constraints. It turns out
that the number of faces of the bipartite graph $G$ is equal to $c-s$,
and the matching polytope lives in a space of $G$ dimensions
\footnote{There is an extra relation among the vectors forming the
  polytope.  Indeed as we will show in section \ref{COPLANAR} the
  vectors are all coplanar and the polytope is actually a $G-1$
  dimensional object.}.

In its original definition \cite{PostnikovCorto} the matching polytope
was defined as follows.  Every PM $\sigma_i$, $i=1,\dots,c$, is
associated to a set of edges.  One associates a $P_{c,E}$ matrix to
the bipartite diagram, where $c$ labels the PMs and $E$ the edges.
The element $P_{c_i,E_j}$ of the matrix $P$ is $1$ if the edge $E_j$
belongs to the $c_i$-th PM, $0$ otherwise.

The two definitions of matching polytope are equivalent because the
PMs are not all independent, and one can reduce the $P$ matrix by the
relations between the PMs.  There are $G$ remaining components. This
polytope is usually referred as the toric diagram of the Master Space
if the bipartite diagram is drawn on a torus \cite{Forcella:2008bb}.
This terminology was used also in \cite{Franco:2012mm} in the field
theory interpretation of the bipartite diagrams on the disk.  Here we
prefer to keep on referring to the matching polytope also for the
projection of the matching polytope defined in \cite{PostnikovCorto}.

\subsubsection*{The matroid polytope}

A second natural polytope that one can define on the bipartite diagram
is the \emph{matroid polytope}.  In appendix \ref{GRPL} we will review
the basic aspects of the matroids.  Here we just give a computative
definition of the matroid polytope.

The matroid polytope is obtained from the matching polytope, by
considering all the internal faces $G_{int}$ and by adding extra
linear relations. These relations are the identification of PMs that
differ by an internal face.  This procedure projects the $G$
dimensional matching polytope on a $G- G_{int}$ dimensional polytope
called the matroid polytope \footnote{As in the case of the matching
  polytope also the vectors that form the matroid polytope are
  coplanar and the actual dimension of the polytope is reduced by
  one.} .

In the language of the POs one can define a matroid polytope as
follows.  First one starts by identifying the POs associated to the
same source set.  A source is the external label associated to an
external line having an arrow pointing on the interior of the graph.
Then one associates a $s\times E$ matrix $Q$ to the bipartite diagram
where $s$ is the number of different sources and $E$ is the number of
external edges. The entry of the matrix $Q_{s_i,E_j}$ is $1$ if, after
the identification, the $E_j$-th edge is in the $s_i$-th PO, $0$
otherwise.  As shown in \cite{Franco:2012wv} this construction is
equivalent to the projection of the master space on another polytope,
usually called the toric diagram of the moduli space in the brane
tiling interpretation.  The projection is usually done by modding the
master space by the internal face variables.  Since POs associated to
the same source set differ by internal faces, modding the matching
polytope by them reproduces the projection on the matroid polytope.
In the rest of the paper we will refer to the toric diagram of the
moduli space as the matroid polytope.

\subsubsection{A remark on polytopes and toric geometry}

At this point of the discussion we want to distinguish the relevant
aspects of the toric geometry associated to the bipartite field
theories (BFTs) defined in \cite{Franco:2012mm} and the one relevant
for the study of the scattering amplitudes.

As we have just reviewed the matching and the matroid polytopes
correspond to the toric diagram of the master and of the moduli space
in the field theoretical interpretation (in the case of the torus) and
this interpretation can be extended to the case of the disk.  It was
done in \cite{Franco:2012mm}, where it was observed that these
theories are not generically SCFTs, and the external edges are not
dynamical fields. Another interpretation of these BFTs was given in
\cite{Xie:2012mr,Heckman:2012jh} and the relation between the two
interpretations was explained in \cite{Franco:2013pg}.

In the case of the BFTs one associates to every vector a coordinate in
$\mathbb{C}^{c}$ and the master space is an affine variety obtained
from the symplectic quotient with the linear relations between the
PMs.  The moduli space is obtained with an extra quotient by the
internal faces.

In the case of the scattering amplitudes the associated toric variety
is projective and non affine and its construction will be explained in
section \ref{toriccell}.

It is maybe worthwhile to state that even if the polytopes in the BFT
and in the scattering amplitude are the same, the associated toric
varieties are different. Indeed in the first case the linear relations
among the PMs are implemented as algebraic polynomial equations on the
$\pi_i$, while in the second case they are implemented as symplectic
quotient relations.

\section{Coordinates on $Gr(k,n)^{tnn}$ from bipartite graph}
\label{mainres}

In this section we first review the parameterization of the totally
non negative Grassmannian in terms of ratios of  Pl\"ucker  coordinates as
discussed in \cite{Talaska}.  We refer the reader to appendix \ref{GRPL}
for the basic definitions of the Pl\"ucker  embedding and the
Grassmannian.  Then we show that these coordinates
become simpler if expressed in terms of the PMs.

The connection between the bipartite graphs and $Gr(k,n)^{tnn}$ is
obtained through the boundary measurement. We review the mathematical
aspects of this connection and the definition of the measurement in
the appendix \ref{bipGR}, while here we try to keep the discussion more
simple and intuitive.

The boundary measurement is a map that assigns a subset in
$Gr(k,n)^{tnn}$ to a bipartite graph. Explicitly one fixes a perfect
orientation.  This defines a set of sources and a set of sinks in the
set of the external edges.  A source is an external edge directed
inside the graph in the PO, while a sink is directed through the
boundary, as in  (\ref{POref1}).  
For every PO one has $k$ sources and $n-k $ sinks.
From this definition of sources and sinks one can define a $k \times
n$ matrix $A$.  The $k$ row entries correspond to the sources and the
$n$ columns are all the possible external edges.

It is possible to assign a positive real variable to every edge and
then express the entries in $A$ as the product of the edge
variables connecting the external edge in the $i$-th row of $A$ with
the external edge in the $j$-th column.  Every minor of $A$ is a
Pl\"ucker coordinate of $Gr(k,n)^{tnn}$.  The final matrix $A$ will be
composed by a $k\times k$ identity matrix corresponding to the paths
from the sources to the sources, and a $k\times n-k$ matrix
corresponding to the paths from the sources to the sinks.  If the
bipartite diagram is associated to the full Grassmannian the $k\times
n-k$ sub matrix have all non zero positive entries, If instead the
bipartite diagram is associated to a sub space of the Grassmannian
then some of the entries of the $k\times n-k$ matrix are zero and they
provide the specification of the cell.  This procedure gives a set of
local coordinates of a patch of $Gr(k,n)^{tnn}$. The peculiar patch
is singled out by the determinant of the $k\times k$ identity matrix
that fixes to $1$ the value of one to the Pl\"ucker coordinates
associated to the selected PO. There are $n \choose k$ of these
patches. One can glue the patches associated to each PO, obtaining a
global coordinatization of $Gr(k,n)^{tnn}$ or of the closure of the
cell associated to the bipartite graph.

\subsection{Coordinates from the \emph{flows}}

An explicit algorithm to obtain the local coordinates of
$Gr(k,n)^{tnn}$ from a bipartite graph was discussed in
\cite{Talaska}. This procedure provide the expression for the ratios
of the $n \choose k$ $-1$ Pl\"ucker coordinates over the Pl\"ucker
coordinate associate to the selected PO.

First of all one has to define the notion of \emph{flow}.  By fixing
the source set $I_{\mathcal{O}}$ and another set of external vertices
$J$ with the same length as $I_{\mathcal{O}}$ a flow from
$I_{\mathcal{O}}$ to $J$ is a collection of disjoint and self avoiding
paths from $I_O-I_O \cap J$ to $J-I_O \cap J$.  A flow from $I_O$ to
$I_O$ is called \emph{conservative}. One can then construct a ratio
between the edge variables in every flow by putting in the numerator
the edges directed from the white to the black vertices and in the
flow and in the denominator the edges from black to white. This ratio
is defined as weight of the flow, $w(F)$.  From now on we will call
$\Delta_J(A)$ the ratio of the $k$-minor associated to $J$ and the
$k$-minor associated to $I$ (this last one is equal to one in this
patch): the set of $k$ sources provided by the selected PO. A a set of
local coordinates for this patch is given by:
\begin{equation} \label{Talaskaa}
\Delta_J(A)=
\frac{\sum_{F:I_O\rightarrow J} w(F)}{\sum_{F':I_O\rightarrow I_O} w(F')} 
=  \frac{p_J}{p_I}
\end{equation}
This is an explicit realization of the connection between
$Gr(k,n)^{tnn}$ and a bipartite graph. $\Delta_J(A)$ are well defined
coordinates in the patch associated to the $I$ sources, and they are
equal to the Pl\"ucker coordinates $p_J$ in this patch. However they
only provide a set of local coordinates in this patch and the
Pl\"ucker coordinates $p_J$ are in general different from the
$\Delta_J(A)$ outside this patch. The $p_J$ provide instead a set of
global coordinates, somehow similar to the homogeneous coordinates of
the projective space \footnote{Note that the $p_I$ that we are
  defining in this paper differ from the $p_I$ of
  \cite{Talaska,PostnikovCorto} since they are global coordinates here
  and local coordinates there. In other words we are not defining
  $p_J=\sum_{F:I_O\rightarrow J} w(F)$.}.

\subsection{Coordinates from the PMs}
\label{Sec:NoiCoord}

In this section we explain in detail the simple map (\ref{Grpi}), 
already introduced in section \ref{Sec:ReviewNima}, between
the Pl\"ucker coordinates $p_J$ and the $\pi_i$ toric embedding
coordinates associated to the PMs, and we provide an explicit proof
for it.  This map provides an explicit relation between the embedding
coordinates of the toric variety associated to a given on-shell
diagram and the Pl\"ucker embedding coordinates of the closure, in the
totally non negative Grassmannian, of the cell associated to the same
on-shell diagram.  This is a global map, in the sense that it is
defined on the embedding coordinates and not only on the local coordinates
on a patch, and provides a way to reformulate the scattering amplitudes
in terms of toric varieties.

In $Gr(k,n)^{tnn}$ given a source set of $k$ elements
$I=(i_1,\dots,i_k)$ and another set $J=(j_1,\dots,j_k)$ one can
associate to $I$ a set of PM coordinates labelled by $\pi_I^{\gamma}$
and to $J$ a set of PM coordinates labelled by $\pi_J^{\beta}$.

This represents a new notation associated to the PMs coordinates
$\pi_i$. This new notation has two indexes, a capital latin index that
denotes the source set of the associated PO, and a greek index that
labels the degeneration (multiple PMs can be associated to the same
source set). The PMs associated to the same source set differ by
internal loops in the POs.  We hope that this double notation for the
same quantity does not generate confusion in the reader.

We are going to prove that the Pl\"ucker coordinates $p_J$ and the
toric coordinates $\pi_J^\beta$ are simply related by a linear map:
\begin{equation}\label{NoiG}
p_J  =\sum_{\beta} \pi_{J}^{\beta}
\end{equation}

This result is proven by showing that this formulation is equivalent
to the one in equation (\ref{Talaskaa}) and proven in \cite{Talaska},
in terms of \emph{flows} and \emph{conservative flows}.  We will first
show that the equation (\ref{Talaskaa}) can be equivalently written in
terms of PMs coordinates as:
\begin{equation}\label{NoiL}
\Delta_J=
\frac{p_J}{p_I} 
=\frac{\sum_{\beta} \pi_{J}^{\beta}}{\sum_{\gamma} \pi_{I}^{\gamma}} 
\end{equation}
Once this local expression is obtained it is easy to derive the global
relation (\ref{NoiG}) by simply considering that the $p_J$ and the
$\pi_J^\beta$ coordinates define a point in the Grassmannian and in
the toric variety only modulo a multiplication by a common factor.  We
can then erase the denominators and find the equation (\ref{NoiG}).  More
formally one can consider that (\ref{NoiL}) is a local map valid on
the patch in the closure of the cell in the totally non negative
Grassmannian such that $p_I$ is different from zero, and in the patch
in the totally non negative toric variety such that $\sum_{\gamma}
\pi_{I}^{\gamma}$ is non vanishing. It is then possible to glue the
map defined on the $I$ patch with the one defined on the $K$ patch by
using the transition functions:
\begin{equation}
\label{formtal}
f_{IK}= \frac{p_I}{p_K}= 
\frac{\sum_{\beta} \pi_{I}^{\beta}}{\sum_{\gamma} \pi_{K}^{\gamma}}
\end{equation}  By iterating this procedure in every patch one obtains
the global map (\ref{NoiG}).

Let us start  with the proof of (\ref{NoiL}).  In
\cite{Talaska} each coordinate is specified by a set $J$,
corresponding to the source set.  By fixing a set $I$ of sources the
numerator of (\ref{Talaskaa}) is identified with the non overlapping
paths (referred as the flows) from $I-\{I \bigcap J\}$ to $J-\{I
\bigcap J\}$.  The denominator is given by the conservative flows.  A
flow is conservative if it is obtained from a fixed perfect
orientation PO$_I^{\alpha}$ by considering all the possible
disconnected loops consistent with the choice of arrows in
PO$_I^{\alpha}$.  Such a PO will be referred as \emph{reference} PO in
the rest of the paper.

The coordinates in $Gr(k,n)^{tnn}$ are expressed in terms of the paths
in the bipartite graph. We want to express them in terms of the PM
coordinates $\pi$.  This can be done by observing that the ratio of
two PM coordinates $\pi_i$ and $\pi_j$ is a path in the bipartite
diagram. The orientation of the these paths is fixed as follows.  One
starts by assigning an orientation to the edges in each PM (for
example from black to white). Then in the ratio of two PM coordinates
(the ratio of the edges that compose the PM themselves) one keeps the
black to white orientation in the numerator and reverse the
orientation in the denominator. In this way one obtains an oriented
path in the bipartite diagram. In terms of the edges this is a ratio
of edges, the ones from black to white in the numerator and the ones
from white to black in the denominator.

There are two possible types of paths in the disk, the \emph{closed
  paths} that surround an internal face and the \emph{open paths} from
the boundary to the boundary.  Note that a ratio of PMs can also be the
composition of disconnected open and/or closed paths.

The description of the paths is simplified once one expresses them in
terms of the POs.  Indeed for every choice of reference
PM$_I^{\alpha}$ the possible open and closed paths on the disk are
identified by the direction of the arrows in PO$_I^{\alpha}$.  The
ratio of the PM coordinates associated to the source set $J$,
$\pi_J^{\beta}$ (with $\beta$ running over the possible multiplicity,
different PMs with sharing the same source set) by the reference
coordinate $\pi_I^{\alpha}$ (where $\alpha$ is fixed) is obtained as
follows.  We keep fixed the orientation of the arrows in
PO$_I^{\alpha}$ and reverse the orientation of the arrows of
PO$_J^{\beta}$ for each $\beta$.  In terms of the edges we divide the
product of edges in PM$_J^{\beta}$ by the edges in PM$_I^{\alpha}$.
This generates a path, either open, either closed or a combination of
both, on the disk.  This is exactly the same construction described
above in terms of the PMs.

Formula (\ref{NoiL}) is equivalent to fix the source set $I$ and
obtain the $J$-th coordinate as the sum of all the possible non
overlapping loops from $I-\{I \bigcap J\}$ to $J-\{I \bigcap J\}$
divided by the sum of the disconnected closed ones in PO$_I^\alpha$.

We start by studying the case with $I \neq J$.  The path connects an
even number (at least two) of points on the boundary of the disk. We
will now prove that this is a flow from $I-\{I \bigcap J\}$ to $J-\{I
\bigcap J\}$.  This is shown in two steps.
\begin{enumerate}
\item \underline{Sources and sinks}\\
  First one has to find the source and the sink set of the flow.
  There are two possible types of edges in the source set of the flow.
  The first type of source edges are the incoming edges of
  PO$_I^{\alpha}$ highlighted in red in PM$_I^{\alpha}$ and not
  highlighted in red in any PM$_J^{\beta}$.  The second type of source
  edges in the flow are the source edges of PO$_I^{\alpha}$ not
  highlighted in red in PM$_I^{\alpha}$ but highlighted in red in
  every PM$_J^{\beta}$.  An analogous discussion holds for the sink
  set.  This description of the source and sink set of the flow is
  equivalent to the definition of \cite{Talaska}: the sources of the
  flow are in $I-\{I \bigcap J\}$ and the sinks are in $J-\{I \bigcap
  J\}$.  For example we consider PO$_{3}$ as the reference PO in
  $Gr(2,4)^{tnn}$ and we look for the Pl\"ucker coordinate associated to
  PO$_{7}$
  \begin{center}
    \begin{tabular}{ccc}
      \begin{tabular}{c} Perfect matching \\ and\\ perfect orientation
      \end{tabular} &
      \begin{tabular}{c}
        \includegraphics[width=3cm]{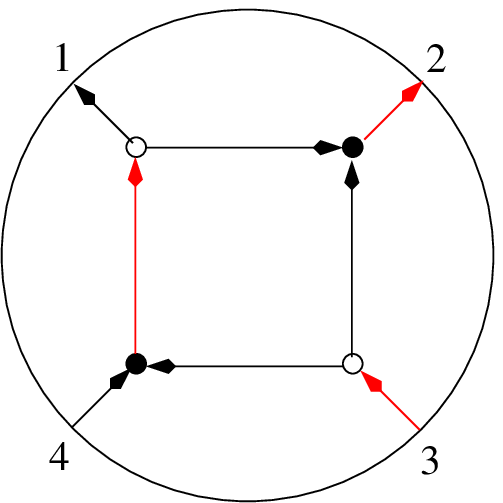}
      \end{tabular}&
      \begin{tabular}{c}
        \includegraphics[width=3cm]{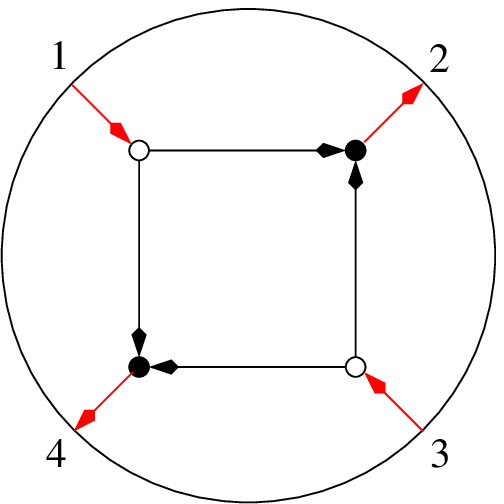}
      \end{tabular}\\ Parametrization&$\pi_3$&$\pi_7$\\ Source set &
      $\{3,4\}$&$\{1,3\}$\\
    \end{tabular}
  \end{center}
  \begin{equation}
    \label{POref1}
  \end{equation} The source set of the flow is obtained from the rules
  given above
  \begin{itemize}
  \item The only edge in the source set of the reference perfect
    orientation and highlighted in PM$_3$ is the external edge
    connected to the external vertex labelled by $3$.  This edge is
    also in PM$_7$ and it cannot be in the flow.
  \item The only edge in the source set of the reference perfect
    orientation and not highlighted in PM$_3$ is the external edge
    connected to the external vertex labelled by $4$.  This edge is
    also in PM$_7$ and it is the source set of the flow.
  \end{itemize}
  One can find that in the same way the sink set of the flow is given
  by the external edge connected to $3$. The final flow is the same
  path obtained with the prescription of \cite{Talaska}.
\item \underline{Non overlapping paths}\\
  Since the path is given by a ratio of PMs every vertex is crossed
  either zero or two times by each path.  This shows that every path
  has no overlap.
\end{enumerate}
The denominator in (\ref{NoiL}) is obtained by considering all the
possible non overlapping closed loops in the reference perfect
orientation PO$_{I}^{\alpha}$ and summing over them.  This is
equivalent to consider all the possible PO$_I^{\gamma}$ with $\gamma
\neq \alpha$.  In terms of the PM coordinates the denominator is
\begin{equation}\label{den}
\frac{\pi_I^{\alpha} + \sum_I \pi_I^{\gamma}}{\pi_I^{\alpha}}
\end{equation} 
The first term corresponds to the trivial path, the absence of loops.

The other terms are obtained from the reference PO labelled by $\alpha$
by inverting the orientation of some of the internal loops. This
procedure generates a new PO labelled by $\gamma$, with the same sink
and source set of the reference PO.
Every PM labelled by $\gamma$ with the same source set of the PM
labelled by $\alpha$ is obtained in this way: one starts from the
reference $\alpha$ and inverts one or more disconnected closed loops,
obtaining a new perfect orientation with the same source set $I$.  At
the level of the coordinates for every $\gamma$ the ratio
$\frac{\pi_I^\gamma}{\pi_I^\alpha}$ corresponds to one of the possible
disconnected loops in PO$_I^\alpha$.  All the loops are obtained by
this prescription because, by fixing $I$, all the possible choices of
internal arrows differ by a closed loops. By starting from one perfect
orientation and by inverting all the (possibily disconnected) internal
loops all the POs corresponding to the same source set $I$ are
obtained.

We give an explicit example of this construction in (\ref{loop36})
\begin{equation}
\label{loop36}
\includegraphics[width=13cm]{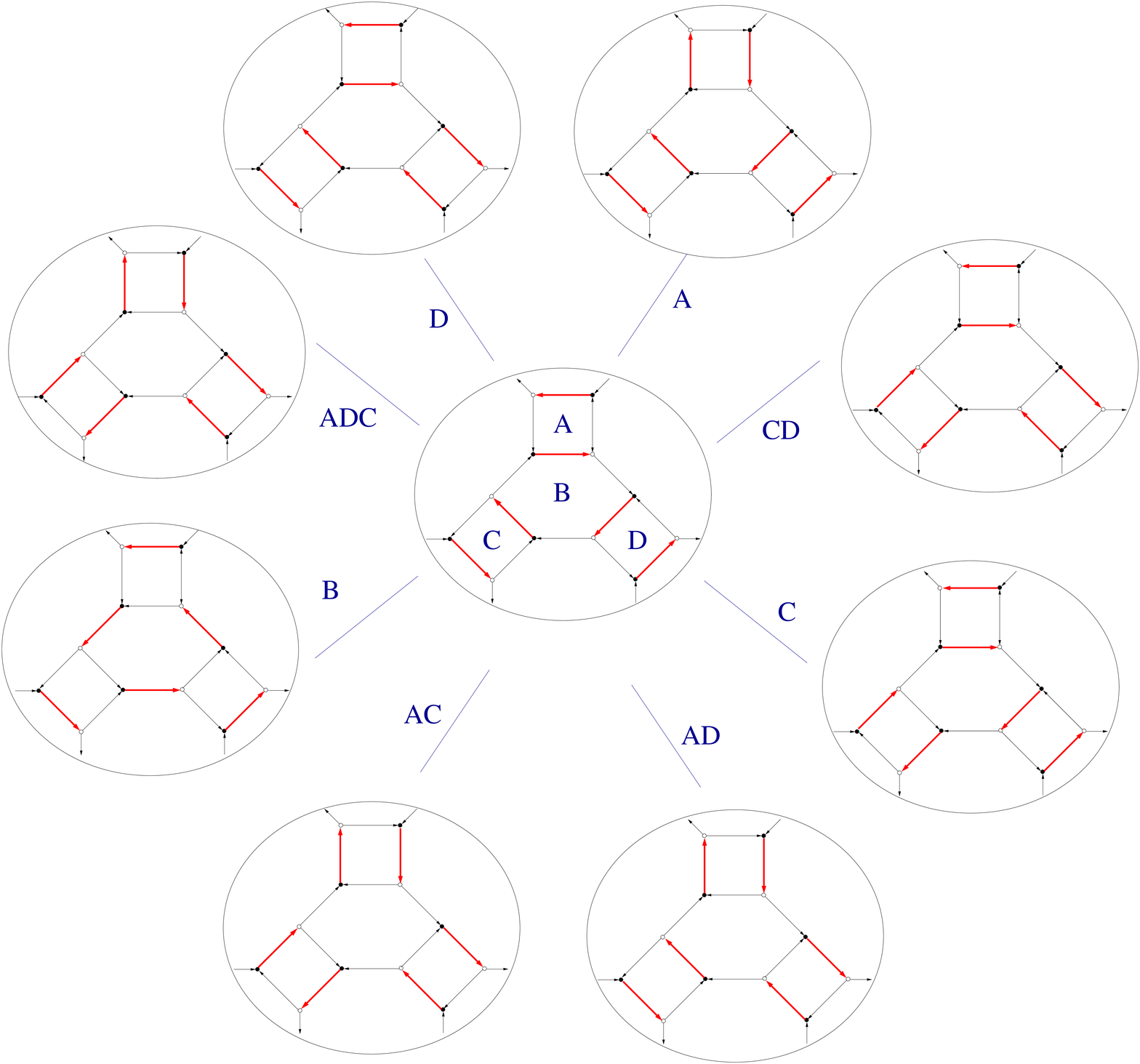}
\end{equation}
\begin{center}
A subcell of $Gr(3,6 )^{tnn}$
\end{center}
This is the case of a subcell of $Gr(3,6)^{tnn}$.  All the possible
PMs sharing the same source set of the reference PM are obtained as
follows. One fixes the external sources and one PM corresponding to
that source. This also sets the reference PO.  This PO may have some
internal loops. For example here we label these loops as $A$, $B$, $C$
and $D$ \footnote{In this case we identify with the capital latin
  letters $A$, $B$, $C$ and $D$ also the coordinates associated to the
  loop, namely the ratio of the edges from black to white divided by
  the edges from white to black, such that the loop is
  counterclockwise.}  in (\ref{loop36}).  The other PMs are found by
reversing the orientation of all the combinations of the internal
loops. In this case for example we have 8 possibilities as shown in
(\ref{loop36}).  Every link connecting two different POs labels the
loops that are reversed.  The ratio of the first PM and one of the
others specifies a disconnected set of loops. In terms of coordinates
we have
\begin{equation}
1+\sum_{\gamma \neq \alpha}\frac{\pi_{I}^{\gamma}}{\pi_{I}^\alpha}=
1+A+B+C+D+AD+AC+CD+ACD 
\end{equation}
This is the same denominator obtained by applying the prescription of
\cite{Talaska}.

To complete the proof of (\ref{NoiL}) we must discuss the case of
$I=J$.  In this case the numerator and the denominator coincide and
one obtains $\Delta_I=1$, that fixes the patch in this local coordinate
system.

\section{Toric Geometry and cell decomposition}
\label{toriccell}

In this section, elaborating on recent results in the mathematical and
physical literature, we discuss the relation between the bipartite
diagrams on disks and toric varieties, and we explain the cell
decomposition of $Gr(k,n)^{tnn}$.

A bipartite diagram is associated to a matching polytope $P$ as
explained in section \ref{poly}.  $P$ is the convex hull of a set of
$c$ vectors $v_{\sigma_i}$ in $G$ dimensions: namely a $G$ dimensional
toric diagram. This toric diagram implements the linear relations
among the PMs $\sigma_i$ as linear relations among the vectors
$v_{\sigma_i}$ that define it.

The polytope $P$ defines a projective toric variety as explained in
\cite{Cox,Sottile}.  The G-vectors $v_{\sigma_i}$, with components:
$v_{\sigma_i} = (v_{\sigma_i}^1, ..., v_{\sigma_i}^G)$, define a
projective toric variety $X_P$ as the closure of the image of the map
$\phi : (\mathbb{C}^*)^{G} \rightarrow \mathbb{CP}^{c-1}$
\begin{equation}
\label{mapphi}
\phi:
{\bf t} = (t_1,\dots,t_{G}) \rightarrow [{\bf t}^{v_{\sigma_1}},\dots, {\bf t}^{v_{\sigma_c}}] = [\pi_1,\dots, \pi_c]
\end{equation}
where ${\bf t}^{v_{\sigma_i}}= t_1^{v_{\sigma_i}^1}
t_2^{v_{\sigma_i}^2} \dots t_G^{v_{\sigma_i}^G}$.

The real part of $X_P(\mathbb{R})$ is defined to be the intersection
of this variety with $\mathbb{RP}^{c-1}$; the positive part of this
real projective variety $X_P^{>0}$ is defined to be the image of
$(\mathbb{R}_{>0})^{G}$ under $\phi$; its non negative part $X_P^{\geq
  0}$ is defined to be the closure of $X_P^{>0}$ in $X_P(\mathbb{R})$.
$X_P^{\geq 0}$ is the main actor in our reinterpretation of scattering
amplitudes in terms of toric geometry.  Indeed (\ref{NoiG}) provides a
map from $X_P^{\geq 0}$ to $Gr(k,n)^{tnn}$, and hence to the
scattering amplitudes.

The procedure to obtain $X_P$ is somehow intuitively equivalent to
interpreting the matching polytope $P$ as the dual toric cone of an
associated toric variety. However the variety defined in this way is
in general non normal, meaning that the vectors defining $P$ generate
the associated cone on the real numbers but not, in general, on the
integers.  The toric ideal $I_P$ of a projective toric variety $X_P$ is
the homogeneous ideal of all the binomials in $\pi_i$ that vanish on
the closure of the image of $\phi$.

One very easy example in the polytope in $\mathbb{R}^3$ defined by the four vectors $v_{\sigma_i}$ satisfying the 
linear relation: 
\begin{equation}
v_{\sigma_1} + v_{\sigma_4} =v_{\sigma_2} + v_{\sigma_3} 
\end{equation}
In this case the toric ideal is generated by the binomial $\pi_1 \pi_4
- \pi_2 \pi_3$, and the non negative real projective toric variety
$X_P^{\geq 0}$ is defined as the zero locus of this binomial,
i.e. $\pi_1 \pi_4 = \pi_2 \pi_3$, in $\mathbb{RP}_{\geq 0}^{3}$.

The Pl\"ucker coordinates $p_J$ are assigned to combination of the
$\pi_i$ as explained in section \ref{Sec:NoiCoord}.  There is a global
way to assign the coordinates: one specifies all the possible POs and
groups the ones with the same source set.  The sources set specifies
the Pl\"ucker coordinates $p_J$, every PO specifies one $\pi_i$: the
$p_J$ is given as the sum of the $\pi_i$ coordinate assigned to POs
with the same sources.  The constraints on the Pl\"ucker coordinates
follow from the constraints on the $\pi_i$ in the toric ideal and they
reproduce the usual Pl\"ucker relations.

This map from $X_P^{\geq 0}$ to $Gr(k,n)^{tnn}$ is an isomorphism for
strictly positive values of the coordinates on the toric variety and
on the Grassmannian, while it in general fails to be a one to one map
for the full non negative part of the toric variety.  However this map
allows the description of the cell decomposition of $Gr(k,n)^{tnn}$
and hence of the singularities of the integrands for the scattering
amplitudes.

The restriction to the positive values of $\pi_i$ for the toric
variety is isomorphic to the top  dimensional cell of the
Grassmannian, namely the Grassmannian itself with all the $p_J$
strictly positive.  The subcells are obtained by deleting the
removable edges, defined in section \ref{Sec:ReviewNima}, in the
associated on-shell diagram.  The removal of an edge corresponds to
remove the PM $\sigma_j$ containing the edge itself.  In the polytope
$P$ it implies that we are restricting to the codimension one external
sub-polytope described by the vectors $v_{\sigma_i}$ that are the ones
that are not associated to the $\sigma_j$ just removed.  Starting from
$P$ one can erase the row corresponding to the edge that has been
removed and the column that corresponds to the removed PM.  By
removing one edge on a graph $\mathcal{G}$ one obtains a subgraph
$\mathcal{H}$.

The new graph $\mathcal{H}$ is associated to a sub-cell of the
Grassmannian \cite{PostnikovLungo,PostnikovCorto}.  The cell is
parameterized as before. One assigns the Pl\"ucker coordinates from
the POs and the Pl\"ucker relations follow from the constraints on the
toric ideal of the sub-polytope.  By iterating this procedure one
obtains the decomposition of $Gr(k,n)^{tnn}$ in positive cells
\footnote{Maybe it is worthwhile to remind the reader that the cells
  are subspaces where the non zero coordinates are strictly
  positive, and are isomorphically associated to the positive part of
  the toric variety associated to the sub-polytope. The closure of
  these cells provides some non trivial topological spaces that are
  the relevant ones for the scattering amplitudes.}.

In \cite{PostnikovLungo,PostnikovCorto} the details of the positroid
cell decomposition of the totally non negative Grassmannian are
discussed. Here we report the main results.  The closure of each
cell is homeomorphic to a closed ball, and the boundary of each cell
is homeomorphic to a sphere.  It follows (see {\bf Theorem 18.5} of
\cite{PostnikovLungo}) that the closure of a cell in $Gr(k,n)^{tnn}$
is the union of the cell itself with the lower dimensional cells.

For a given bipartite graph the image of the parametrization of the
cell under the Pl\"ucker embedding is described by a map to the
projective space.  This map gives origin to a rational map from the
toric variety to $Gr(k,n)^{tnn}$.  This map is well defined on the
totally non negative part of the toric variety and the image of this
map is the closure of the corresponding cell in the totally non
negative Grassmannian.  Since the totally non negative part of the
toric variety is homeomorphic to the matching polytope the cell is
parameterized by the polytope as well.

To conclude: the totally non negative part of the toric variety gets
mapped to the closure of the cell and the totally positive part maps
homeomorphically to the cell.

Two comments are in order.

First, by reducing the polytope on its faces one may find more
subspaces than cells with a fixed dimensionality. Indeed it may happen
that some of the sub-diagrams $\mathcal{H}$ give a different
parameterization of the same cell.  For example in this construction
each $0$-dimensional cell corresponds to a single PM. On the other
hand each $0$-dimensional cell is parameterized by a single Pl\"ucker
coordinate.  In general there are more PMs than Pl\"ucker coordinates:
some of the PMs are associated to the same cell.  They are the PMs
that in the top cell differ by internal loops.

Second, as discussed in {\bf Corollary 7.4} of \cite{PostnikovCorto} in this
reduction there are not cells that are not associated to at least one
sub-diagram.

We can conclude that the projection of the matching polytope on its
faces can be interpreted as a cell decomposition of $Gr(k,n)^{tnn}$ in
positive cells.

\section{Examples}
\label{ESE}

In this section we study examples of increasing complexity to
elucidate the relation among the Grassmannian, the bipartite diagrams
on the disks, and toric geometry.  We start by revisiting in detail,
by using the technology that we introduced in the last sections, the two
examples already proposed in sections \ref{3Gr} and \ref{4Gr}:
namely the simple case of $Gr(1,3)^{tnn}$, and then the bit more
involved case of $Gr(2,4)^{tnn}$ where the top cell is identified by a
box bipartite graph.  We study in great detail this last example, in
particular we use the POs and PMs to assign the coordinates to the
various cells and we study its cell decomposition.  We then move to
study some aspects of some subcells of new higher dimensional totally
non negative Grassmannians.

\subsection{$Gr(1,3)^{tnn}$ and $Gr(2,3)^{tnn}$}

The first two examples correspond to the graphs
in  (\ref{figu13230})
\begin{equation}
\label{figu13230}
\begin{array}{cc}
\includegraphics[width=4cm]{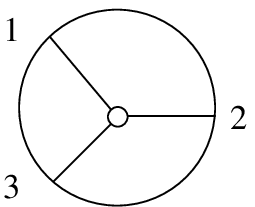}
&
\includegraphics[width=4cm]{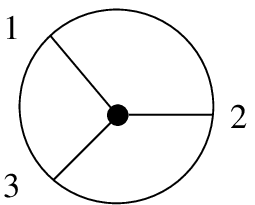}
\end{array}
\end{equation}
These graph are associated to the top cell of $Gr(1,3)^{tnn}$ and
$Gr(2,3)^{tnn}$ respectively.  Notice that since the Grassmannian is
symmetric for $k \rightarrow n-k$, by studying one case one should be
able to recover the other one.  In both cases there are three PMs and
POs as shown in (\ref{figu1323}).
\begin{equation}
\label{figu1323}
\begin{array}{c}
\includegraphics[width=10cm]{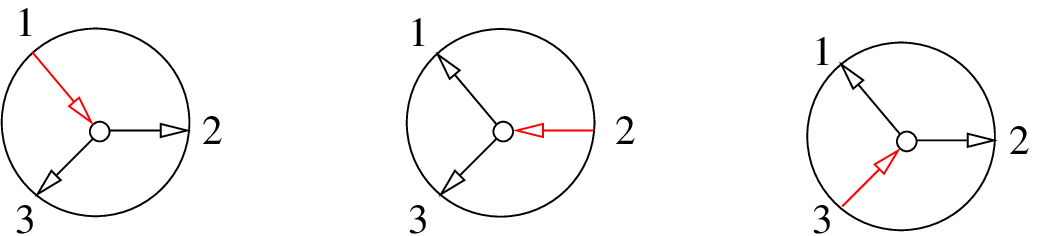}
\\
\includegraphics[width=10cm]{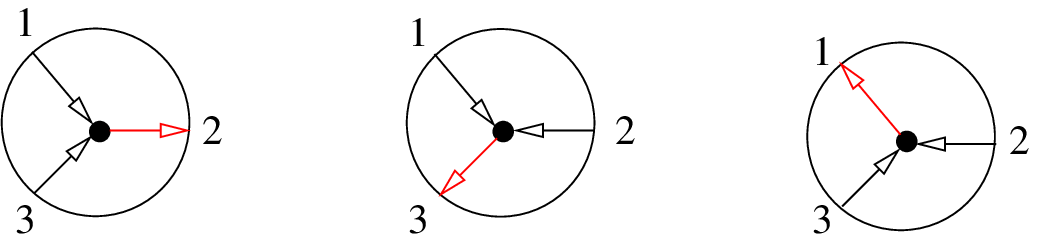}
\end{array}
\end{equation}
and there are no relations between the PMs in each case.  The matching
polytope is obtained from three vectors $v_{\sigma_i} = e_i$ and it
coincides with the matroid polytope because there are not internal
faces.  The local Pl\"ucker coordinates are obtained as explained
above.  First one fixes a reference PO, the $i$-th, and then one gives
the local description of the coordinates by setting to one the
coordinate associated to the choice of the $i$-th PO.  In the first
case the $i$-th PM corresponds to the source set $I_{i} = \{i\}$.  In
every patch selected by the $i$-th PM the $j$-th Pl\"ucker coordinate
(namely the ratio of the $j$-th Pl\"ucker coordinate over the $i$-th
Pl\"ucker coordinate, that however it is equal to $1$ in the $i$-th
patch) is given by $\Delta_{j}=\frac{\pi_j}{\pi_i}$.  A similar
situation holds in the second case. We have
\begin{center}
\begin{tabular}{c||c||c|c|c}
Reference PO &Source & $\Delta_{12}$&$\Delta_{13}$& $\Delta_{23}$\\
\hline
first & $\{1,3\}$ & $\frac{\pi_2}{\pi_1}$&$1$&$\frac{\pi_3}{\pi_1}$\\
second& $\{1,2\}$ & $1$&$\frac{\pi_1}{\pi_2}$&$\frac{\pi_3}{\pi_2}$\\
third & $\{2,3\}$ & $\frac{\pi_2}{\pi_3}$&$\frac{\pi_1}{\pi_3}$&$1$
\end{tabular}
\end{center}
These are the three charts of $\mathbb{RP}_{\geq 0}^2$. The transition
functions between the $i$-th and the $k$-th chart is given by the
function $\frac{\pi_i}{\pi_k}$. In this way it is possible to
reconstruct the full $\mathbb{RP}_{\geq 0}^2$ and reproduce its global
description as presented in section \ref{3Gr}.

\subsection{$Gr(2,4)^{tnn}$}

As we have previously explained in section \ref{4Gr} the top cell of
$Gr(2,4)^{tnn}$ is identified by the box diagram in 
(\ref{24top}).  This diagram describes the tree level scattering of
four particles, two with positive helicities.  In this section we
study this diagram with the help of the toric technology we have
introduced in the previous sections, with the aim to relate the toric
description with the standard description of $Gr(2,4)^{tnn}$ and its
decomposition in positive cells as previously studied in
\cite{ArkaniHamed:2012nw}
\footnote{See \cite{} for an  older study 
of the cell decomposition of $Gr(2,4)$}.

First of all we have to identify the POs and the PMs $\sigma_i$.
They are summarized together in (\ref{24POPM})
\begin{equation}
\label{24POPM}
\includegraphics[width=10cm]{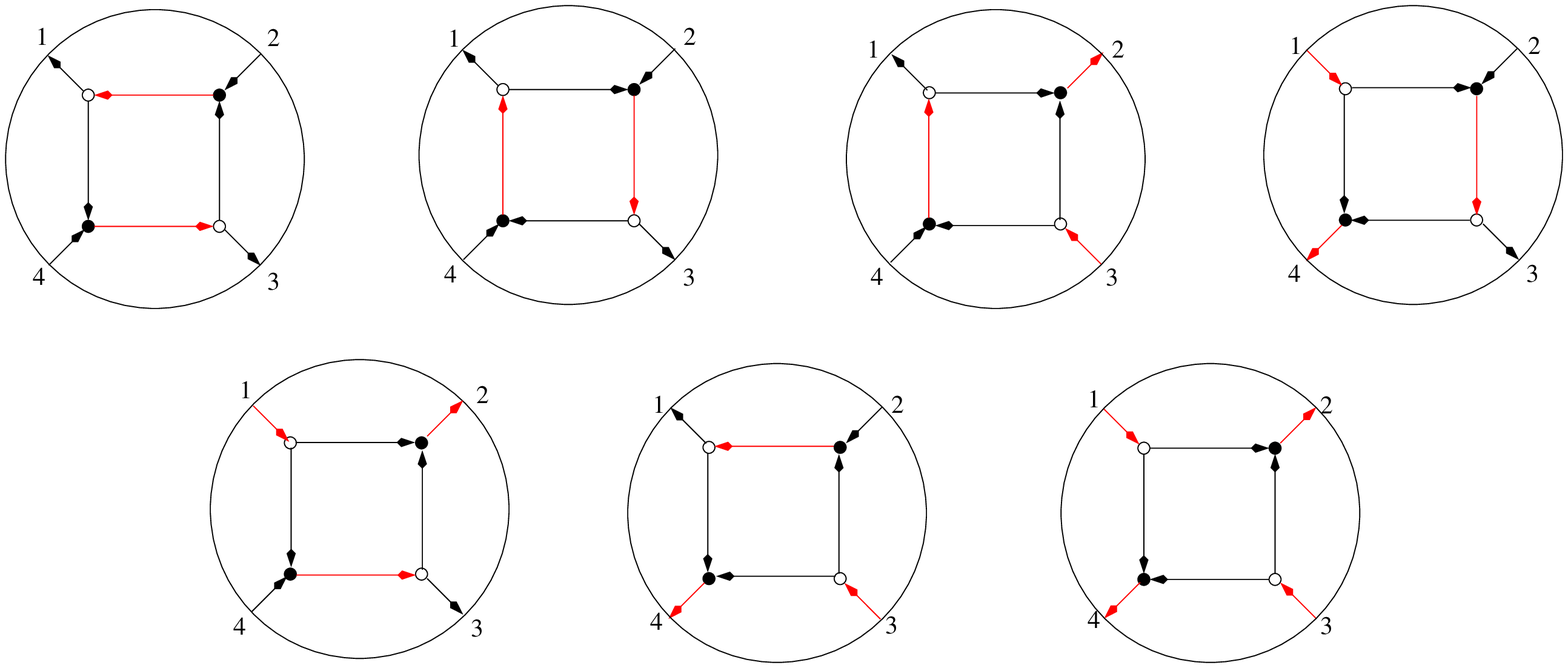}
\end{equation}
The relations (\ref{PMs24}) among the $\sigma_i$ generate the toric
diagram defined by the vectors in equation (\ref{Mtch24}). By following
the procedure explained in the previous sections one can construct the
relations (\ref{dualrel}) among the $\pi_i$ coordinates and hence the
algebraic geometric description of the toric variety as intersection in
$\mathbb{RP}_{\geq 0}^6$, that we re-propose here for completeness:
\begin{equation}
\label{relPM}
\pi_1 \, \pi_7 = \pi_5 \, \pi_6
\quad , \quad
\pi_2 \, \pi_7 = \pi_3 \, \pi_4
\end{equation}
and from here provide a global description of the associated
$Gr(2,4)^{tnn}$ as we have already done in section \ref{4Gr}.

In this section we want instead take an alternative path and start
from a local description, from patch to patch of the Grassmannian and
the toric variety, and then reproduce the global description, by
gluing together all the patches. This procedure should help us to gain
a better insight in the correspondence between toric varieties and
scattering amplitudes.

At the local level the Pl\"ucker coordinates are associated to the
graph in the following way. As explained above every PM identifies a
PO and consequently a source set. The graph in (\ref{24top})
represents the highest dimensional cell in the Grassmannian: namely
the Grassmannian itself with strictly positive coordinates.  Gluing
together the different parametrizations of the top cell of the
Grassmannian defined by all the possible set of sources, it is
possible to reconstruct the closure of the cell and hence the totally
non negative Grassmannian itself.  Indeed by fixing a graph and a PO
we give a description of the cell such that the coordinate associated
to the source set is $1$.  This is a local patch of the totally
non negative Grassmannian and it was proven in \cite{PostnikovLungo}
that in this patch the map between the underling toric variety and the
Grassmannian can be uniquely extended, in a continuous way, to a map
from the closure of the toric variety (namely allowing some of the
coordinates of the perfect matching to go to zero) to the totally non
negative Grassmannian.  There are ${n \choose k }=6$ local maps that
can be glued together to give a global map from $X_P^{\geq 0}$ to
$Gr(k,n)^{tnn}$ that provides a global description of the closure of
the cell itself.

We can indeed assign to every pair $(\mathcal{G},PO_{\mathcal {G}})$,
representing a graph with a given orientation, the coordinates as
explained in section \ref{Sec:NoiCoord}. Furthermore it can be useful
to translate this assignment to the face variables

The face variables are another useful set of coordinates that one can
use to connect the bipartite diagrams with the Grassmannian.  They are
real non negative variables $f_i$ assigned to every face in the
bipartite diagram that come with a counterclockwise orientation of the
graph.  Their product covers the whole disk, and it is normalized to
be $1$: namely $\prod_k f_k=1$, and there are indeed only $G-1$ free
coordinates corresponding to the dimension of the closure of the cell
associated to the on-shell diagram.

It is possible to connect the $\pi_i$ coordinates and the face
variables $f_i$ as follows.  First one fixes a reference PM, say $
\sigma_i$, and the associated coordinate $\pi_i$, then one considers
the ratios of the various PM coordinates $\pi_j$ divided by the
reference PM coordinate $\pi_i$. The ratio $\pi_j/\pi_i$ is then the
variable on the toric variety associated to a loop in the disk of the
on-shell diagram (it can be a product of internal faces and loops
ending on the boundary) and can be expressed as a product or ratio of
face variables.  Hence the face variables $f_i$ are a set of local
coordinates for both the toric variety $X_P^{\geq 0}$ and the closure
of the cell in the $Gr(k,n)^{tnn}$, that are valid in the toric patch
with $\pi_i \neq 0$. In addition they are quite special coordinates,
because once we normalize $G-1$ of these coordinates $f_j$ dividing
them by a reference face coordinate $f_i$: $\tilde{f_j}=f_j/f_i $, the
invariant top form in (\ref{dafare}), used to link the Grassmannian to
the scattering amplitudes, trivializes to a simple product $\prod_j d
\tilde{f_j}/ \tilde{f_j}$. For this reason the tilde face coordinates
are also an interesting example of coordinates that are called
canonical coordinates.

Let now see in detail, in the specific case of $Gr(2,4)^{tnn}$, how
the local map between the toric variety and the Grassmannian works,
both in terms of the embedding coordinates $\pi_j$, and in terms of the
local coordinates $f_j$.
This map is explicitly shown in (\ref{coordass})

Let us comment on the coordinates assignation in (\ref{coordass}).
Every patch has been fixed by a PO. In this case there are seven
possibilities, corresponding to the seven natural patches of the
embedding space $\mathbb{RP}_{\geq 0}^6$ of the toric variety defined
in (\ref{relPM}). A PM, and hence a toric coordinate, corresponds to a
PO. This coordinate is $\pi_i$ where the label $i$ represents the
$i$-th toric patch defined by $\pi_i \neq 0$.  At the same time one
associates a set sources in the on-shell diagram to a PO. As
previously explained the set of sources is in general smaller than the
set of POs and hence in general more POs correspond to the same set
source.  A Pl\"ucker coordinate $p_I$ and a patch in the Pl\"ucker
embedding space are associated to every source.  In this case we have
$\mathbb{RP}_{\geq 0}^5$, for which $p_I \neq 0$.  The local
coordinates $\Delta_J=p_J/p_I$.  are defined in this $I$-th patch.
The local maps shown in (\ref{coordass}) are maps between the $i$-th
patch in the toric variety and the $I$-th patch in the Grassmannian.
The numerators and the denominators of the local coordinates
$\Delta_J$ can be read from the procedure explained in section
\ref{mainres} as follows. One first looks at the structure of the
incoming arrows in every PO listed in (\ref{coordass}). If there is
more than one PO with the same source set one has to consider the sum
of the associated coordinates $\pi_j$.  For every choice of $J$ one
puts in the numerator the sum of the coordinates $\pi_j$ associated to
the same source set.  In the denominator the procedure works in the
same way.  This means that if the reference $i$-th PO is the only one
associated to a choice $I$ of incoming indexes, than one divides by
$\pi_i$.  Otherwise if the $i$-th reference is not the only one
associated to the set of incoming indexes parameterized by $I$, than
one has to divide by the sum of $\pi_i$ with all the other coordinates
$\pi_k$, with $k\neq i$ such that the related $k$-th POs share the
same set of sources, namely the same structure of external indexes
$I$.  One can also parameterize the local coordinates $\Delta_J$ with
the face variables. They are obtained by the ratios of the $\pi_j$
coordinates in the parameterization of $\Delta_J$ as previously
explained

\begin{center}
\begin{tabular}{ccc}
$
\begin{array}{c}
\includegraphics[width=2cm]{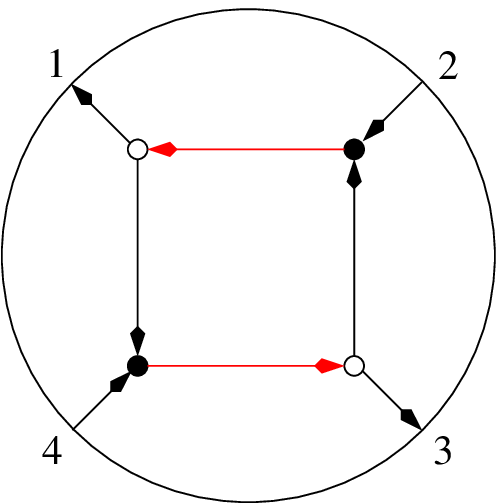}
\end{array}
$
&
$
\begin{array}{cc}
\Delta_{12}=\frac{\pi_4}{\pi_1+\pi_2}&
\Delta_{13}=\frac{\pi_7}{\pi_1+\pi_2}\\
\Delta_{14}=\frac{\pi_5}{\pi_1+\pi_2}&   
\Delta_{23}=\frac{\pi_6}{\pi_1+\pi_2}\\
\Delta_{24}=1&
\Delta_{34}=\frac{\pi_3}{\pi_1+\pi_2}\\
\end{array}
$
&
$
\begin{array}{cc}
\Delta_{12}=\frac{f_2 f_3 f_4}{1+f_1^{-1}}&
\Delta_{13}=\frac{f_2 f_4}{1+f_1^{-1}}\\
\Delta_{14}=\frac{f_2}{1+f_1^{-1}}&
\Delta_{23}=\frac{f_4}{1+f_1^{-1}}\\
\Delta_{24}=1&
\Delta_{34}=\frac{f_2 f_4 f_5}{1+f_1^{-1}}\\
\end{array}
$
\\
$
\begin{array}{c}
\includegraphics[width=2cm]{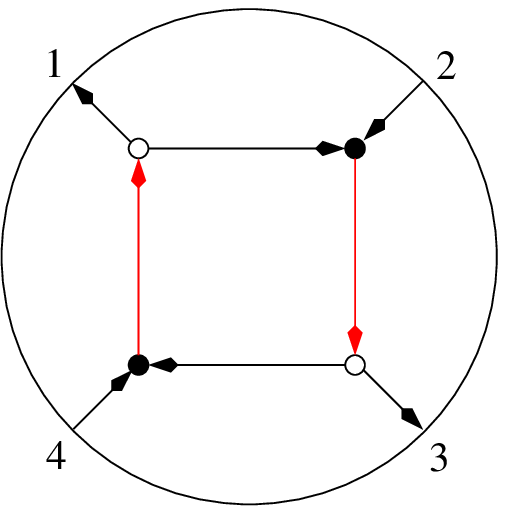}
\end{array}
$
&
$
\begin{array}{cc}
\Delta_{12}=\frac{\pi_4}{\pi_1+\pi_2}&
\Delta_{13}=\frac{\pi_7}{\pi_1+\pi_2}\\
\Delta_{14}=\frac{\pi_5}{\pi_1+\pi_2}&
\Delta_{23}=\frac{\pi_6}{\pi_1+\pi_2}\\
\Delta_{24}=1&
\Delta_{34}=\frac{\pi_3}{\pi_1+\pi_2}\\
\end{array}
$&
$
\begin{array}{cc}
\Delta_{12}=\frac{1}{f_5(1+f_1)}&
\Delta_{13}=\frac{1}{f_3 f_5(1+f_1)}\\
\Delta_{14}=\frac{1}{f_3 f_4 f_5(1+f_1)}&
\Delta_{23}=\frac{1}{f_2 f_3 f_5(1+f_1)}\\
\Delta_{24}=1&
\Delta_{34}=\frac{1}{f_3(1+f_1)}\\
\end{array}
$
\\
$
\begin{array}{c}
\includegraphics[width=2cm]{PO3.eps}
\end{array}
$
&
$
\begin{array}{cc}
\Delta_{12}=\frac{\pi_4}{\pi_3}&
\Delta_{13}=\frac{\pi_7}{\pi_3}\\
\Delta_{14}=\frac{\pi_5}{\pi_3}&
\Delta_{23}=\frac{\pi_6}{\pi_3}\\
\Delta_{24}=\frac{\pi_1+\pi_2}{\pi_3}&
\Delta_{34}=1\\
\end{array}
$
&
$
\begin{array}{cc}
\Delta_{12}=\frac{f_3}{f_5}&
\Delta_{13}=\frac{1}{f_5}\\
\Delta_{14}=\frac{1}{ f_4 f_5}&
\Delta_{23}=\frac{1}{f_2 f_5}\\
\Delta_{24}=f_3+\frac{1}{f_2 f_4 f_5}&
\Delta_{34}=1\\
\end{array}
$
\\$
\begin{array}{c}
\includegraphics[width=2cm]{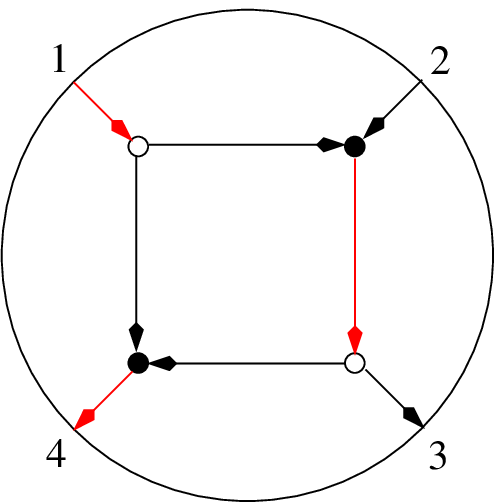}
\end{array}
$
&
$
\begin{array}{cc}
\Delta_{12}=1&
\Delta_{13}=\frac{\pi_7}{\pi_4}\\
\Delta_{14}=\frac{\pi_5}{\pi_4}&
\Delta_{23}=\frac{\pi_6}{\pi_4}\\
\Delta_{24}=\frac{\pi_1+\pi_2}{\pi_4}&
\Delta_{34}=\frac{\pi_3}{\pi_4}\\
\end{array}
$&
$
\begin{array}{cc}
\Delta_{12}=1&
\Delta_{13}=\frac{1}{f_3}\\
\Delta_{14}=\frac{1}{f_3 f_4}&
\Delta_{23}=\frac{1}{f_2 f_3}\\
\Delta_{24}=f_5+\frac{1}{f_2 f_3 f_4}&
\Delta_{34}=\frac{f_5}{f_3}\\
\end{array}
$
\\$
\begin{array}{c}
\includegraphics[width=2cm]{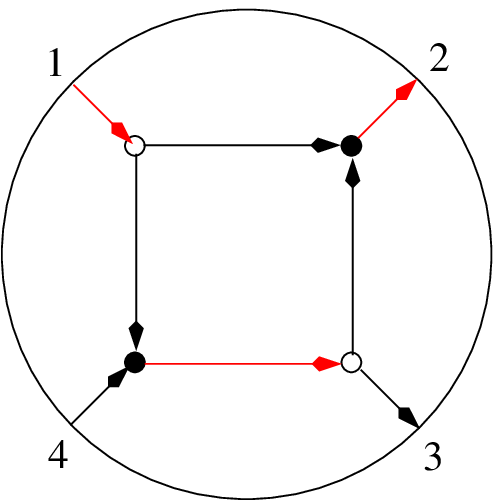}
\end{array}
$
&
$
\begin{array}{cc}
\Delta_{12}=\frac{\pi_4}{\pi_5}&
\Delta_{13}=\frac{\pi_7}{\pi_5}\\
\Delta_{14}=1&
\Delta_{23}=\frac{\pi_6}{\pi_5}\\
\Delta_{24}=\frac{\pi_1+\pi_2}{\pi_5}&
\Delta_{34}=\frac{\pi_3}{\pi_5}\\
\end{array}
$&
$
\begin{array}{cc}
\Delta_{12}=f_3 f_4&
\Delta_{13}=f_4\\
\Delta_{14}=1&
\Delta_{23}=\frac{f_4}{f_2}\\
\Delta_{24}=\frac{1}{f_2}+f_3 f_4 f_5&
\Delta_{34}=f_4 f_5\\
\end{array}
$
\\$
\begin{array}{c}
\includegraphics[width=2cm]{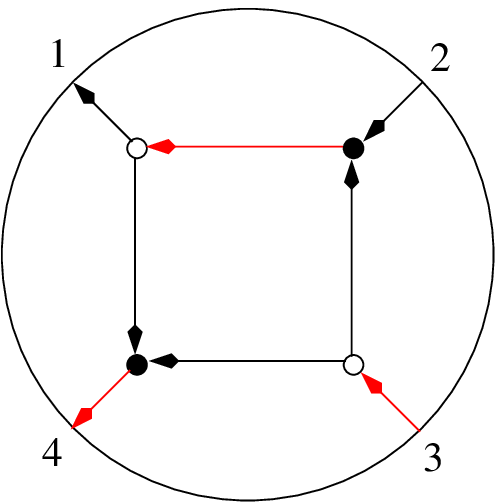}
\end{array}
$
&
$
\begin{array}{cc}
\Delta_{12}=\frac{\pi_4}{\pi_6}&
\Delta_{13}=\frac{\pi_7}{\pi_6}\\
\Delta_{14}=\frac{\pi_5}{\pi_6}&
\Delta_{23}=1\\
\Delta_{24}=\frac{\pi_1+\pi_2}{\pi_6}&
\Delta_{34}=\frac{\pi_3}{\pi_6}\\
\end{array}
$&
$
\begin{array}{cc}
\Delta_{12}=f_2 f_3&
\Delta_{13}=f_2\\
\Delta_{14}=\frac{f_2}{f_4}&
\Delta_{23}=1\\
\Delta_{24}=\frac{1}{f_4}+f_2 f_3 f_5&
\Delta_{34}=f_2 f_5\\
\end{array}
$
\\$
\begin{array}{c}
\includegraphics[width=2cm]{PO7.eps}
\end{array}
$
&
$
\begin{array}{cc}
\Delta_{12}=\frac{\pi_4}{\pi_7}&
\Delta_{13}=1\\
\Delta_{14}=\frac{\pi_5}{\pi_7}&
\Delta_{23}=\frac{\pi_6}{\pi_7}\\
\Delta_{24}=\frac{\pi_1+\pi_2}{\pi_7}&
\Delta_{34}=\frac{\pi_3}{\pi_7}\\
\end{array}
$&
$
\begin{array}{cc}
\Delta_{12}=f_3&
\Delta_{13}=1\\
\Delta_{14}=\frac{1}{f_4}&
\Delta_{23}=\frac{1}{f_2}\\
\Delta_{24}=\frac{1}{f_2 f_4}+ f_3 f_5&
\Delta_{34}=f_5\\
\end{array}
$
\\
\end{tabular}
\end{center}
\begin{equation}
\label{coordass}
\end{equation}

\subsubsection{Gluing the patches}

In the last subsection we discussed how to assign a local system of
toric coordinates to a bipartite diagram that parameterizes a patch
of the closure of a cell in the totally non negative Grassmannian.
The system is local because every patch is selected by a reference PO
and an associated set of non-zero coordinates, $\pi_i$ and $p_I$, both
in the toric variety and in the cell in the Grassmannian. The
$\Delta_J$ coordinates are ratios of the (sums of) $\pi_j$ coordinates
assigned to the PMs. Every patch has one $\Delta_J$ coordinate, namely
the one that fix the patch: $\Delta_I=p_I/p_I$, fixed to $1$. This is
the coordinate related to the reference PO that fixes the patch
itself.

The maps in (\ref{coordass}) are in principle maps between the
positive part of the toric variety (\ref{relPM}), associated to the
on-shell diagram, to the positive part of the Grassmannian
$Gr(2,4)$. Namely both the $\pi_j$ and the $\Delta_J$ are positive non
zero variables.

To have global map between the non negative toric variety and the non
negative Grassmannian, we should proceed in two steps: first of all
extend the maps in (\ref{coordass}) to their closure in every patch,
and then glue all the patches together.  As previously stated in
\cite{PostnikovLungo} it was proven that these maps have a unique
continuous extension.  Hence following the discussion in section
\ref{mainres} one can glue the different patches by using the
transition maps in formula (\ref{formtal}) to obtain a global
description of the parameterization of the closure of the cell
associated to the bipartite diagram in the totally non negative
Grassmannian, in this particular case of the full $Gr(2,4)^{tnn}$.

The transition map is obtained as explained in section
(\ref{Sec:NoiCoord}) and considering that here $\Delta_J=p_J/p_I$ in
the $I$-th patch.  With the aim to clarify the discussion, let us
introduce the notation $\Delta_J^{(I)}$ for the $p_J$ coordinate in
the patch where $p_I \neq 0$: $\Delta_J^{(I)}=p_J/p_I$. Then the
transition map between the $I$-th patch and the $K$-th patch is:
\begin{equation}
\Delta_{J}^{(K)}=\frac{\sum_{\beta} \pi_{I}^{\beta}}{\sum_{\gamma} \pi_{K}^{\gamma}}\Delta_{J}^{(I)}
\end{equation}

Of course the $\pi_j$ coordinates in the previous transition function
can be potentially defined on different toric patches of the toric
variety, and when this happens, it is understood that every time it is
considered the transition function $\pi_i/ \pi_k$ between the $i$-th
and the $k$-th toric patch.

One can assemble the transition functions $f_{I,K}$ in a matrix as
follows:
\begin{equation}
  \label{trm}
  f_{I,K}=
  \left(
    \begin{array}{ccccccc}
      & \{2,4\} & \{3,4\} & \{1,2\} & \{1,4\} & \{2,3\} & \{1,3\} \\
      \{2,4\} & 1 & \frac{\pi _1+\pi _2}{\pi _3} & \frac{\pi _1+\pi _2}{\pi _4} & \frac{\pi _1+\pi _2}{\pi _5} 
      & \frac{\pi _1+\pi _2}{\pi _6} & \frac{\pi _1+\pi _2}{\pi _7} \\
      \{3,4\} & \frac{\pi _3}{\pi _1+\pi _2} & 1 & \frac{\pi _3}{\pi _4} & \frac{\pi _3}{\pi _5} & \frac{\pi _3}{\pi _6} & \frac{\pi _3}{\pi _7} \\
      \{1,2\} & \frac{\pi _4}{\pi _1+\pi _2} & \frac{\pi _4}{\pi _3} & 1 & \frac{\pi _4}{\pi _5} & \frac{\pi _4}{\pi _6} & \frac{\pi _4}{\pi _7} \\
      \{1,4\} & \frac{\pi _5}{\pi _1+\pi _2} & \frac{\pi _5}{\pi _3} & \frac{\pi _5}{\pi _4} & 1 & \frac{\pi _5}{\pi _6} & \frac{\pi _5}{\pi _7} \\
      \{2,3\} & \frac{\pi _6}{\pi _1+\pi _2} & \frac{\pi _6}{\pi _3} & \frac{\pi _6}{\pi _4} & \frac{\pi _6}{\pi _5} & 1 & \frac{\pi _6}{\pi _7} \\
      \{1,3\} & \frac{\pi _7}{\pi _1+\pi _2} & \frac{\pi _7}{\pi _3} & \frac{\pi _7}{\pi _4} & \frac{\pi _7}{\pi _5} & \frac{\pi _7}{\pi _6} & 1 \\
    \end{array}
  \right)
\end{equation}
Each column and row of this matrix is associated to a set
$I,K=\{.,.\}$ that specifies a patch and is associated to a reference
PM or PO.

For example we can consider the transition function for the coordinate
$\Delta_{34}$, between the second patch and the third patch in
(\ref{coordass}), defined respectively by the set of sources: $(24)$
and $(34)$: $\Delta_{34}^{(24)}$ and $\Delta_{34}^{(34)}$.  In this
case the transition map is obtained from the source set identified by
the third and the second PO and we have:
\begin{equation}
  \Delta_{34}^{(34)}=\frac{\pi_1+\pi_2}{\pi_3}
  \Delta_{34}^{(24)}=1
\end{equation}
One can use these transition maps to glue the various patches.

These rules can be translated from the $\pi_i$ to the face variables
as follows. First one defines the face variables in terms of the
$\pi_i$ as
\begin{eqnarray}
  f_1=\frac{\pi_1}{\pi_2},\quad
  f_2=\frac{\pi_5}{\pi_1}=\frac{\pi_7}{\pi_6},\quad
  f_3=\frac{\pi_2}{\pi_3}=\frac{\pi_4}{\pi_7},\quad
  f_4=\frac{\pi_7}{\pi_5}=\frac{\pi_6}{\pi_1},\quad
  f_5=\frac{\pi_3}{\pi_7}=\frac{\pi_2}{\pi_4}
\end{eqnarray}
that automatically enforce the constraint $\prod f_i=1$.  This
constraint is useful in the gluing of the local patches.  The
transition map in (\ref{trm}) becomes
\begin{equation}
  \label{trmf}f_{I,K}=
  \left(
    \begin{array}{ccccccc}
      0 & \{2,4\} & \{3,4\} & \{1,2\} & \{1,4\} & \{2,3\} & \{1,3\} \\
      \{2,4\} & 1 & f_3 \left(f_1+1\right) & f_5 \left(f_1+1\right) & \frac{f_1+1}{f_1 f_2} & \frac{f_1+1}{f_1 f_4} & f_3 f_5 \left(f_1+1\right) \\
      \{3,4\} & \frac{1}{f_3 \left(f_1+1\right)} & 1 & \frac{f_5}{f_3} & f_4 f_5 & f_2 f_5 & f_5 \\
      \{1,2\} & \frac{1}{f_5 \left(f_1+1\right)} & \frac{f_3}{f_5} & 1 & f_3 f_4 & f_2 f_3 & f_3 \\
      \{1,4\} & \frac{f_1 f_2}{f_1+1} & \frac{1}{f_4 f_5} & \frac{1}{f_3 f_4} & 1 & \frac{f_2}{f_4} & \frac{1}{f_4} \\
      \{2,3\} & \frac{f_1 f_4}{f_1+1} & \frac{1}{f_2 f_5} & \frac{1}{f_2 f_3} & \frac{f_4}{f_2} & 1 & \frac{1}{f_2} \\
      \{1,3\} & \frac{1}{f_3 f_5 \left(f_1+1\right)} & \frac{1}{f_5} & \frac{1}{f_3} & f_4 & f_2 & 1 \\
    \end{array}
  \right)
\end{equation}
For example if we want to glue the patch identified by the fifth PM
with the patch identified by the fourth one, we proceed as follows.
First we write the $\Delta$ coordinates in each patch.  The fifth PM
selects the fourth row of the matrix, because it fixes the patch
$I=\{1,4\}$.  The third column is fixed by the fourth PM, associated
to the sources $K= \{1,2\}$.

One starts by writing the Pl\"ucker coordinates $\Delta_{I}^{(1,4)}$
in terms of the faces
\begin{equation}
  \label{patch5}
  \{\Delta_{12}^{(1,4)},\Delta_{13}^{(1,4)},\Delta_{14}^{(1,4)},\Delta_{23}^{(1,4)},
  \Delta_{24}^{(1,4)},\Delta_{34}^{(1,4)}\}
  =\{f_3 f_4,f_4,1,\frac{f_4}{f_2},
  \frac{1+f_1}{f_1 f_2},f_4 f_5\}
\end{equation}
in the new patch fixed by $\pi_4$ they become
\begin{equation}
  \label{patch4}
  \{\Delta_{12}^{(1,2)},\Delta_{13}^{(1,2)},\Delta_{14}^{(1,2)},\Delta_{23}^{(1,2)},
  \Delta_{24}^{(1,2)},\Delta_{34}^{(1,2)}\}
  =\{
  1,\frac{1}{f_3},\frac{1}{f_3 f_4},\frac{1}{f_3 f_2},
  f_5(1+f_1),\frac{f_5}{f_3}\}
\end{equation}
The transition map is the element $f_{4,3}$ of the matrix
(\ref{trmf}).  It implies the relation
\begin{equation}
  \label{trans45}
  \Delta_{J}^{(1,2)} = \frac{1}{f_3 f_4} \Delta_{J}^{(1,4)} 
\end{equation}
for every $J$.  One can observe that this relation holds for every
pair of coordinates in (\ref{patch5}) and (\ref{patch4}).  The case
$J=\{2,4\}$ needs an additional comment. In this case the equation
(\ref{trans45}) holds thanks to the constraint $\prod f_i =1$.

\subsubsection{Cell decomposition}

Up to now we discussed the parameterization of a cell and of its
closure in the totally non negative Grassmannian $Gr(k,n)^{tnn}$ in
terms of the PM coordinates $\pi_j$ of a bipartite diagram on a disk.
As we discussed in section \ref{Sec:ReviewNima} to connect these
algebraic geometric varieties and the on-shell diagrams to the
scattering amplitude processes it is important
to study the decomposition of $Gr(k,n)^{tnn}$ in subcells, and the
related process in terms of toric geometry.  The subcells are obtained by
sending to zero some of the Pl\"ucker coordinates $p_J$, while forcing
the remaining Pl\"ucker coordinates to be strictly positive, and
preserving the positivity constraints.    The union of the cells
reconstructs the space itself.  In appendix \ref{SubSec:CellDec} we
will provide a more rigorous definition of the cell
decomposition. Here we prefer to keep the discussion more intuitive
because our main aim is to provide the cell decomposition in terms of
bipartite diagrams and toric geometry. We refer the reader to the
appendix and to \cite{PostnikovLungo,PostnikovCorto} for a more
rigorous and detailed discussion.

We already discussed the cell decomposition of $Gr(2,4)^{tnn}$ in
section \ref{4Gr}. Here we will provide a detailed discussion of it in
terms of toric geometry and the previously explained map between the
PM coordinates and the Pl\"ucker coordinates.

As already explained in section \ref{Sec:ReviewNima} the reduction of
the closure of the top cell, namely the totally non negative
Grassmannian itself, into its subcells is related to the notion of
removability of an edge.  On the bipartite diagrams the notion of edge
removability is defined in terms of the zig-zag paths. First one draws
the zig-zag paths and then looks for the edges 
that correspond to single intersections of the paths.
By removing one of those edges in the 
on-shell diagram, one identifies the new on-shell diagram associated to a
subcell of the Grassmannian.  At the level of the PMs it corresponds
to erase the PMs $\sigma_i$ containing the removed edge. 
This procedure correspond to go on some facet of the matching
polytope $P$, described by the  vectors $v_{\sigma_i}$

The selected facet is associated to a toric variety in the usual way
and this variety maps to the subcell of the Grassmannian associated to
the same new on-shell diagram. For example in the on-shell diagram
associated to the top cell of $Gr(2,4)^{tnn}$ one can eliminate one of
the four edges inside the box. There are four different graphs arising
from this elimination as shown in (\ref{24sub1}).
\begin{equation}
\label{24sub1}
\includegraphics[width=10cm]{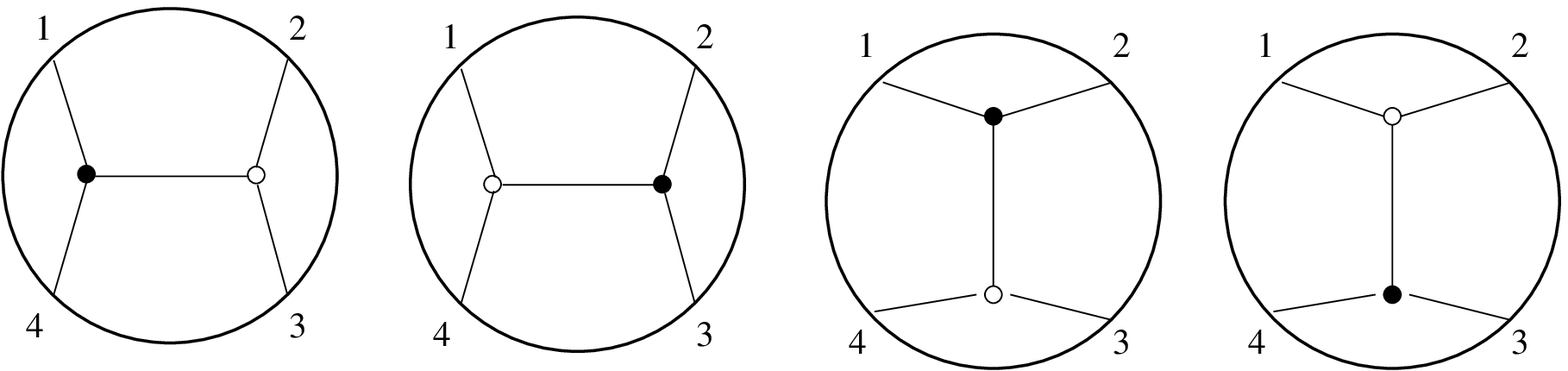}
\end{equation}
This process is usually formally interpreted in terms of matroids.
Even if we will review the definition of the matroid in the appendix
\ref{APPE}, we prefer to give a short introduction to this object here.
A matroid $\mathcal{M}$ of rank $k$ on the set $[n]$ is a non empty
collection of $k$-element subset in $[n]$ such that for any $I$ and
$J$ in $\mathcal{M}$ and $i\in I$ there is $j \in J$ such that
$(I\setminus i) \bigcup j \in \mathcal{M}$.  This property is called
the exchange axiom.  A matroid can be represented by the matroid
polytope defined above.

The top cell identifies the single matroid with non-vanishing
Pl\"ucker coordinates.  In this case the matroid is called positroid.
In principle there can be six matroids with five elements, such that
in each of them all of the Pl\"ucker coordinates don't vanish but
one. But only four of them satisfy the exchange axiom of the
matroids. Indeed two of them should have had some negative or zero
coordinates because of the Pl\"ucker relations.  For example the
matroid with $p_{13}=0$ would impose the constraint \footnote{ Note
  that from now on we will work with the global coordinates $p_I$
  without referring anymore to $\Delta_I$ coordinates and to the
  patches identified by some PO.}
$$
p_{12} p_{34} + p_{14} p_{23}=0
$$
that cannot be satisfied for strictly positive values.  The four
choices that correspond to a matroid are the four graphs found by
removing an internal edge each time.

Let us now fix one of the subcells, for example the one defined by
$p_{23}=0$.  In this case there are five surviving POs. In terms of
the $\pi_i$ coordinates they are :
$\pi_2,\pi_3,\pi_4,\pi_5,\pi_7$. There is one relation among them
\begin{equation}
\pi_2 \pi_7 = \pi_3 \pi_4
\end{equation}
and all the edges are removable.  

Erasing one of them 
corresponds to eliminate some PMs. We can go on a subcell by eliminating
one of the following sets of coordinates.
\begin{equation}
\pi_2 \, \& \, \pi_3 
\quad \hbox{or} \quad 
\pi_2 \, \& \, \pi_4 
\quad \hbox{or} \quad
\pi_3 \, \& \, \pi_7 
\quad \hbox{or} \quad 
\pi_4 \, \& \, \pi_7 
\quad \hbox{or} \quad
\pi_5 
\end{equation}
A similar situation holds on the other three subcells.

By eliminating $\pi_2$ and $\pi_3$ we set both $p_{24}$ and $p_{34}$
to zero.  The non zero coordinates in this local parameterization of
the subcell are $p_{12}$, $p_{13}$ and $p_{14}$, that satisfy the
exchange axiom.  By eliminating $\pi_2$ and $\pi_4$ the three non zero
coordinates are $p_{13}$, $p_{14}$ and $p_{34}$, still a matroid.  By
eliminating $\pi_3$ and $\pi_7$ the three non zero coordinates are
$p_{12}$, $p_{23}$ and $p_{24}$ and by eliminating $\pi_4$ and $\pi_7$
the three non zero coordinates are $p_{23}$, $p_{24}$ and $p_{34}$.
The last possibility, $p_{14}=0$ is still a matroid.  Differently from
the first reduction from the top cell here we observe that some of the
subcells have more than one extra vanishing Pl\"ucker coordinates to
respect of the higher dimensional cell. In any case all of them satisfy
the exchange axiom of the matroids.

By considering all the possibilities we have
\begin{center}
\begin{tabular}{cc}
\begin{tabular}{c}
$\pi_1=0$\\
$\pi_5=0$\\
$p_{14}=0$
\end{tabular}
&
\begin{tabular}{c|c|c}
PM& Vanishing $p$  &Non-vanishing $p$ \\
\hline
$\pi_2 = \pi_3=0$& $p_{24}= p_{34}=0$
&$\{p_{12}, p_{13}, p_{23}\}$\\
$\pi_2 = \pi_4=0$& $p_{24}= p_{12}=0$
&$\{p_{13}, p_{23}, p_{34}\}$\\
$\pi_3=\pi_7=0$ & $p_{34}= p_{13}=0$ &
$\{p_{12}, p_{23}, p_{24}\}$ \\
$\pi_4 = \pi_7=0$ & $p_{12}= p_{13}=0$ 
&$\{p_{23},p_{24},p_{34}\}$\\
$\pi_6=0$& $p_{23}=0$& 
$\{p_{12}, p_{13}, p_{24}, p_{34}\}$
\end{tabular}
\end{tabular}
\\
~
\\
\begin{tabular}{cc}
\begin{tabular}{c}
$\pi_1=0$\\
$\pi_6=0$\\
$p_{23}=0$
\end{tabular}
&
\begin{tabular}{c|c|c}
PM& Vanishing $p$  &Non-vanishing $p$ \\
\hline
$\pi_2 = \pi_3=0$& $p_{24}= p_{34}=0$
&$\{p_{12}, p_{13}, p_{14}\}$\\
$\pi_2 = \pi_4=0$& $p_{24}= p_{12}=0$
&$\{p_{13}, p_{14}, p_{34}\}$\\
$\pi_3=\pi_7=0$ & $p_{34}= p_{13}=0$ & 
$\{p_{12}, p_{14}, p_{24}\}$\\
$\pi_4 = \pi_7=0$ & $p_{12}=p_{13}=0$ &
$\{p_{14}, p_{24}, p_{34}\}$\\
$\pi_5=0$& $ p_{14}=0$& 
$\{p_{12}, p_{13}, p_{24}, p_{34}\}$
\end{tabular}
\end{tabular}
\\
~
\\
\begin{tabular}{cc}
\begin{tabular}{c}
$\pi_2=0$\\
$\pi_3=0$\\
$p_{34}=0$
\end{tabular}
&
\begin{tabular}{c|c|c}
PM& Vanishing $p$  &Non-vanishing $p$ \\
\hline
$\pi_1 = \pi_5=0$& $p_{24}= p_{14}=0$
&$\{p_{12}, p_{13}, p_{23}\}$\\
$\pi_1 = \pi_6=0$& $p_{24}= p_{23}=0$
&$\{p_{12}, p_{13}, p_{14}\}$\\
$\pi_5=\pi_7=0$ & $p_{14} =p_{13}=0$ & 
$\{p_{12}, p_{23}, p_{24}\}$\\
$\pi_6 = \pi_7=0$ & $p_{23} =p_{13}=0$ &
$\{p_{12}, p_{14}, p_{24}\}$\\
$\pi_4=0$& $p_{12}=0$& 
$\{p_{13}, p_{14}, p_{23}, p_{24}\}$
\end{tabular}
\end{tabular}
\\
~
\\
\begin{tabular}{cc}
\begin{tabular}{c}
$\pi_2=0$\\
$\pi_4=0$\\
$p_{12}=0$
\end{tabular}
&
\begin{tabular}{c|c|c}
PM& Vanishing $p$  &Non-vanishing $p$ \\
\hline
$\pi_1 = \pi_5=0$& $p_{24}= p_{14}=0$
&$\{p_{34}, p_{13}, p_{23}\}$\\
$\pi_1 = \pi_6=0$& $p_{24}= p_{23}=0$
&$\{p_{34}, p_{13}, p_{14}\}$\\
$\pi_5=\pi_7=0$ & $p_{14} =p_{13}=0$ &
$\{p_{23}, p_{24}, p_{34}\}$ \\
$\pi_6 = \pi_7=0$ & $p_{23} =p_{13}=0$ &
$\{p_{14}, p_{24}, p_{34}\}$\\
$\pi_4=0$& $p_{34}=0$& 
$\{p_{13}, p_{14}, p_{23}, p_{24}\}$
\end{tabular}
\end{tabular}
\end{center}
There are 20 matroids but not all of them are different.  Two of them have
four non zero coordinates, and both of them are obtained in two
different ways.  The other eight distinct matroids have three elements
and each one appears twice.  One can iterate this procedure to reach
the lower dimensional cells until the zero dimensional cells, where
just a single coordinate is non vanishing.

This section provided an explicit map from the cell decomposition of
the Grassmannian to the partial facet decomposition of the matching
polytope and hence to the partial decomposition of the non negative
toric variety in some toric totally positive sub varieties.

\subsection{A subcell of $Gr(3,6)^{tnn}$}
\label{Gr36cell}

In the previous sections, by making use of the toric geometry, we
revisited the two examples: $Gr(1,3)^{tnn}$ and $Gr(2,4)^{tnn}$, that
we have already discussed at the beginning of the paper.  In this
section we would like to study another more involved example, already
discussed in \cite{Franco:2012mm}: a particular subcell of
$Gr(3,6)^{tnn}$.  The aim of the section is to apply the algorithm that we
developed to map the toric PM coordinates to the Pl\"ucker coordinates
to a complex example and check its convenience.

The top cell of $Gr(3,6)^{tnn}$ can be pictorially represented by
the following diagram
\begin{equation}
\label{top36fr}
\includegraphics[width=5cm]{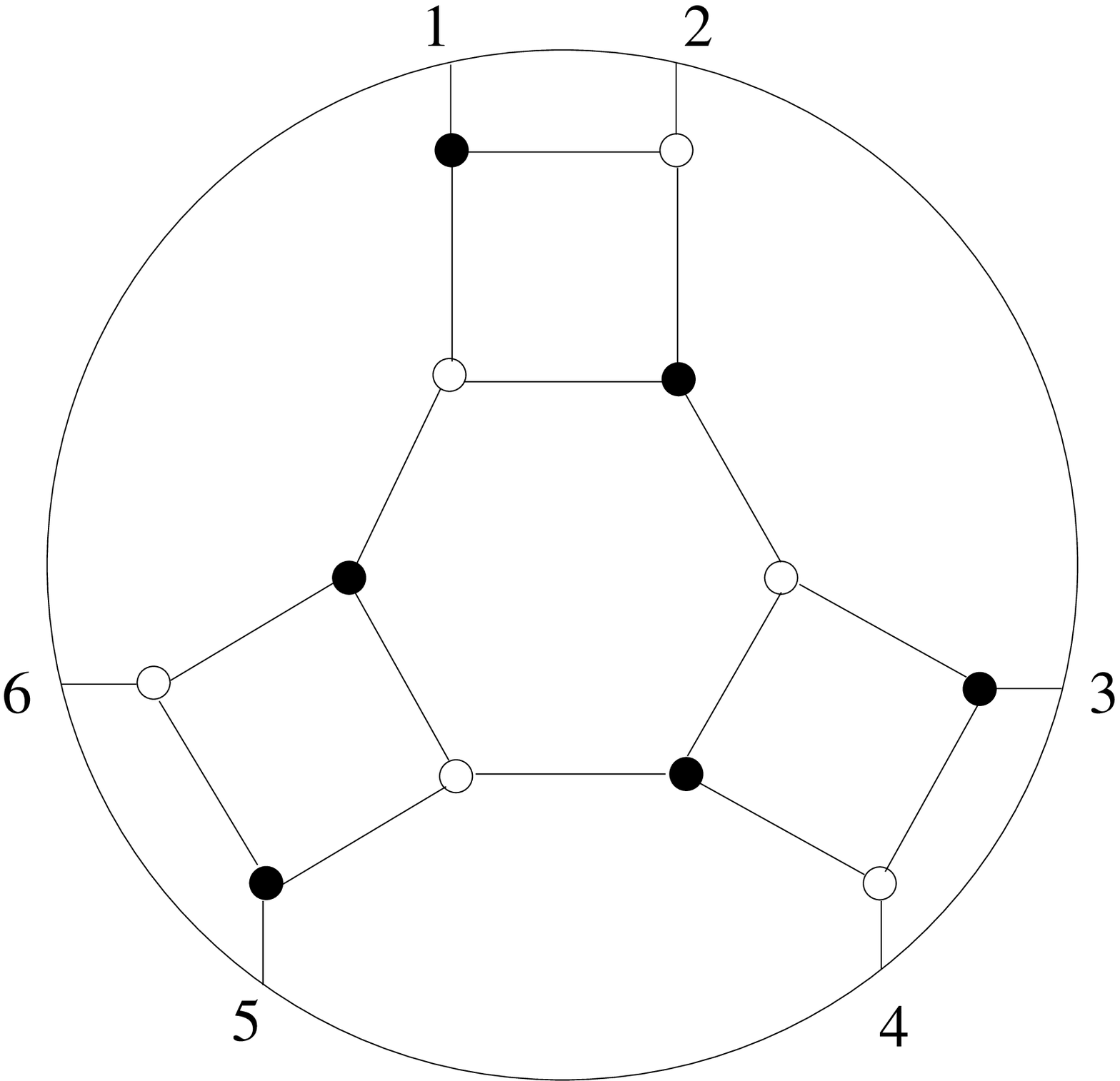}
\end{equation}
Here we are interested in a subcell, obtained by setting some edge
variable to zero. After we do this some of the Pl\"ucker coordinates
vanish and one ends up in a subcell.  For example here we want to set
to zero $p_{156}$ e $p_{345}$ and obtain the subcell. This is done by
considering the two POs associated to these Pl\"ucker
coordinates. These two POs are represented in (\ref{sub36f})
\begin{equation}
\label{sub36f}
\includegraphics[width=9cm]{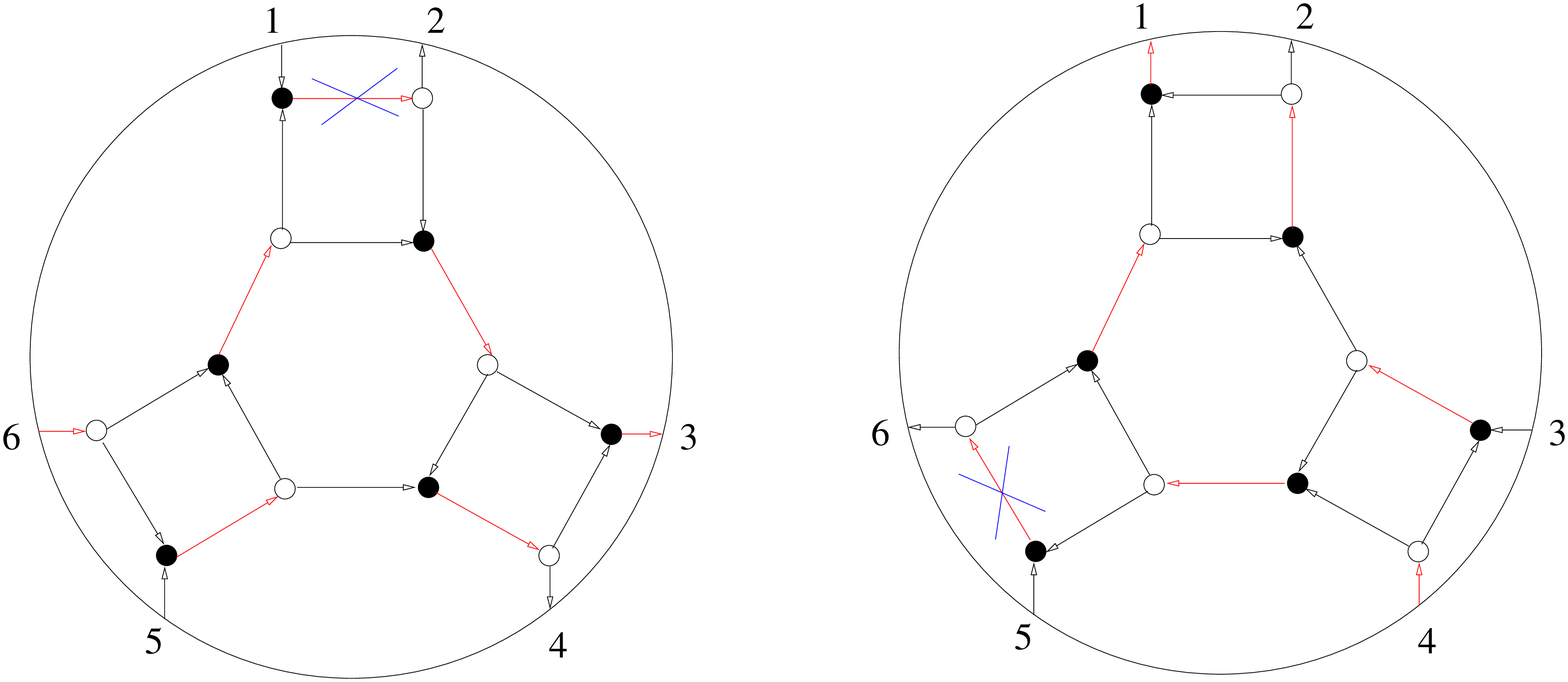}
\end{equation}
In (\ref{sub36f}) we also marked two edges in the top cell that have to be
erased to obtain the subcell that we want to analyze.  We eventually
obtain the diagram in (\ref{sub36})
\begin{equation}
\label{sub36}
\includegraphics[width=6cm]{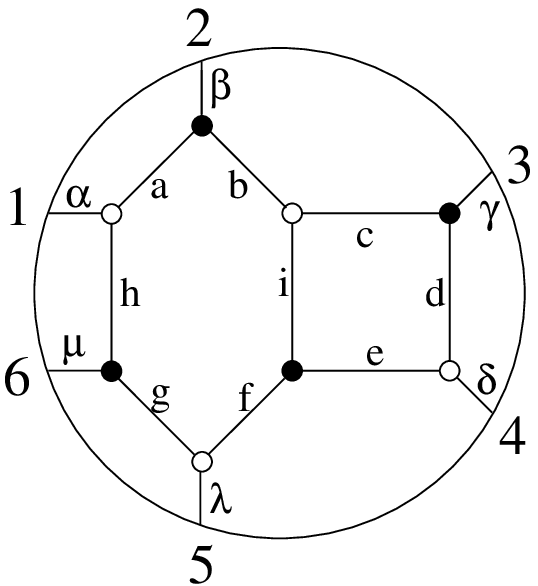}
\end{equation} 
referred in \cite{Franco:2012mm} as the hexagon-box
diagram.  
This on-shell diagram has $25$ PMs (and POs). They can be represented
graphically together as in (\ref{POPM36}).

The relations between the global coordinates on $Gr(3,6)^{tnn}$ and the PM
coordinates is
\begin{center}
  \begin{tabular}{cc|cc}
    Pl\"ucker &PM&Pl\"ucker &PM\\
    $p_{123}$&$\pi_1$&
    $p_{124}$&$\pi_2$\\
    $p_{125}$&$\pi_3$&
    $p_{126}$&$\pi_4$\\
    $p_{134}$&$\pi_5$&
    $p_{135}$&$\pi_6+\pi_7$\\
    $p_{136}$&$\pi_8+\pi_9$&
    $p_{145}$&$\pi_{10}$\\
    $p_{146}$&$\pi_{11}$&
    $p_{156}$&0\\
    $p_{234}$&$\pi_{12}$&
    $p_{235}$&$\pi_{13}+\pi_{14}$\\
    $p_{236}$&$\pi_{15}+\pi_{16}+\pi_{17}$&
    $p_{245}$&$\pi_{18}$\\
    $p_{246}$&$\pi_{19}+\pi_{20}$&
    $p_{256}$&$\pi_{21}$\\
    $p_{345}$&0&
    $p_{346}$&$\pi_{22}$\\
    $p_{356}$&$\pi_{23}+\pi_{24}$&
    $p_{456}$&$\pi_{25}$
  \end{tabular}
\end{center}
\newpage
\begin{equation}
  \label{POPM36}
  \includegraphics[width=16cm]{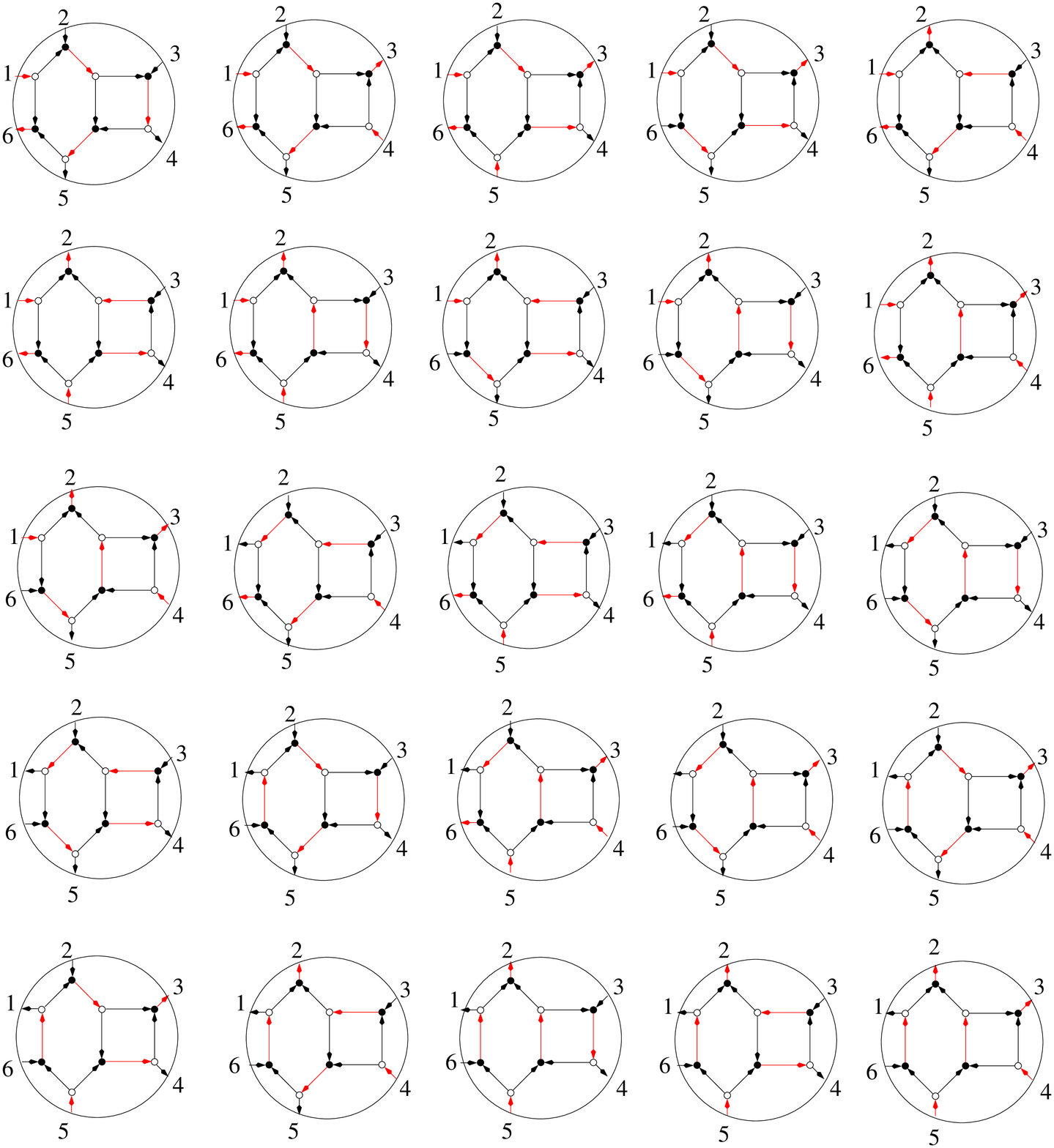}
\end{equation}
\newpage
From the defining equations of the toric variety associated to the
on-shell diagram, one can also obtain the Pl\"ucker relation among the
$p_I$ coordinates. This works as follows.  First one has to construct
the matching polytope.  This is obtained by specifying the matrix of
the PMs.  In every row one fixes an edge in the diagram in
(\ref{sub36}) while in each column one fixes a PM, ordered as in
(\ref{POPM36}).  Each entry is $1$ if the edge belongs to the
PM, $0$ otherwise
\begin{equation}
  P=
  \left(
    \begin{array}{c|ccccccccccccccccccccccccc}
      a& 1 & 1 & 1 & 1 & 1 & 1 & 1 & 1 & 1 & 1 & 1 & 0 & 0 & 0 & 0 & 0 & 0 & 0 & 0 & 0 & 0 & 0 & 0 & 0 & 0 \\
      b&0 & 0 & 0 & 0 & 1 & 1 & 1 & 1 & 1 & 1 & 1 & 0 & 0 & 0 & 0 & 0 & 0 & 0 & 0 & 0 & 0 & 1 & 1 & 1 & 1 \\
      c&0 & 1 & 1 & 1 & 0 & 0 & 0 & 0 & 0 & 1 & 1 & 0 & 0 & 0 & 0 & 0 & 0 & 1 & 1 & 1 & 1 & 0 & 0 & 0 & 1 \\
      d&0 & 1 & 0 & 0 & 1 & 0 & 0 & 0 & 0 & 1 & 1 & 1 & 0 & 0 & 0 & 0 & 0 & 1 & 1 & 1 & 0 & 1 & 0 & 0 & 1 \\
      e&0 & 0 & 1 & 0 & 0 & 1 & 1 & 0 & 0 & 1 & 0 & 0 & 1 & 1 & 0 & 0 & 0 & 1 & 0 & 0 & 1 & 0 & 1 & 1 & 1 \\
      f&1 & 1 & 1 & 0 & 1 & 1 & 1 & 0 & 0 & 1 & 0 & 1 & 1 & 1 & 0 & 0 & 0 & 1 & 0 & 0 & 0 & 0 & 0 & 0 & 0 \\
      g&0 & 0 & 0 & 0 & 0 & 0 & 0 & 0 & 0 & 0 & 0 & 0 & 0 & 0 & 0 & 0 & 1 & 0 & 0 & 1 & 1 & 1 & 1 & 1 & 1 \\
      h&0 & 0 & 0 & 0 & 0 & 0 & 0 & 0 & 0 & 0 & 0 & 1 & 1 & 1 & 1 & 1 & 0 & 1 & 1 & 0 & 0 & 0 & 0 & 0 & 0 \\
      i&1 & 1 & 1 & 1 & 0 & 0 & 0 & 0 & 0 & 0 & 0 & 0 & 0 & 0 & 0 & 0 & 1 & 0 & 0 & 1 & 1 & 0 & 0 & 0 & 0 \\
      \alpha&0 & 0 & 0 & 0 & 0 & 0 & 1 & 0 & 1 & 1 & 1 & 0 & 0 & 1 & 1 & 0 & 0 & 1 & 1 & 0 & 0 & 0 & 1 & 0 & 1 \\
      \beta& 1 & 1 & 0 & 0 & 1 & 0 & 0 & 0 & 0 & 0 & 0 & 1 & 0 & 0 & 0 & 0 & 1 & 0 & 0 & 1 & 0 & 1 & 0 & 0 & 0 \\
      \gamma& 0 & 0 & 0 & 1 & 0 & 0 & 0 & 1 & 1 & 0 & 1 & 0 & 0 & 0 & 1 & 1 & 0 & 0 & 1 & 0 & 0 & 0 & 0 & 0 & 0 \\
      \delta&0 & 0 & 0 & 0 & 1 & 1 & 0 & 1 & 0 & 0 & 0 & 1 & 1 & 0 & 0 & 1 & 0 & 0 & 0 & 0 & 0 & 1 & 0 & 1 & 0 \\
      \lambda&1 & 0 & 0 & 0 & 0 & 0 & 1 & 0 & 1 & 0 & 0 & 0 & 0 & 1 & 1 & 0 & 1 & 0 & 0 & 0 & 0 & 0 & 1 & 0 & 0 \\
      \mu&0 & 0 & 1 & 1 & 0 & 1 & 0 & 1 & 0 & 0 & 0 & 0 & 1 & 0 & 0 & 1 & 0 & 0 & 0 & 0 & 1 & 0 & 0 & 1 & 0 \\
    \end{array}
  \right)
\end{equation}
Not all the columns of this matrix are linearly independent.  Indeed the
kernel of this matrix is non empty and it identifies a set of linear
relations among the PMs $\sigma_i$ and hence among the vectors
$v_{\sigma_i}$ that define the matching polytope.  A possible basis of
relations is:
\begin{eqnarray}
&&
v_{\sigma_{1}}+v_{\sigma_{25}}=v_{\sigma_{7}}+v_{\sigma_{20}},\quad
v_{\sigma_{2}}+v_{\sigma_{24}}=v_{\sigma_{3}}+v_{\sigma_{22}},\quad
v_{\sigma_{1}}+v_{\sigma_{23}}=v_{\sigma_{7}}+v_{\sigma_{17}}
\nonumber \\
&&
v_{\sigma_{1}}+v_{\sigma_{21}}=v_{\sigma_{17}}+v_{\sigma_{3}},\quad
v_{\sigma_{1}}+v_{\sigma_{20}}=v_{\sigma_{17}}+v_{\sigma_{2}},\quad
v_{\sigma_{5}}+v_{\sigma_{19}}=v_{\sigma_{11}}+v_{\sigma_{12}}
\nonumber \\
&&
v_{\sigma_{2}}+v_{\sigma_{16}}=v_{\sigma_{4}}+v_{\sigma_{12}},\quad
v_{\sigma_{5}}+v_{\sigma_{15}}=v_{\sigma_{9}}+v_{\sigma_{12}},\quad
v_{\sigma_{5}}+v_{\sigma_{14}}=v_{\sigma_{7}}+v_{\sigma_{12}}
\nonumber \\
&&
v_{\sigma_{2}}+v_{\sigma_{13}}=v_{\sigma_{3}}+v_{\sigma_{12}},\quad
v_{\sigma_{3}}+v_{\sigma_{11}}=v_{\sigma_{4}}+v_{\sigma_{10}},\quad
v_{\sigma_{1}}+v_{\sigma_{10}}=v_{\sigma_{2}}+v_{\sigma_{7}}
\nonumber \\
&&
v_{\sigma_{3}}+v_{\sigma_{9}}=v_{\sigma_{4}}+v_{\sigma_{7}},\quad \,\,\,
v_{\sigma_{2}}+v_{\sigma_{8}}=v_{\sigma_{4}}+v_{\sigma_{5}},\quad\,\,\,\,
v_{\sigma_{2}}+v_{\sigma_{6}}=v_{\sigma_{3}}+v_{\sigma_{5}}
\nonumber \\
&&v_{\sigma_{1}}+v_{\sigma_{22}}=v_{\sigma_{17}}+v_{\sigma_{5}},\quad
v_{\sigma_{5}}+v_{\sigma_{18}}=v_{\sigma_{10}}+v_{\sigma_{12}}
\end{eqnarray}
These relations translate in product relations among the
real positive PM coordinates $\pi_i$ of the associated cell. The 
closure of these product relations (the toric ideal discussed in
section \ref{toriccell}) provide a set of polynomial equations for the
toric variety associated to the on-shell diagram.  From the relations
between $\pi_i$ and $p_I$ one obtains the Pl\"ucker relations among
the Pl\"ucker coordinates that parameterize the closure of this
subcell of $Gr(3,6)^{tnn}$.

\subsection{Toric ideals and cell decomposition}

In this section we would like to discuss two further examples:
$Gr(3,5)^{tnn}$ and $Gr(3,6)^{tnn}$. They are simple bipartite diagrams without
internal loops. The aim of the section is to provide some explicit
examples for the correspondence between the toric variety and the
Grassmannian sub variety associated to the same on-shell diagram. The
absence of internal loops allows a 1-1 identification between the
Pl\"ucker coordinates and the PM. In addition it is possible to represent
their toric diagram on a three dimensional lattice and their reduction
in subcells can be explicitly observed on the polytope.

We start by defining the bipartite diagrams, and then we derive the the
linear relations among the $v_{\sigma_i} $ variables defying the
associated polytope. We then describe the multiplicative relations
induced on the real positive PM coordinates $\pi_i$. In both the cases
we show that the original linear relations among the PM are not enough
to describe the projective toric variety as an intersection in the
projective space. Namely new relations are generated in the process to
take the closure of the map between the polytope and the $\pi_i$.  We
will then add the additional relations among the $\pi_i$ that allow to
generate the toric ideal in both cases, and we explicitly show the
toric interpretation of the associated scattering processes.  

In the two examples we will present in the following, the matching
polytope should be four dimensional: in both cases $G=4$. However the
base of the cones lies on an hyperplane of codimension one: in both
case $G-1=3$. This fact will allow us to explicitly draw the picture
of the polytope as the three dimensional base of the associated
cone. The fact that the vectors defining the matching polytope are
coplanar is a generic phenomenon, as observed in \cite{Franco:2012mm}, 
and proven in appendix \ref{COPLANAR}.

\subsubsection{Triangular basis} 
The first example is a subcell of $Gr(3,5)^{tnn}$ represented by the
toric diagram and the PMs in (\ref{torpm}).
\begin{equation}
\label{torpm}
\includegraphics[width=14cm]{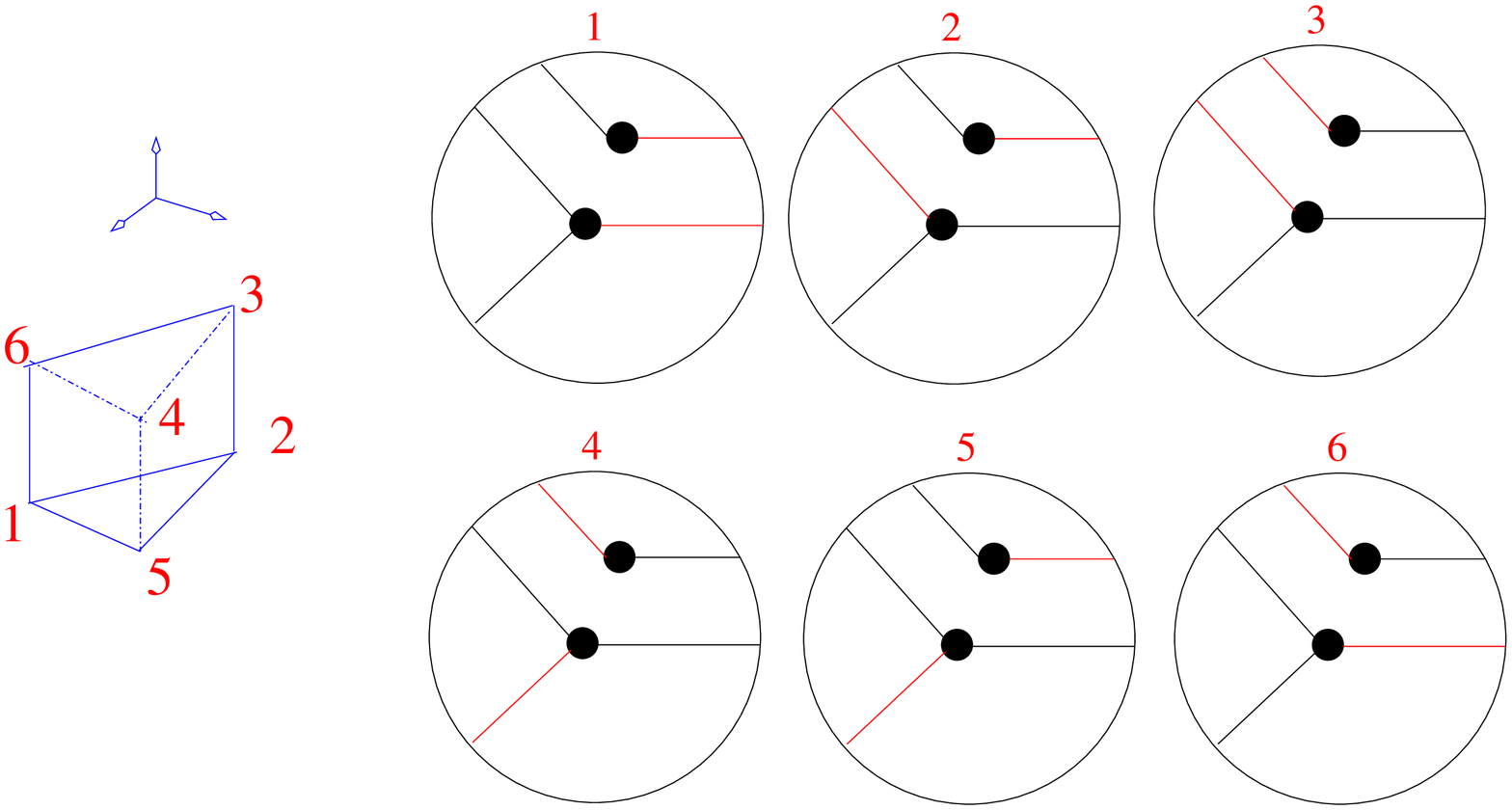}
\end{equation}
The toric diagram can be defined by two relations among the perfect
matchings. In terms of the vectors forming the toric diagram we have
\begin{equation}
\label{relpm}
v_{\sigma_1} + v_{\sigma_3} = v_{\sigma_2} +v_{\sigma_6}
\quad
\quad
v_{\sigma_2} + v_{\sigma_4} = v_{\sigma_3} + v_{\sigma_5}
\end{equation}
Even if there are two relations among the vectors in the toric diagram
the closure of the relations among the associated PM coordinates
$\pi_i$ requires an additional relation. Indeed the toric ideal is
generated by the three binomial relations:
\begin{equation}
\label{pirel35}
\pi_1  \pi_3 = \pi_2 \pi_6
\quad
\quad
\pi_2 \pi_4 =\pi_3  \pi_5 
\quad
\quad
\pi_5 \pi_6 = \pi_1 \pi_4
\end{equation}
The fact that there are three and not two relations in the ideal is
crucial in understanding the decomposition in subcells.

The decomposition can be performed by erasing the edges. By following
the procedure explained in \cite{ArkaniHamed:2012nw} one observes that
the subcells are in one to one correspondence with the five edges in
(\ref{torpm}).  By erasing one edge and the PMs containing
the edge itself one obtains the subcells.  One can observe from
(\ref{torpm}) that by erasing one edge the only PMs that survive are
always the ones associated to one of the five faces of the toric
diagram: $F_{1456}$, $F_{4523}$, $F_{2316}$, $F_{643}$ and $F_{125}$,
as expected.

One starts by sending to zero one $\pi_i$ in the three relations in
(\ref{pirel35}).  Some of these relations fail to be satisfied by
positive numbers, and one is forced to send to zero also another
$\pi_i$. One has to iterate the procedure until all the relations are
satisfied or vanishing and the remaining PM coordinates $\pi_i$ define
a codimension one polytope.

One can observe that this procedure gives the five faces in 
(\ref{torpm}).  In principle there could be internal faces in the
polytope and this procedure may also end on some of them, that are not
in principle associated to a subcell in the Grassmannian.  In the next
subsection we will study a case with internal faces.

\subsubsection{The cube} 

The second example we want to study in this section
is of a subcell of  $Gr(3,6)^{tnn}$.
The bipartite on diagram is shown in (\ref{bip36})
\begin{equation}
\label{bip36}
\includegraphics[width=4cm]{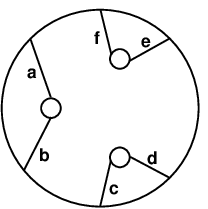}
\end{equation}
The PMs and the POs are shown in  (\ref{P3O6}) 
\begin{equation}
\label{P3O6}
\includegraphics[width=10cm]{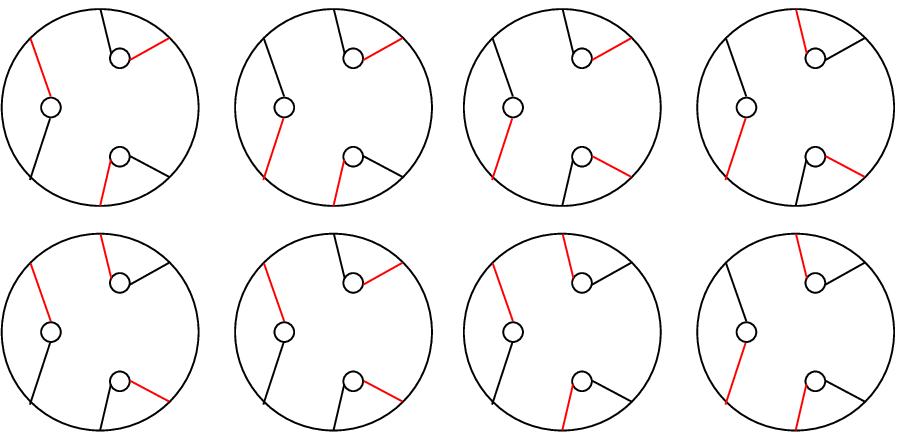}
\end{equation}
The matching polytope is
\begin{equation}
P=
\left(
\begin{array}{c|cccccccc}
a&1&0&0&0&1&1&1&0\\
b&0&1&1&1&0&0&0&1\\
c&1&1&0&0&0&0&1&1\\
d&0&0&1&1&1&1&0&0\\
e&1&1&1&0&0&1&0&0\\
f&0&0&0&1&1&0&1&1\\
\end{array}
\right)
\end{equation}
where in the rows are represented the edges and in the column the PMs.
The base of the toric diagram corresponding to the matching polytope can be represented as 
\begin{equation}
\includegraphics[width=10cm]{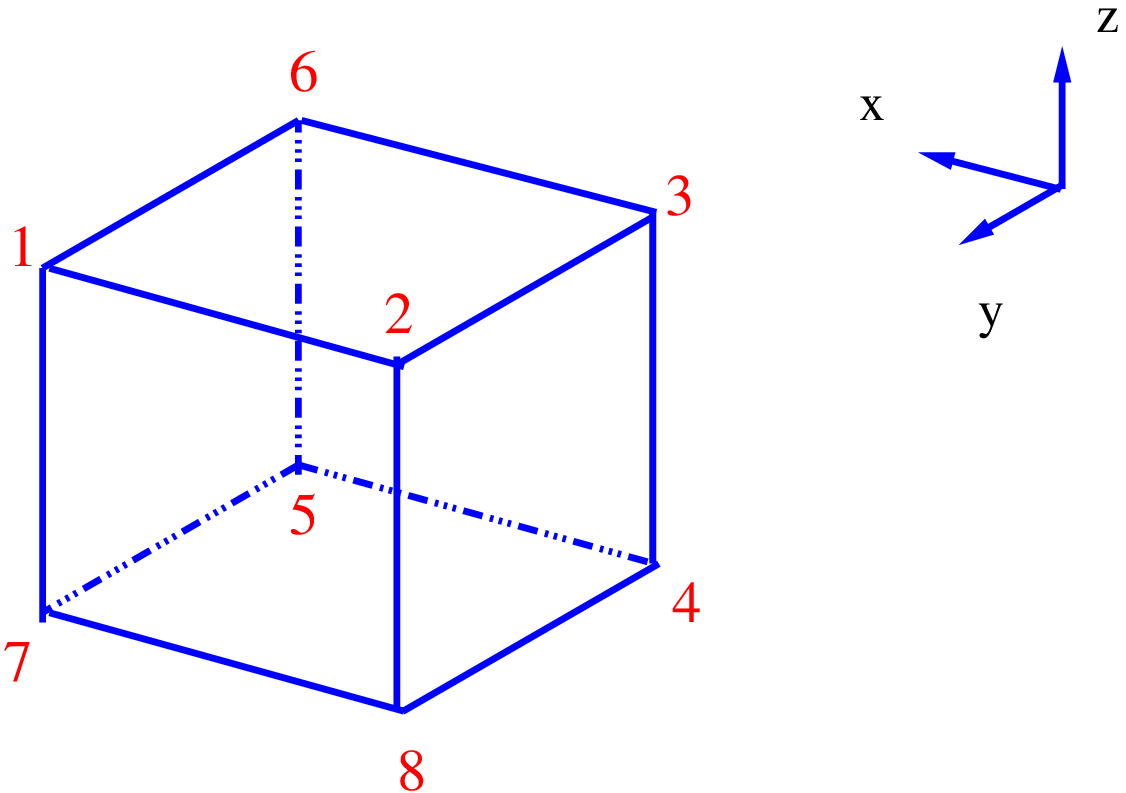}
\end{equation}
where the numbers are related to the PM ordered as in  (\ref{P3O6}).
The information contained in this diagram can be encoded in the
following relations among the vectors forming the diagram
\begin{equation}\label{v36rel}
v_{\sigma_8} + v_{\sigma_3} = v_{\sigma_2} + v_{\sigma_4}, \quad
v_{\sigma_7} + v_{\sigma_3} = v_{\sigma_1} + v_{\sigma_4}, \quad
v_{\sigma_6} + v_{\sigma_2} = v_{\sigma_1} + v_{\sigma_3}, \quad
v_{\sigma_5} + v_{\sigma_2} = v_{\sigma_1} + v_{\sigma_4} 
\end{equation}
As in the case studied above once we use the relations (\ref{v36rel})
to define the multiplicative relations among the real positive
coordinates $\pi_i$, and we then take the closure of this map,
allowing the $\pi_i$ to have zero values, we discover that the
original relations are not enough and new relations are generated by
the closure of the map.  In this case indeed the toric ideal is
generated by the following set of binomial quadratic relations that
define the projective toric variety associated to the on-shell
diagram:
\begin{eqnarray}
\label{relideal}
&&\pi_{4} \pi_{7} = \pi_{3} \pi_{8}, \quad
\pi_{3} \pi_{7} = \pi_{6} \pi_{8}, \quad
\pi_{2} \pi_{7} = \pi_{1} \pi_{8}, \quad
\pi_{3} \pi_{5} = \pi_{4} \pi_{6}, \quad
\pi_{2} \pi_{5} = \pi_{6} \pi_{8}\nonumber \\
&&
\pi_{1} \pi_{5} = \pi_{6} \pi_{7} , \quad
\pi_{2} \pi_{4} = \pi_{3} \pi_{8} , \quad
\pi_{1} \pi_{4} = \pi_{6} \pi_{8} , \quad
\pi_{1} \pi_{3} = \pi_{2} \pi_{6} 
\end{eqnarray}
As in the previous example the subcells are obtained either by erasing
an edge in the on-shell diagram or by setting to zero the PM
coordinates with the procedure explained above.

In this case there are internal faces in the matching polytope, but by
sending to zero the PM coordinates $\pi_i$ in a way that preserve the
relations (\ref{relideal}) one ends up only on the external faces that
describe the subcells.

\section{Dual Diagrams and Cluster Transformations}
\label{CLUSTER}

Until this point in this paper we have been discussing in detail the
relation between an on-shell diagram and a cell in the totally non
negative Grassmannian, the relation between an on-shell diagram and a
projective toric variety, and the relation between the projective
toric variety and the cell in the Grassmannian associated to the same
on-shell diagram.  However, as we have already commented at the very
beginning of section \ref{Sec:ReviewNima}, physically equivalent
scattering processes are not associated to a single on-shell diagram,
but to a class of on-shell diagrams: diagrams that can be mapped in
other diagrams with mergers and square moves are physically
equivalent, see section \ref{Sec:ReviewNima}.  At this point it is
maybe worthwhile to remind a series of correspondences proven in the
mathematical literature (see for example \cite{PostnikovLecture}):
decorated $(n,k)$ permutations are in one to one correspondence with
cells in the non negative Grassmannian $Gr(k,n)^{tnn}$; decorated
$(k,n)$ permutation are in one to one correspondence with moves
equivalent classes of on-shell diagrams.

Hence there should exist a well defined transformation among the
geometric objects that we associate to different on-shell diagrams
identified by the same permutation. Only the geometric structures that
are invariant under such transformation are physically relevant to
reconstruct the associated scattering process. This transformation is
called cluster transformation\footnote{The origin of this name comes
  from cluster algebra \cite{Fomin}.  Here we will not discuss any
  mathematical detail of cluster algebra and refer the reader to the
  original paper for reference.  Related discussions on the connection
  between the moves and the cluster transformations on bipartite
  diagrams on disks already appeared in
  \cite{Heckman:2012jh,Franco:2013pg}.}.

At the local level, namely considering strictly positive values for
the coordinates of the cells in Grassmannian and of the toric variety
associated to an on-shell diagram, an allowed move in the on-shell
diagram corresponds to a particular change of coordinates in the cell
in the Grassmannian and in the toric variety: the particular
transformation is called cluster transformation.  It works in the
following way.

A suitable set of coordinates for a cell in $Gr(k,n)^{tnn}$ is
provided by, for example, the face variables in the on-shell
diagram. Face  variables are indeed a good set of local coordinates
both for the cell in Grassmannian and for the positive toric
variety. To a single (reduced) on-shell diagram one can associate a
cell in Grassmannian by using for example the map (\ref{Talaskaa})
between the edge or face variables for the on-shell diagram and the
Pl\"ucker coordinates for the cell.  Acting with mergers and square
moves on the defined on-shell diagram one in general ends on another
on-shell diagram. As we have just reminded a permutation is in one to
one correspondence with a cell in Grassmannian and it is also in one
to one correspondence with a class of equivalence of on-shell
diagrams.  This fact implies that the two cells in the Grassmannian
provided for example by (\ref{Talaskaa}), associated to these two
on-shell diagrams linked by a set of moves, should be isomorphic and
hence the same cell.  Indeed the on-shell diagram obtained by moves is
in general different from the original one, for example the structure
of the faces and the number of edges and PMs change.  But with the use
of (\ref{Talaskaa}) one can assign to this new diagram the same set of
Pl\"ucker coordinates as the original one. This new diagram describes
indeed the same point in the cell in $Gr(k,n)^{tnn}$, but the
Pl\"ucker coordinates are written in terms of different combinations
of face variables. The transformation between the two sets of face
variables are called cluster transformation.

We would like to link this local discussion with the global picture of
the geometric structures associated to an on-shell diagram that we
have been exploring in the rest of the paper.  In this section indeed
we would like to better understand what happen to this global picture
under the allowed moves of the on-shell diagrams.

In particular to every on-shell diagram we have associated a matching
polytope $P$ and a non negative real toric variety $X_P^{\geq 0}$, and
a map from this variety to the closure of the cell in $Gr(k,n)^{tnn}$.
In this section we would like to provide a preliminary analysis of the
cluster transformation in this setup. 
A matching polytope $P$ is associated to every on-shell diagram.
Matching polytopes  associated to on-shell diagrams related
by mergers or square moves  are not in general  toric
equivalent: namely there could not exist an $SL(G,\mathbb{Z})$
transformation that maps the two polytopes one into the other, as
extensively studied in the case of bipartite diagrams on the torus in
\cite{Forcella:2008ng}.  However the projective toric varieties or
their subsets that map to the cell in the Grassmannian should be
related by some well defined transformation that locally reproduces the
cluster transformation, that we have just discussed, among the face
variables. In the following two examples we look for such
correspondence.

Observe that even if the matching polytopes of two on-shell
diagrams related by moves are not toric equivalent, their matroid
polytope should be toric equivalent modulo
$SL(G-G_{int},\mathbb{Z})$. Indeed the matroid polytope could also be
defined only in terms of matroid and Pl\"ucker coordinates that are
blind to the moves of the on-shell diagrams (see for example
\cite{PostnikovLecture}). Moreover the quiver field theory
interpretation of the on-shell diagrams provided in
\cite{Franco:2012mm,Franco:2012wv} supports this intuition.  Hence
locally on the toric projective variety the cluster transformations
should behave somehow in a similar fashion of the ones discussed in
\cite{Amariti:2012dd}: namely a toric transformation on the
matroid polytope and a non toric transformation on the fibers
of the matching polytope, seen as the total space with base the matroid polytope. 
Exploring this connections would be very
nice, but we leave it for future research. In this section empowered
by this intuition we would limit ourselves to the exploration of a
couple of examples.

Before starting our analysis we remind the reader how a cluster
transformation acts on the face variables in a square move.  The
cluster transformation associated to a square move can be locally
described on a piece of the bipartite on-shell diagram.  One considers
an internal face with four edges and represents the square move on the
face variable $f$ as in (\ref{Seedmut}) \footnote{The transformation
  in (\ref{Seedmut}) is apparently different from the one in
  (\ref{mosse}). The reason is that here we are representing the most
  general effect of the square move on the face variables while in
  section \ref{Sec:ReviewNima} we just represented a simplified version, valid for
  some sub-cases, like $Gr(2,4)^{tnn}$, for pedagocial reasons.}
\begin{equation}
\label{Seedmut}
\includegraphics[width=10cm]{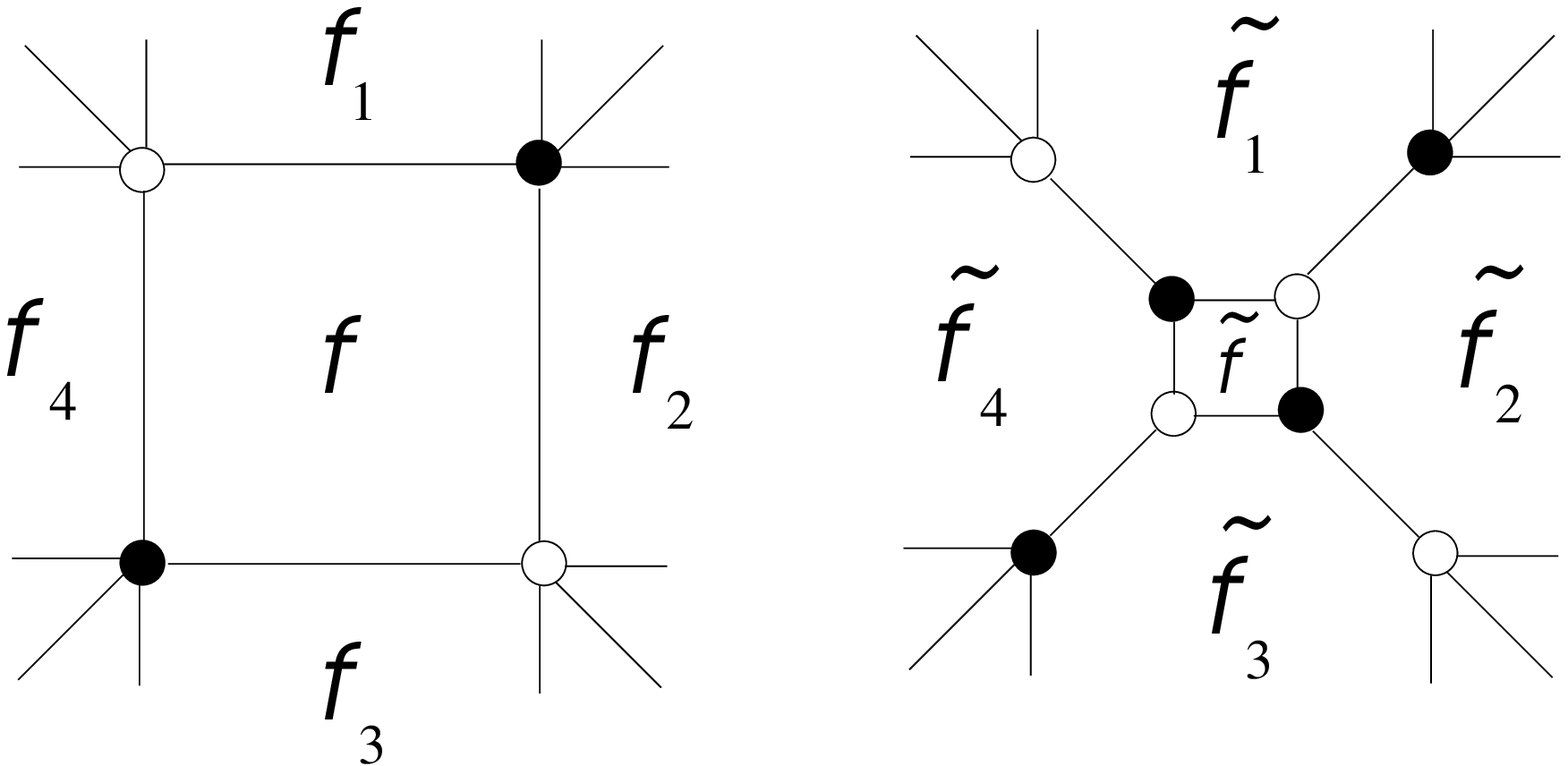}
\end{equation}
The corresponding cluster transformation is 
\begin{equation} \label{faces}
\tilde f = \frac{1}{f},\quad
\tilde f_1= \frac{f_1}{1+f^{-1}},\quad
\tilde f_2= f_2 (1+f),\quad
\tilde f_3= \frac{f_3}{1+f^{-1}},\quad
\tilde f_4= f_4 (1+f)
\end{equation}
In the rest of the section we 
parameterize the cells and their closure in the
totally non negative Grassmannian in terms of the PM coordinates
$\pi_i$ as explained in the previous sections. 
This allows us 
to derive the cluster transformations (\ref{faces}) and deduce
a set of "equivalence" transformations among the $\pi_i$ or subsets
of them.  Indeed we will observe that by imposing the equivalence of the
Pl\"ucker coordinates for moves related on-shell diagrams in terms of
the PM coordinates $\pi_i$ in the two phases we obtain a set of
equations among (sums of) $\pi_i$ in the two phases. By translating
these sets of equations in terms of the face variables in a local patch
we will obtain the cluster transformations (\ref{faces}).

We will now apply these ideas to a pair of examples. We start by
studying two diagrams associated to the top cell of $Gr(2,4)^{tnn}$.
Then we move to the subcell of $Gr(3,6)^{tnn}$ referred as the hexagon-
box diagram, and study its connection with its dual theory,
obtained by applying a square move on the box.

\subsection{Cluster transformations in $Gr(2,4)^{tnn}$}

In the first example we by study the cluster transformation associated
to a square move in the box diagram associated to the top cell of
$Gr(2,4)^{tnn}$, represented graphically in  (\ref{24top}).  By acting
with a square move on the internal face of the box diagram we generate
another box diagram with four external edges, but with the black and
white vertexes exchanged, see  (\ref{Seedmut2}).  We would like to study the cluster
transformation on the face variables by analyzing the relation between
the PM coordinates in the two phases and the associated Pl\"ucker coordinates.  We call
$\pi_i$ the PM coordinates associated to the original box on-shell diagram, and $\tilde
\pi_i$ the PM coordinates of the on-shell box diagram obtained after the square move. 
We keep the convention that $\pi_i$ and $\tilde \pi_i$ are ordered as the
associated PMs in  (\ref{Seedmut2}) from $1$ to $7$.
\begin{equation}
\label{Seedmut2}
\includegraphics[width=15cm]{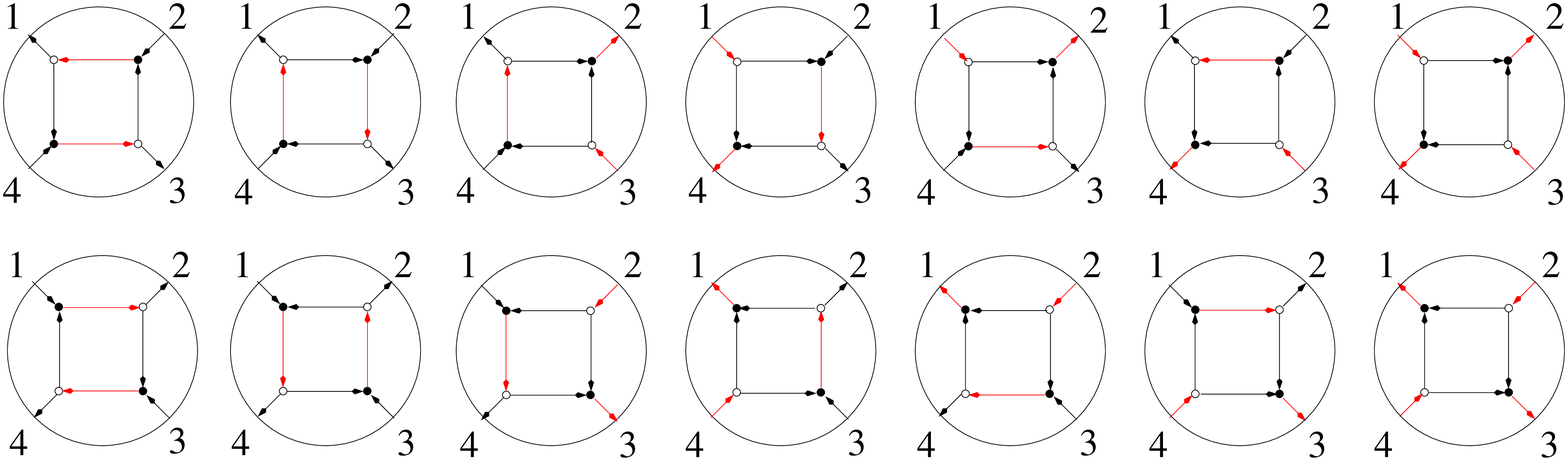}
\end{equation}
From the PMs one can read the POs and the corresponding source sets.
Using equation (\ref{NoiG}) one then obtains the Pl\"ucker coordinates in terms of combinations of the 
PM coordinates for the two phases.
If we impose that the two phases correspond to the same point in the top cell of $Gr(2,4)^{tnn}$, as it should be thanks to the mathematical 
theorems stated in the previous section, we obtain a set of $6$ equations relating the PMs in the two
diagrams.
Namely equating the (sums of) PMs coordinates associated to the same Pl\"ucker
coordinate in the two move equivalent diagrams we obtain: 
\begin{equation}
\label{Plupmpm}
\begin{array}{ccccc}
p_{12} &=& \pi_{4} &=& \tilde \pi_{3}\\
p_{13} &=& \pi_{7} &=& \tilde \pi_{1}+\tilde \pi_{2}\\
p_{14} &=& \pi_{5} &=& \tilde \pi_{6}\\
p_{23} &=& \pi_{6} &=& \tilde \pi_{5}\\
p_{24} &=& \pi_{1}+\pi_{2} &=& \tilde \pi_{7}\\
p_{34} &=& \pi_{3} &=& \tilde \pi_{4}
\end{array}
\end{equation}
This system of equations provides an equivalence relation for the
toric varieties associated to the same scattering process.  The next
step consists of expressing the face variables in the two phases in
terms of the PM coordinates. 
This is done by restricting to the totally positive part of the
closure of the cell, i.e. all the coordinates are strictly positive
and non zero.  We have
\begin{equation}
\label{facepm}
\begin{array}{ccccccc}
  f &=& \frac{\pi_1}{\pi_2}& \rightarrow& 
  \tilde f &=& \frac{\tilde \pi_2}{\tilde \pi_1} \\
  f_1 &=& \frac{\pi_5}{\pi_1}& \rightarrow& \tilde f_1 &=&  \frac{\tilde \pi_1}{\tilde \pi_5}  \\
  f_2 &=& \frac{\pi_2}{\pi_3} & \rightarrow& \tilde f_2 &=& \frac{\tilde \pi_3}{\tilde \pi_2}  \\
  f_3 &=& \frac{\pi_6}{\pi_1}& \rightarrow& \tilde f_3 &=& \frac{\tilde \pi_1}{\tilde \pi_6}  \\
  f_4 &=& \frac{\pi_2}{\pi_4}& \rightarrow& \tilde f_4 &= &\frac{\tilde \pi_4}{\tilde \pi_2}  
\end{array}
\end{equation}
The relations (\ref{facepm}) and (\ref{Plupmpm}) are still not enough
to obtain the cluster transformations on the local face variables. We
still have to consider the relations between the PMs coordinates in
both the diagrams. They are:
\begin{eqnarray}\label{rel2phas}
  \pi_7 \pi_1 =  \pi_5 \pi_6
  ,\quad 
  \pi_7 \pi_2 =  \pi_3 \pi_4
  ,\quad
  \tilde \pi_7 \tilde \pi_1 =  \tilde \pi_5 \tilde \pi_6
  ,\quad 
  \tilde \pi_7 \tilde \pi_2 =  \tilde \pi_3 \tilde \pi_4
  \nonumber 
\end{eqnarray}
Using equation (\ref{facepm}), (\ref{Plupmpm}) and (\ref{rel2phas}) we
can now reproduce the local cluster transformations among the face
variables for the two phases (\ref{faces}).  Explicitly we have:
\begin{equation}
\begin{array}{ccccccccccc}
  \tilde f &=& \frac{\tilde \pi_2}{\tilde \pi_1} &=& 
  \frac{\tilde \pi_3 \tilde \pi_4}{\tilde \pi_5 \tilde \pi_6 }
  &=& 
  \frac{ \pi_3 \pi_4}{ \pi_5  \pi_6 } &=& \frac{\pi_2}{\pi_1}&=&
  \frac{1}{f} \nonumber \\
  \tilde f_1 &=& \frac{\tilde \pi_1}{\tilde \pi_5}
  &=& \frac{ \tilde \pi_6 }{\tilde \pi_7} &=& \frac{\pi_5}{\pi_1+\pi_2}
  &&=&& f_1 (1+f^{-1})^{-1} \nonumber \\
  \tilde f_2 &=& \frac{\tilde \pi_3}{\tilde \pi_2}
  &=& \frac{ \tilde \pi_7 }{\tilde \pi_4} &=& \frac{\pi_1+\pi_2}{\pi_3}
  &&=&& f_2 (1+f) \nonumber \\
  \tilde f_3 &=& \frac{\tilde \pi_6}{\tilde \pi_1}
  &=& \frac{ \tilde \pi_5 }{\tilde \pi_7} &=& \frac{\pi_6}{\pi_1+\pi_2}
  &&=&& f_3 (1+f^{-1})^{-1} \nonumber \\
  \tilde f_4 &=& \frac{\tilde \pi_4}{\tilde \pi_2}
  &=& \frac{ \tilde \pi_7 }{\tilde \pi_3} &=& \frac{\pi_1+\pi_2}{\pi_4}
  &&=&& f_4 (1+f) \nonumber 
\end{array}
\end{equation}

\subsection{A subcell of $Gr(3,6)^{tnn}$}

The previous example mapped two diagrams with substantially the same
structure in terms of faces, edges and PMs.

Here we study a more complex example.  Namely
we apply the square move to the hexagon-box diagram, already
discussed in section \ref{Gr36cell} and shown in (\ref{sub36}),
associated to a subcell of $Gr(3,6)^{tnn}$.  In this case the dual
diagram, as already discussed in \cite{Franco:2012mm} is obtained by
acting on the internal box with a square move. The dual diagram is a
double box diagram with six external edges. This diagram has a
different structure of faces, edges and PM with respect to the
starting hexagon-box diagram and it is shown in (\ref{sub236})
\begin{equation}
\label{sub236}
\includegraphics[width=5cm]{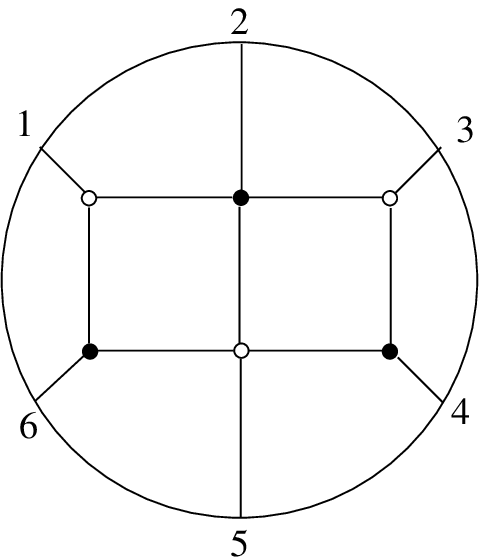}
\end{equation}
This diagram has $22$ PMs and POs represented in (\ref{PO36db}),
while the hexagon-box has $25$ PMs and POs as shown in (\ref{POPM36}).
\begin{equation}
\label{PO36db}
\includegraphics[width=16cm]{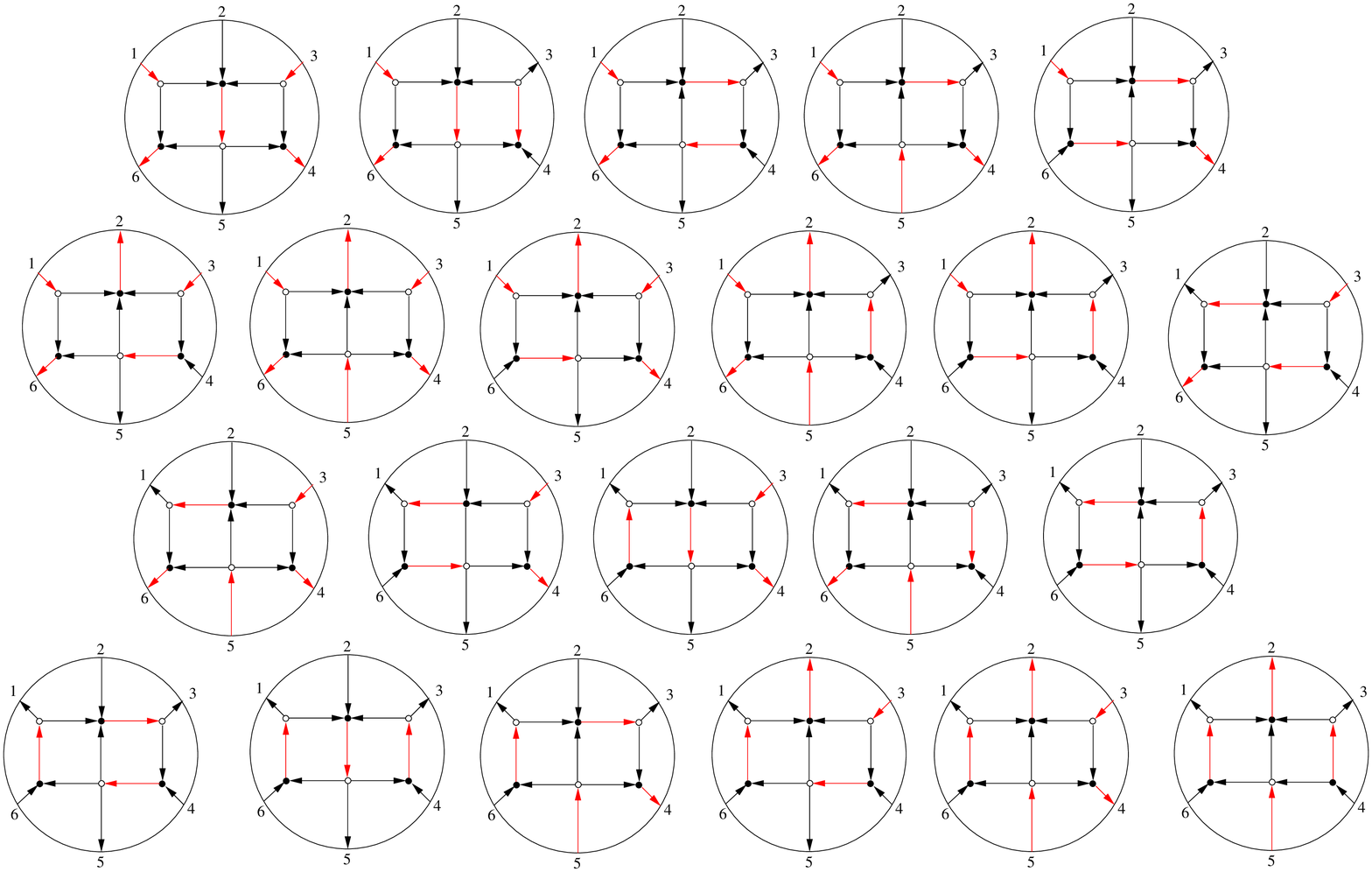}
\end{equation}
Each PM $\widetilde \sigma_i$ is associated to a vector $v_{\widetilde
  \sigma_i}$ in the matching polytope $\widetilde P$. These vectors satisfy
the set of linear relations:
\begin{eqnarray}
  &&
  v_{\widetilde \sigma_1}+  v_{\widetilde \sigma_{22}}=  v_{\widetilde \sigma_2}+  v_{\widetilde \sigma _{21}}
  ,\quad
  v_{\widetilde \sigma_3}+  v_{\widetilde \sigma_{21}}=  v_{\widetilde \sigma_4}+  v_{\widetilde \sigma _{20}}
  ,\quad
  v_{\widetilde \sigma_1}+  v_{\widetilde \sigma_{20}}=  v_{\widetilde \sigma_6}+  v_{\widetilde \sigma _{14}}
  \nonumber \\ &&
  v_{\widetilde \sigma_1}+  v_{\widetilde \sigma_{19}}=  v_{\widetilde \sigma_4}+ v_{\widetilde \sigma _{14}}
  ,\quad
  v_{\widetilde \sigma_1}+  v_{\widetilde \sigma_{18}}=  v_{\widetilde \sigma_2}+  v_{\widetilde \sigma _{14}}
  ,\quad
  v_{\widetilde \sigma_1}+  v_{\widetilde \sigma_{17}}=  v_{\widetilde \sigma_3}+  v_{\widetilde \sigma _{14}}
  \nonumber \\ &&
  v_{\widetilde \sigma_6}+  v_{\widetilde \sigma_{16}}=  v_{\widetilde \sigma_{10}}+  v_{\widetilde \sigma _{11}}
  ,\quad
  v_{\widetilde \sigma_6}+  v_{\widetilde \sigma_{15}}=  v_{\widetilde \sigma_9}+  v_{\widetilde \sigma_{11}}
  ,\quad
  v_{\widetilde \sigma_3}+  v_{\widetilde \sigma_{13}}=  v_{\widetilde \sigma_5}+  v_{\widetilde \sigma_{11}}
  \nonumber \\ &&
  v_{\widetilde \sigma_3}+  v_{\widetilde \sigma_{12}}=  v_{\widetilde \sigma_4}+  v_{\widetilde \sigma_{11}}
  ,\quad
  v_{\widetilde \sigma_1}+  v_{\widetilde \sigma_{10}}=  v_{\widetilde \sigma_2}+  v_{\widetilde \sigma_8}
  ,\quad
  v_{\widetilde \sigma_1}+  v_{\widetilde \sigma_9}=  v_{\widetilde \sigma_2}+  v_{\widetilde \sigma_7}
  \nonumber \\ &&
  v_{\widetilde \sigma_3}+  v_{\widetilde \sigma_8}=  v_{\widetilde \sigma_5}+  v_{\widetilde \sigma_6}
  ,\quad
  v_{\widetilde \sigma_3}+  v_{\widetilde \sigma_7}=  v_{\widetilde \sigma_4}+  v_{\widetilde \sigma_6}
\end{eqnarray}
As usual these relations translate in product relations among the $\widetilde \pi_i$
coordinates, and their closure generates the toric ideal.

Using equation (\ref{NoiG}) we obtain the relation between the global
coordinates on the Grassmannian and the PMs coordinates in these
square move equivalent on-shell diagrams:
\begin{equation}
\begin{array}{cc|cc|cc}
\text{ Pl\"ucker }&\text{PM}&\text{ Pl\"ucker}&\text{PM}
&\text{ Pl\"ucker}&\text{PM}\\
p_{123} &\tilde \pi_{1} & p_{124} &\tilde \pi_{2}+\tilde \pi_{3}&
p_{125} &\tilde \pi_{4} \\ p_{126} &\tilde \pi_{5}
p_{134} &\tilde \pi_{6} & p_{135} &\tilde \pi_{7}\\
p_{136} &\tilde \pi_{8} & p_{145} &\tilde \pi_{9}&
p_{146} &\tilde \pi_{10} \\ p_{234} &\tilde \pi_{11}& 
p_{235} &\tilde \pi_{12} & p_{236} &\tilde \pi_{13}+\tilde \pi_{14}\\
p_{245} &\tilde \pi_{15} & p_{246} &\tilde \pi_{16}+\tilde \pi_{17}+\tilde \pi_{18}&
p_{256} &\tilde \pi_{19} \\ p_{346} &\tilde \pi_{20}&
p_{356} &\tilde \pi_{21} & p_{456} &\tilde \pi_{22}\\
\end{array}
\end{equation}
Now we proceed as before. We impose the equivalence between the (sums
of) PM coordinates related to the same Pl\"ucker coordinate. These
equations provide an equivalence relation between the two toric
varieties associated to the same scattering process.  We can then
restrict the PM coordinates to have non zero positive values and
express the face variables for the two phases in terms of ratios of PM
coordinates.  We can then study the relation between the face
variables in the two phases as in the previous section.  We order the
faces as in (\ref{Seedmut}) and we finally we find the expected
cluster transformations among the face variables.  Explicitly we have
\footnote{ Note that we obtained (\ref{mapPM}) thanks to the
  relations among the $\pi_i$ and $\tilde \pi_i$ variables in the
  toric ideal.}
\begin{eqnarray}
\label{mapPM}
\tilde f &=& \frac{\tilde \pi_2}{\tilde \pi_3}=\frac{\tilde \pi_1 \tilde \pi_9}
{\tilde \pi_4 \tilde \pi_6} 
= \frac{ \pi_1 \pi_{10}}{ \pi_5 \pi_3}=
\frac{\pi_7}{ \pi_6}= \frac{1}{ f}
\nonumber \\
\tilde f_1&=&\frac{\tilde \pi_4}{\tilde \pi_7}
=\frac{\pi_3}{ \pi_6+\pi_7}
= \frac{ \pi_3}{ \pi_6}\left(\frac{1}{1+\frac{ \pi_7}
{ \pi_6}} \right)=
f_1 (1+ f^{-1})^{-1}
\nonumber \\
\tilde f_2&=&\frac{\tilde\pi_{21}}{\tilde\pi_{22}}=\frac{\pi_{23}
+ \pi_{24}}{ \pi_{25}}=
\frac{\pi_{23}}{ \pi_{25}} \left(
1+\frac{ \pi_{24}}{ \pi_{23}}
\right)= f_2 (1+ f)
\nonumber \\
\tilde f_3&=&\frac{\tilde\pi_6}{\tilde\pi_7}
=\frac{\pi_5}{ \pi_6+\pi_7}
= \frac{\pi_5}{ \pi_6}\left(\frac{1}{1+\frac{ \pi_7}
{ \pi_6}} \right)=
f_3 (1+ f^{-1})^{-1}
\nonumber \\
f_4 &=& \frac{\tilde \pi_{13}}{\tilde \pi_{14}}=
\frac{\tilde \pi_{5}\tilde \pi_{6} \tilde \pi_{12}}
{\tilde \pi_{1}\tilde \pi_{4}\tilde \pi_{20}}=
\frac{\pi_4 \pi_5 (\pi_{13}+\pi_{14})}
{\pi_1 \pi_3 \pi_{22}}=
\frac{\pi_5 }{\pi_{17}}
\frac{\pi_{13}+\pi_{14}}{\pi_{14}}
= f_4 (1+f)
\end{eqnarray}

\section{Conclusions}
\label{CONC}

In this paper we started the investigation of the toric geometry
underling the scattering amplitude processes in QFT. Our main aim was
to provide an alternative description with the hope that it could
bring some new insights in this fascinating field of research.  In
particular we explored the connection among the totally non negative
Grassmannian, its cell decomposition, bipartite diagrams on disks and
projective toric varieties. This was motivated by the connection
between the on-shell diagrams representing the scattering amplitudes
and the cell decomposition of the totally non negative Grassmannians.
We believe that an interesting result of this paper is a map from the
non negative real toric projective variety to the closure of the cells
in $Gr(k,n)^{tnn}$ associated to the same on-shell diagram, in terms
of the global embedding coordinates $\pi_i$ and $p_I$.  Thanks to this
parameterization we studied the cell decomposition of $Gr(k,n)^{tnn}$
in terms of toric geometry.  Eventually we obtained the cluster
transformations associated to the square moves in terms of toric
geometry. We feel that there are various possible interesting
developments to investigate.  The most interesting line of research
would be to investigate if the toric picture could help to reproduce
the computation of already known scattering amplitudes or provide new results in
the field.

Motivated by the connection between the top form 
of $Gr(k,n)^{tnn}$ and the scattering amplitudes, and the relation between the toric 
varieties and the Grassmannians, one may think to some
relation between the volumes of submanifolds of the toric variety and the Grassmannian
integrand.  For example, by interpreting the bipartite diagram on the
disk as a supersymmetric field theory, as in
\cite{Franco:2012mm,Xie:2012mr,Heckman:2012jh,Franco:2012wv,Franco:2013pg},
one can imagine a connection of the Grassmannian integrand with some toric
localization procedures \cite{Martelli:2006yb,Butti:2006au}.

We feel that another interesting topic regards the relation between the cluster
transformations and the square moves. Indeed this is reminiscent of
the gluing of the patches in the cluster Poisson varieties associated to the integrable systems 
living on the bipartite diagrams on
the torus \cite{Goncharov:2011hp,Eager:2011dp}. A first question is if a similar
construction may be valid in the case of the disk. In the case of the
torus it was observed that the master space, parameterized in terms of
all the possible $U(1)$ symmetries, mesonic and baryonic, anomalous
and non anomalous, played a prominent role in the integrability of the
dynamical system \cite{Amariti:2012dd}.  Here one may wonder about a similar
role played by the matching polytope and in particular if the Poisson structure could 
play a role in the scattering processes \cite{Golden:2013xva}.

It may be also interesting to study the connection between the
removability of the edges and the toric geometry.  Indeed here we
still kept the usual definition of removability: an edge is removable
if two zig-zag paths crossing on it do not cross on other edges.  In
principle it may be possible to understand the removability directly
in terms of the toric variety, 
the constraints among
the PMs and in terms of the relations between the PMs and the
Pl\"ucker coordinates. Indeed after erasing a removable edge one ends
up on a facet of the matching polytope, where some of the constraints
among the PMs and some of the Pl\"ucker coordinates vanish. This
suggests that the notion of removability maybe directly derived from the
reduction of the matching polytope on its facets.

We could expect that an interesting role could be
played by the specular duality \cite{Hanany:2012vc} in the QFT interpretation 
of the on-shell diagrams. The specular
duality, defined in the case of bipartite diagrams on the torus,
relates the bipartite diagram with the so called mirror dual. After an
untwisting procedure one construct the mirror diagram by assigning a
face to each zig-zag path. This diagram is called mirror because it
describes the IIA mirror geometry probed by a stack of D6 branes.  The
duality exchanges the baryonic and the mesonic symmetries, namely the
external and the internal points of the toric diagram of the moduli
space.  It follows that it does not preserve the moduli space but it
preserves the master space.  In the case of reflexive polygons
\cite{Hanany:2012hi} this duality maps theories on the torus with
theories on the torus. But in general, in non reflexive cases, it maps
theories on the torus with theories on higher genus Riemann surfaces.
It is natural to wonder if the specular duality plays any role in the
case of the disk, by preserving the matching polytope. And, maybe more
interesting, if it plays any role in the study of the non planar
limit.

Another quite unexplored line of research in the quiver field theory interpretation 
of bipartite diagrams on disks is the relation between the zig-zag paths and the matroid
polytope. Indeed in the case of bipartite diagrams on torus, 
the zig-zag paths are
related to the mesonic moduli space of the associates quiver field theories living on D3 
branes, through the $(p,q)$-web diagram.  The
mesonic moduli space is described by a toric diagram and the primitive normals
of this diagram are related to the difference among consecutive
external points, i.e. the zig-zag paths on the bipartite diagram.  On the
contrary one can build an inverse algorithm: from the toric diagram
one can reconstruct the zig-zag paths and hence 
the bipartite
diagram itself.  In the case of the disk, one could wonder if a
similar role is played by the zig-zag paths, but this interpretation is
still lacking.

We expect many exciting new lines of research both in the toric
geometric description of scattering amplitudes and in the development 
of BFT.

\section*{Acknowledgments}

We would like to thank Eduardo Conde Pena for his valuable
participation in the early stage of this project and for many
enlightening discussions.  We would also like to kindly
acknowledge F.~Cachazo, E.~C.~Pena, and A.~Zaffaroni, for their
relevant comments and insights on the draft of the present paper. We
are also grateful to R.~Argurio, M.~Bianchi, A.~Kashani-Poor 
for discussions and comments.  A.~A.~ is grateful to the
Institut de Physique Th\'eorique Philippe Meyer at the \'Ecole Normale
Sup\'erieure for fundings.  D.~F.~ is a ``Charg\'e de recherches" of the
Fonds de la Recherche Scientifique--F.R.S.-FNRS (Belgium), and his
research is supported by the F.R.S.-FNRS and partially by IISN -
Belgium (conventions 4.4511.06 and 4.4514.08), by the ``Communaut\'e
Fran\c{c}aise de Belgique" through the ARC program and by the ERC
through the ``SyDuGraM" Advanced Grant.

\appendix 

\section{The totally non negative Grassmannian}
\label{APPE}

In the main part of the paper we studied the decomposition of the
totally non negative Grassmannian in subspaces from the
toric geometry perspective.
The relation between toric diagrams and the
totally non negative Grassmannians has been 
first noticed in \cite{PostnikovCorto}.

Since we discussed many aspects of the Grassmannian that are not
commonly used in physics we found it useful to review some basic
definitions in this appendix.

We first review the basic definitions
of the totally non negative
Grassmannian and its cell decomposition.  We then describe its
connection to bipartite graphs  as originally
discussed in \cite{PostnikovLungo,PostnikovCorto}.

\subsection{Grassmannians}
\label{GRPL}

The Grassmannian $Gr(k,n)$, with integers $n > k > 0$, is the set
of $k$-dimensional linear subspaces in $\mathbb{R}^n$ passing
through the origin.  Every point can be represented by a full rank $k \times n$
matrix $A$ modulo $GL(k)$ transformation.
There exists a natural embedding of this space in a big enough projective space. 
First define $[n]\equiv \{1,\dots,n\}$, and then a subset $I$ of $k$
elements of $[n]$.
Then define $A_I$ the $k\times k$ submatrix of $A$ where the
columns are labeled by $I$, there are $n\choose k$ of them. Then consider the determinants of this sub matrix. 
The Pl\"ucker  coordinates are the determinants  $p_I(A)=\det A_I$. 
These are projective coordinates, because the action of  $GL(k)$ multiplies all the coordinates by a common 
factor.
the map $A\rightarrow A_I$ induces the Pl\"ucker  embedding of the real
Grassmannian in $\mathbb{RP}^{{n\choose k}-1}$. These coordinates
are constrained by quadratic polynomial relations, called Pl\"ucker  relations.

For example we consider $Gr(2,4)$. In this case the matrix A is $4 \times 2$, and it can be represented in a local patch, thanks to the action 
of a $GL(2)$ transformation, as:  
\begin{equation}
A = 
\left(
\begin{array}{cccc}
1 & a_{12} & 0 & -a_{14}\\ 
0 & a_{32} & 1 & a_{34}
\end{array}
\right)
\end{equation}
The minors ($\Delta_J(A)= p_J(A)/ p_{13}(A)$, with $p_{13}(A)=1$ in this patch) are
\begin{eqnarray}
\Delta_{12} = a_{32},\quad 
\Delta_{13} = 1,\quad
\Delta_{14} = a_{34},\quad
\Delta_{23} = a_{12}, \quad
\Delta_{24} = a_{12} \,a_{34}+a_{14}\,a_{32},\quad 
\Delta_{34} = a_{14}\nonumber\\
\end{eqnarray}
and they are related by
\begin{equation}
\Delta_{13}\Delta_{24} = \Delta_{12}\Delta_{34} + \Delta_{14}\Delta_{23}
\end{equation}
There is a subset of $Gr(k,n)$ called the totally non negative
Grassmannian $Gr(k,n)^{tnn}$ represented by the matrix $A$ such that
all the Pl\"ucker  coordinates (the minors) are non-negative.  
This space admits a nice quite simple cells decomposition. Indeed a simple way to decompose the totally non 
negative Grassmannian in
cells is the so called positroid stratification \footnote{It is maybe
  important to underline that the positroid stratification is different
  from the usual Schubert cell decomposition of the Grassmannian as
  explained for example in \cite{PostnikovLecture}.}

The positroid stratification can be defined as follows.  First one
defines the matroid.  A matroid $\mathcal{M}$ of rank $k$ on the set
$[n]$ is a non empty collection of $k$-element subset in $[n]$ such
that for any $I$ and $J$ in $\mathcal{M}$ and $i\in I$ there is $j \in
J$ such that $(I\setminus i) \bigcup j \in \mathcal{M}$.  A positive
cell $C_{\mathcal{M}}$ is the subset of elements in $Gr(k,n)^{tnn}$
represented by all the matrices $A$ such that
\begin{equation}
\left\{
\begin{array}{rl}
p_{I}>0 &I\in \mathcal{M}\\
p_I=0& otherwise
\end{array}
\right.
\end{equation}
If the cell is non empty the associated matroid is called positroid.
$Gr(k,n)^{tnn}$ is a disjoint union of its cells.
This cell decomposition of the totally non negative Grassmannian
is crucial for the interpretation of planar bipartite
graphs on disks in terms of scattering amplitudes.

\subsection{Bipartite Diagrams and $Gr(k,n)^{tnn}$}
\label{bipGR}

A bipartite diagram on a disk can be connected with a subset in the
totally non negative Grassmannian by making use of the
\emph{boundary measurement}.

Consider a PO in a graph with $n$ external vertices. This PO defines a
set of $k$ sources, edges directed from the boundary to the internal,
and $n-k$ sinks.  For a given graph $k$ is independent from the
choice of the PO. Assign a real positive variable $x_e$ to every edge in the diagram.
If $i$ is in the source set and $j$ in the sink set one defines the
boundary measurement as a series in $x_e$. Formally
it is
\begin{equation}
M_{i,j} \equiv
\sum_{P(i\rightarrow j)} (-1)^{wind(P)} x^P
\end{equation}
where $P$ are the directed paths from $i$ to $j$.  The monomials $x^P$
are the ratios of the edge variables in every path: in the numerator
there are edges directed from the white to the black vertices in
the path and in the denominator edges from black to white.  The
winding index is the number of self intersections of a path modulo
$2$.  If every edge variable is a positive real number the measurement is
a positive function.

For a fixed PO, say $\mathcal{O}$, if one names the source set
$I_\mathcal{O}$, one can construct a $k \times n$ matrix
$A_\mathcal{O}$ with two properties.  First the source set identifies
a $k \times k$ minor corresponding to the identity.  Second for any
$i$ in the source set and $j$ in the sink set
$\Delta_{(I_{\mathcal{O}}\setminus i) \bigcup
  j}(A_{\mathcal{O}})=M_{ij}$.  This construction defines the boundary
measurement map $Meas_{\mathcal{G}}$  that maps a perfectly oriented bipartite
graph to a subset in the Grassmannian.  Actually every PO underlies the
same subset in $Gr(k,n)$, i.e. the boundary measurement $Meas_{\mathcal{G}}$ does
not depend on the PO.  The edge set is redundant, indeed there are
gauge symmetries that one can use to reduce the boundary map
$Meas_G:R^{E(\mathcal{G})}_{>0}\rightarrow Gr(k,n)^{tnn}$ to
$\widetilde{Meas}_{\mathcal{G}}:R^{E(\mathcal{G})-V(\mathcal{G})}_{>0}\rightarrow Gr(k,n)^{tnn}$,
or
$\widetilde{Meas}_{\mathcal{G}}:R^{G(\mathcal{G})-1}_{>0}\rightarrow Gr(k,n)^{tnn}$,
where $E$, $V$ and $G$ are respectively the number of
edges, vertices and faces in the diagram.

\subsubsection{Cell decomposition and bipartite diagrams}
\label{SubSec:CellDec}

In the study of the scattering amplitudes it has been shown that the 
decomposition in positive cells of $Gr(k,n)^{tnn}$ and the closure of 
these cells are the relevant geometric objects underling the scattering process.
Here we briefly review this decomposition in terms of bipartite graphs.

One can associate to a bipartite graph $\mathcal{G}$ with a perfect orientation
$\mathcal{O}$ a cell in $Gr(k,n)^{tnn}$. This cell is parameterized
by the Pl\"ucker coordinates as in \cite{Talaska}, such that the
Pl\"ucker  coordinate associated to the source set of $\mathcal{O}$ is set to 1.
This is a local parameterization but one can move from one patch to
another with a basis exchange: take an oriented paths in the PO and
switch the orientation of every edge in the path. 

The closure of the positive cell $C_\mathcal{M}$ associated to the
graph $\mathcal{G}$ is obtained gluing together all these patches:
\begin{equation}
\overline{C_\mathcal{M}} = \bigcup_{(\mathcal{G},\mathcal{O})}
\overline{Meas}_{(\mathcal{G},\mathcal{O})} (\mathbb{R}_{\geq 0} ^{E(\mathcal{G})})
\end{equation}
The closure of every boundary measurement
$\overline{Meas}_{(\mathcal{G},\mathcal{O})} (\mathbb{R}_{\geq 0} ^{E(\mathcal{G})})$ is
obtained from a perfectly oriented graph by allowing zero values for some edge(s).
The union of all the measurement of those perfectly oriented subgraphs
is the closure of the measurement of $(\mathcal{G},\mathcal{O})$ and correspond
to the closure of the cell $\overline{C}_{\mathcal{M}}$.

The edges that can be removed from $\mathcal{G}$ to construct a subgraph are
named \emph{removable}. An edge is removable if the two zig-zag paths
that intersect on it do not intersect on any other edge in the graph.
There is a 1-1 correspondence between the removable edges and the
$(d-1)$-subcells covered by a $d$-cell.  One can show that by further
reducing the subgraph one arrives at the lower dimensional cells. This
procedure is called the cell decomposition.
The closure of the cell is the union of all its subcells.

\section{Some further considerations}
\label{COPLANAR}
 
In this appendix we group a couple of observations that come from our analysis.

First of all we argue that the vectors defining matching polytope are
always coplanar and that as a consequences the matching polytope and
the matroid polytope can be represented as polytopes in $G-1$ and
$G-G_{int} - 1$ dimensions respectively.  This coplanarity condition
has already been observed in \cite{Franco:2012mm} in all the examples
that have been considered so far.  We will prove this result in this
section.  This notion of coplanarity is well known in the case of
bipartite diagrams on tori. Indeed in that case the toric diagram of
the master space (the analogous of the matching polytope that we are
considering here) is formed by vectors that lie on the same plane, as
proven in \cite{Forcella:2008bb}. This condition has been used to
prove that the master space is a CY variety.  Here we adapt the proof
of the coplanarity of \cite{Forcella:2008bb} to our case.

We start by observing that the PMs on the disk can be divided into two
classes of edges
\begin{itemize}
\item {\bf Internal:} edges connecting the  vertices of the
  bipartite diagram.
\item {\bf External:} edges connecting a vertex in the
  bipartite diagram with the boundary.
\end{itemize}
Every PM covers all the  vertices in the diagram.  By
assigning a weight $1$ to the internal edges and $1/2$ to the external
ones the weighted sum of the edges in every PM
corresponds to the number of white (or black) vertices in the
graph.  This observation can be reformulated mathematically as
follows.  We first define the PM matrix $P$ as
$P=\left( \begin{array}{c}P_{i}\\P_{e}\end{array}\right)$ where $i$
runs over the internal edges and $e$ over the external ones.  Then we
have
\begin{equation}
(1,\dots,1)_E \cdot
\left( \begin{array}{cc}
Id_{i} & 0 \\
0 & \frac{1}{2} Id_{e}
\end{array}
\right)
\cdot P = 
\frac{V}{2} (1,\dots,1)_{c}
\end{equation}
From now on the proof goes along the lines of the one in
\cite{Forcella:2008bb}.  
By defining the matrix $Q$ as the kernel of $P$: $P \cdot Q=0$  we
have
\begin{equation}
(1,\dots,1)_c \cdot Q = 0
\end{equation} 
i.e. the rows of $Q^t$ are traceless, and all the vectors are coplanar.

The second thing we want to observe in this section is that the toric
ideal of a toric variety associated to an on-shell diagram on a disk
could be always generated by a set of quadratic binomials.  This
conclusion is obtained thanks to equation (\ref{NoiG}), that provides a
linear map between the PM coordinates and the Pl\"ucker coordinates,
and the fact that the Pl\"ucker embedding of the Grassmannian in the
projective space is provided by a set of quadratic polynomials.  The
fact that the equations defining this kind of toric varieties are
quadratic equations seems to be a peculiarity of this class of toric
varieties associated to on-shell diagram on a disk.  Indeed it is not
true for  general toric varieties, and for example it is not in
general true for the toric varieties associated to a bipartite diagram
on a torus.

\section{List of Notations}

\begin{center}
\begin{tabular}{ll}
  $ \sigma_i$ & Perfect matching variables with linear relations among them. \\
  $v_{\sigma_i}$ & Vectors associated to the PM. \\
  $\pi_i $ & Perfect matching coordinates with product relations among them.\\ 
  $ Gr(k,n)$ & Grassmannian manifold of $k$-dimensional linear subspaces of 
  $\mathbb{R}^n$.\\
  $ Gr(k,n)^{tnn}$ & Totally non negative Grassmannian. \\
  $\mathcal{C}_{M}$& (Totally positive) cell in the 
  totally non negative Grassmannian \\
  $ \mathcal{G,H}$ & Bipartite graphs\\
  PM& Perfect matching\\
  PO& Perfect orientation\\
  $c$&Number of PMs and POs.\\
  $ G$ & Number of faces of a bipartite graph. \\
  $ E$ & Number of external and internal edges in a bipartite graph.\\
  $ V$ & Number of nodes in  a bipartite graph.\\
  $ F$ & Flow in a perfect orientation of a bipartite graph.\\
  $ w(F)$ & Weight of a flow.\\
  $f_i$ & Face variables, such that $\prod_{i=1}^{G} f_i=1$.\\ 
  $ \mathcal{M}$ & Matroid.\\
  $ P$& Matching polytope, convex hull of the $c$ vectors $v_{\sigma}$.\\
  $ Q$ & Matroid polytope.\\
  $ I=\{i_1,\dots,i_k\}$& Set of indices enumerating $k$ external lines of a 
  bipartite diagram.\\
  $ A$ & Matrix representing a point in $Gr(k,n)^{tnn}$.\\
  $p_I$ & Global  Pl\"ucker coordinates.\\
  $\Delta_I$& Local  Pl\"ucker coordinates.\\
  $X_P$ &Projective Toric Variety.\\
  $X_P(\mathbb{R})$ &Real  part of  $X_P$.\\
  $X_P^{\geq 0}$ &Totally non negative part of $X_P$.\\ 
  $X_P^{> 0}$ & Positive part of  $X_P$.\\ 
  $ I_p$& Toric ideal of  $X_P$.\\
\end{tabular}
\end{center}

\end{document}